\documentstyle[12pt,epsf,rotate]{article}

\catcode`\@=11

\@addtoreset{equation}{section}

\def\@normalsize{\@setsize\normalsize{15pt}\xiipt\@xiipt
           \abovedisplayskip 14pt plus3pt minus3pt%
           \belowdisplayskip \abovedisplayskip
           \abovedisplayshortskip  \z@ plus3pt%
           \belowdisplayshortskip  7pt plus3.5pt minus0pt}
\def\small{\@setsize\small{13.6pt}\xipt\@xipt
           \abovedisplayskip 16pt plus3pt minus3pt%
           \belowdisplayskip \abovedisplayskip
           \abovedisplayshortskip  \z@ plus3pt%
           \belowdisplayshortskip  7pt plus3.5pt minus0pt
             \def\@listi{\parsep 4.5pt plus 2pt minus 1pt
             \itemsep \parsep
             \topsep 9pt plus 3pt minus 3pt}}
\def\underline#1{\relax\ifmmode\@@underline#1\else
        $\@@underline{\hbox{#1}}$\relax\fi}

\@twosidetrue

\long\def\@makecaption#1#2{
 \vskip 10pt \setbox\@tempboxa\hbox{#1: #2}
 \ifdim \wd\@tempboxa >\hsize #1: #2\par \else \hbox
        to\hsize{\box\@tempboxa\hfil} \fi}

\catcode`\@=12

\relax

       \def\dash{\hbox{---}}
\def\tilde{\widetilde}

\def\tr{\mathop{\rm tr}}           
\def\Re{\mathop{\rm Re}}           \def\Im{\mathop{\rm Im}}

\def\gta{\mathrel{\vcenter{\hbox{$>$}\nointerlineskip\hbox{$\sim$}}}}
\def\ltap{\raisebox{-.4ex}{\rlap{$\sim$}} \raisebox{.4ex}{$<$}}
\def\gtap{\raisebox{-.4ex}{\rlap{$\sim$}} \raisebox{.4ex}{$>$}}
\def\bar{\overline}

\def\beq{\begin{equation}}      \def\eeq{\end{equation}}
\def\bea{\begin{eqnarray}}      \def\eea{\end{eqnarray}}
\def\bq{\begin{quote}}          \def\eq{\end{quote}}

\relax

\renewcommand{\(}{\left(}
\renewcommand{\)}{\right)}
\renewcommand{\[}{\left[}
\renewcommand{\]}{\right]}
\renewcommand{\bar}[1]{\overline{#1}}

\def\SMo{Standard Model}
\def\SM{\SMo\ } \def\SMp{\SMo.\ } \def\SMk{\SMo,\ }
\def\EW{electro--weak } 
\def\GBo{Goldstone Boson}
    
\def\GBs{\GBo s\ } \def\GBsp{\GBo s.\ }
\def\RG{renormalization group }

\def\sixtpi{\frac{1}{16\pi^{2}}}
\def\SIXTPI{16\pi^{2}}
\def\lnlambda{\ln \left( \frac{\Lambda^{2}}{\mu^{2}} \right)}
\def\vr2{v_{R}^{2}}
\def\kap2{\kappa^{2}}
\def\kapp2{\kappa'^{2}}
\def\hvr{\hat{v}_{R}}
\def\hvr2{\hat{v}_{R}^{2}}
\def\hkap2{\hat{\kappa}^{2}}
\def\hkapp2{\hat{\kappa'}^{2}}

\begin{document}


\begin{titlepage}
 \renewcommand{\baselinestretch}{1}
 \renewcommand{\thefootnote}{\alph{footnote}}
 \thispagestyle{empty}

\begin{flushright}
  {\hfill  IC/95/126}\\
  {TUM--HEP--222/95}\\
  {MPI--PhT/95--70}\\
  {FTUV/95-36,~IFIC/95-38}\\
  {September 1995}
\end{flushright}

\vspace*{0.5cm}
{\begin{center} {\LARGE\bf
                Dynamical~Left--Right~Symmetry~Breaking}
                \end{center}  }
\vspace*{.8cm}
{\begin{center} {\large{\sc
                Eugeni Akhmedov\footnote{On leave from National
                Research Center ``Kurchatov Institute'', 123182
                Moscow, Russia}\footnote{\makebox[1.cm]{Email:}
                akhmedov@tsmi19.sissa.it},
                Manfred Lindner\footnote{\makebox[1.cm]{Email:}
                Manfred.Lindner@Physik.TU--Muenchen.DE},
                Erhard Schnapka\footnote{\makebox[1.cm]{Email:}
                Erhard.Schnapka@Physik.TU--Muenchen.DE}}\\
                \vspace{0.2cm}
                and
                {\sc
                Jose W. F. Valle\footnote{\makebox[1.cm]{Email:}
                valle@flamenco.ific.uv.es}}}
\end{center}}
\vspace*{0cm}
{\it
\begin{center}  \footnotemark[2]International Centre for
                Theoretical Physics,\\ Strada Costiera 11,
                I--34100 Trieste, ITALY

                \vskip .3cm

                \footnotemark[3]$\ \!\!^,$\footnotemark[4]
                Institut f\"ur Theoretische Physik,
                Technische Universit\"at M\"unchen,\\
                James--Franck--Strasse, D--85748 Garching, GERMANY

                \vskip .3cm

                \footnotemark[5]Instituto de F\'{\i}sica
                Corpuscular - C.S.I.C.\\
                Departament de F\'{\i}sica Te\`orica,
                Universitat de Val\`encia\\
                46100 Burjassot, Val\`encia, SPAIN

\end{center} }
\vspace*{0.5cm}
{\Large \bf
\begin{center} Abstract  \end{center}  }
We study a left--right symmetric model which contains only
elementary gauge boson and fermion fields and no scalars. The
phenomenologically required symmetry breaking emerges dynamically
leading to a composite Higgs sector with a renormalizable effective
Lagrangian. We discuss the pattern of symmetry breaking and
phenomenological consequences of this scenario. It is shown that
a viable top quark mass can be achieved for the ratio of the VEVs
of the bi--doublet $\tan\beta\equiv\kappa/\kappa'\simeq\;$1.3--4.
For a theoretically plausible choice of the parameters the
right--handed scale can be as low as $\sim 20~TeV$; in this case
one expects several intermediate and low--scale scalars in addition
to the \SM Higgs boson. These may lead to observable lepton flavour
violation effects including $\mu\to e\gamma$ decay with the rate
close to its present experimental upper bound.

\renewcommand{\baselinestretch}{1.2}

\end{titlepage}

\newpage

\renewcommand{\thefootnote}{\arabic{footnote}}
\setcounter{footnote}{0}


\section*{Introduction}

Left--right symmetric extensions~\cite{LR1,LR2,LR3} of the \SM are
very attractive. In this class of models parity is unbroken at high
energies and its non--conservation at low energies occurs through
a spontaneous symmetry breakdown mechanism. In addition, it is
remarkable how the known fermions fit very economically and
symmetrically into representations of the underlying gauge group
$SU(2)_L\times SU(2)_R\times U(1)_{B-L}$. For phenomenological
reasons the $SU(2)_R\times U(1)_{B-L}$ symmetry must first be
broken above a few $TeV$ to $U(1)_Y$ and in a second step the
left--over \SM gauge group $SU(2)_L\times U(1)_Y$ is broken as usual
to $U(1)_{em}$. In conventional left--right (LR) models this desired
symmetry breaking sequence is realized in analogy to the \SM with
the help of scalar particles and the Higgs mechanism. Therefore
suitable scalars are introduced and a renormalizable potential
with the required vacuum expectation values (VEV's) is chosen.
But like in the \SM it is desirable to motivate the existence of
such scalars by either adding supersymmetry\footnote{Some examples
are given in Ref.~\cite{LRsusy}.} or to view the Higgs sector as
an effective description of some dynamical symmetry breaking
mechanism in analogy to superconductivity. We will pursue here the
second possibility since the combination of the correct dynamical
LR symmetry breaking sequence with the usual nice features of LR
models appears especially attractive. Composite Higgs scenarios are
moreover more economical and explain nicely why certain Yukawa
couplings are of order unity and how Higgs and fermion masses are
related. However, a phenomenologically acceptable scenario is not
easy to construct and there exist only a few attempts in the
literature~\cite{earlyLRDSB} which try to break the left--right
symmetry dynamically in a phenomenologically acceptable way.

The model which is developed in this paper is to our knowledge
the first complete and successful attempt of this kind which does
not just assume that the left--right breaking dynamics will work
correctly. The underlying Lagrangian is essentially a left--right
symmetric generalization of the BHL model of top
condensation~\cite{BHL} which will be invariant under
local $SU(2)_L\times SU(2)_R\times U(1)_{B-L}$ transformations. The
emerging symmetry breaking pattern can be understood as a sequence
of two steps. First a hybrid bi--fermion condensate in the lepton
sector (equivalent to a Higgs doublet of $SU(2)_R$) breaks the
$SU(2)_R\times U(1)_{B-L}\rightarrow U(1)_Y$ symmetry and the
three \GBs are eaten resulting in right--handed gauge boson masses.
After this dynamical $SU(2)_R\times U(1)_{B-L}\rightarrow U(1)_Y$
breaking, a Dirac condensate for the top quark breaks in a second
step the gauged $SU(2)_L\times U(1)_Y$ symmetry corresponding to
a dynamical $SU(2)_L\times U(1)_Y\rightarrow U(1)_{em}$ breaking.
We will establish that this is the preferred symmetry breaking
sequence of our model by calculating the effective potential in
analogy to the techniques used in the top--condensate approach
to the \SM~\cite{BHL,N,Mir,Mar}.

The dynamics of our model will lead to composite Higgs bosons which
are responsible for the symmetry breaking in the language of the
effective Lagrangian. We show that, unlike in conventional LR
models, whether or not parity is spontaneously broken depends
just on the fermion content of the initial Lagrangian and not on
the choice of parameters of the Higgs potential. Predictions are
obtained for masses of the Higgs bosons in terms of fermion masses
and in addition certain relations between the masses of scalars
are found. The model predicts intermediate--scale Higgs scalars
(with their masses being one or two orders of magnitude lower
than the parity--breaking scale) which are the pseudo--\GBs of
an accidental global $SU(4)$ symmetry of the effective Higgs
potential. It is shown that parity breakdown at a right--handed
scale propagates down and eventually causes the \EW symmetry
breakdown at a lower (electro--weak) scale, i.e. a tumbling
scenario of symmetry breaking is operative.

In Section~\ref{sec:review} we start with reviewing some basic
features of the top--condensate approach and conventional LR models.
Further, we discuss a dynamical LR--symmetry breaking model with
the usual fermion content which features composite triplet scalars
(i.e. reproduces the LR model with the presently most popular
Higgs boson content~\cite{LR3}) and analyze its shortcomings.
In Section~\ref{sec:themodel} we present our model which contains
a new singlet fermion and leads to composite Higgs doublets and a
Higgs singlet. For this model we discuss the four--fermion terms
which are responsible for the dynamical symmetry breakdown.
Section~\ref{sec:effpot} contains results for the effective potential
in bubble approximation and in Section~\ref{sec:bubblepred} we
present our predictions in this approximation. In the following
Section~\ref{sec:fixp} we obtain and discuss the renormalization
group improved predictions, especially for the top quark mass.
Section~\ref{sec:phen} contains some phenomenological considerations
and in Section~\ref{sec:concl} we give our conclusions. Technical
details of our calculations are given in the Appendices.


\section{Preliminary Considerations}
\label{sec:review}
\subsection{The BHL Model of Electro--Weak Symmetry Breaking}
Let us briefly recall some features of the top--condensate
approach~\cite{BHL,N,Mir,Mar} following the BHL model~\cite{BHL}.
The model Lagrangian consists of kinetic terms for fermions and
gauge fields plus attractive four--fermion (4-f) interactions:
\beq
{\cal L} = {\cal L}_{kinetic} + G(\bar{Q}_Lt_R\bar{t}_RQ_L)~,
\label{L4F}
\eeq
where $Q_L$ is the third--generation quark doublet. For large enough
$G$ symmetry breaking occurs and can e.g. be studied in the NJL
approximation\footnote{We use the well known term NJL throughout
this paper even though the work of Vaks and Larkin was received
and published first.}~\cite{VL,NJL}, i.e. in leading order of a
large--$N_c$ expansion with a cutoff $\Lambda$. In the auxiliary
field formalism eq.~(\ref{L4F}) can be rewritten in terms of a
static, non--propagating, scalar doublet $\varphi:=-G\bar{t}_RQ_L$
of mass $G^{-\frac{1}{2}}$ such that the Lagrangian eq.~(\ref{L4F})
becomes~\cite{Eguchi}
\beq
{\cal L}_{aux}= {\cal L}_{kinetic}
               -\bar{Q}_L \varphi t_R-\bar{t}_R \varphi^\dagger Q_L
               -G^{-1} \varphi^\dagger \varphi~.
\label{auxil}
\eeq
Due to the dynamics of the original model further terms (including
kinetic terms) emerge in the effective Lagrangian at low energy
scales  $\mu$ after one integrates out the degrees of freedom with
energies between $\mu$ and $\Lambda$. For large cutoff $\Lambda$
only renormalizable terms are allowed such that we obtain
\beq
{\cal L}_{eff} = {\cal L}_{aux}
  + Z_\varphi \(D_\mu\varphi\)^\dagger
  \(D^\mu\varphi\) + \delta M^2 \varphi^\dagger \varphi -
  \frac{\delta\lambda}{2} \left(\varphi^\dagger \varphi\right)^2
  -\delta g_t \(\bar{Q}_L \varphi t_R+\bar{t}_R \varphi^\dagger Q_L\)
  +\delta{\cal L}_{kinetic}~.
\label{Lcomplete}
\eeq
Note that symmetry breaking occurs when $\delta M^2> G^{-1}$ which
is achieved for $G>G_{\mbox{\scriptsize crit}}$. We can immediately
read off those conditions which express the composite nature of the
effective scalar Lagrangian:
$Z_\varphi \stackrel{\mu^2\rightarrow\Lambda^2}{\longrightarrow}0$,
$\delta M^2 \stackrel{\mu^2\rightarrow\Lambda^2}{\longrightarrow}0$,
and $\delta\lambda
\stackrel{\mu^2\rightarrow\Lambda^2}{\longrightarrow}0$.
This expresses simply the fact that all dynamical effects must
disappear as $\mu$ approaches the scale $\Lambda$. In addition we
have the normalization conditions
$\delta g_t \stackrel{\mu^2\rightarrow\Lambda^2}{\longrightarrow} 0$
and $\delta{\cal L}_{kinetic} \stackrel{\mu^2\rightarrow\Lambda^2}
{\longrightarrow} 0$. We can now use the freedom to rescale the
scalar field $\varphi$ by defining
$\varphi:=\varphi/ \sqrt{Z_\varphi}$ such that the Lagrangian
becomes\footnote{Ignoring fermionic wave function contributions
which do not play a role for the compositeness conditions.}
\bea
{\cal L}_{eff} =  {\cal L}_{kinetic}
+ \(D_\mu\varphi\)^\dagger \(D^\mu\varphi\)
+\frac{\hat\lambda v^2}{2}\varphi^\dagger\varphi
-\frac{\hat\lambda}{2} \( \varphi^\dagger\varphi \)^2
-\hat g_t \(\bar{Q}_L \varphi t_R+\bar{t}_R \varphi^\dagger Q_L\)~.
\label{LSM}
\eea
Here we introduced
$\frac{\hat\lambda v^2}{2} = \frac{\delta M^2 - G^{-1}}{Z_\varphi}$,
$\hat\lambda = \frac{\delta\lambda}{Z_\varphi^2}$ and
$\hat g_t =  \frac{1+\delta g_t}{\sqrt{Z_\varphi}}$ and the effective
Lagrangian has now become the \SMp From the definition of $\hat g_t$,
$\hat\lambda$ and $v$ we see that the compositeness conditions are
\bea
\lim_{\mu^2\to\Lambda^2}\hat g_t^{-2}(\mu^2)= 0~,     \quad
\lim_{\mu^2\to\Lambda^2}\frac{\hat\lambda(\mu^2)}
{\hat g_t^4(\mu^2)}= 0~, \quad
\lim_{\mu^2\to\Lambda^2}\frac{\hat\lambda(\mu^2) v^2(\mu^2)}
{2\hat g_t^2(\mu^2)}=-G^{-1}~,
\label{compcon}
\eea
where $\Lambda$ corresponds to the high energy cutoff of the
BHL--model.

The vanishing of the kinetic term of the composite Higgs field is
thus equivalent to a Landau singularity of the running Yukawa
coupling $\hat{g}_t$. The compositeness conditions
eq.~(\ref{compcon}) can therefore be translated into boundary
conditions of the \RG flow at $\Lambda$. Since the effective
Lagrangian is identical to that of the \SM one can employ the
usual one--loop $\beta$--functions~\cite{oneloopb}. The running
top Yukawa coupling indeed develops the desired Landau pole above
a top mass value of $197~GeV$, and as $m_t$ is increased the Landau
pole moves to lower scales. If the above compositeness condition
$\hat{g}_t\rightarrow\infty$ is imposed on the full \RG equations
at e.g. $\Lambda=10^{15}~GeV$, then one finds $m_t=227~GeV$.
For the above $\Lambda$ already the analysis with one--loop
$\beta$--functions turns out to be rather reliable since running
down from the Landau pole one ends up in the attractive infrared
quasi--fixed point~\cite{quasiIR}\footnote{Note that this
quasi-fixed point is related,  but not identical to the
Pendleton--Ross fixed point \cite{PR}.}.

Note that the solutions of the \RG equations in a limit
corresponding to the fermionic bubble approximation also satisfy
the compositeness conditions. However in this limit one finds with
$\Lambda=10^{15}~GeV$ a much lower top quark mass, $m_t=164~GeV$.
Thus QCD and \EW corrections play a non--negligible role in the
precise mass predictions. This demonstrates that the
$\beta$--functions of the full effective Lagrangian are superior to
those derived in bubble approximation.

Finally we have to comment on the cutoff $\Lambda$ regulating loop
effects of the BHL model. We imagine that such a cutoff is motivated
by new physics which is not specified in detail. In the case of the
BHL model theories have been constructed where some new gauge
interactions with heavy gauge bosons with masses $M_x\sim\Lambda$
motivate the cutoff~\cite{TOPCOLOR,U1,King,Boenisch}. Then, by
integrating out heavy bosons, four--fermion terms as in
eq.~(\ref{L4F}) with effective couplings $G \sim 1/M_x^2$ emerge
as lowest dimensional operators. Throughout this paper we will take
a similar attitude, but we will not try to relate our 4-f structures
to any renormalizable model.

\subsection{Left--right symmetric models}
We will perform steps similar to those described above in LR
symmetric models. In  order to remind the reader of the main
features of LR symmetry and to introduce the notation, we first
describe conventional LR model building before we discuss our
scenario.

In LR  symmetric models based on the gauge group $SU(3)_c\otimes
SU(2)_L \otimes SU(2)_R \otimes U(1)_{B-L}$~\cite{LR1,LR2,LR3}
quarks and leptons of each generation are symmetrically placed into
the doublet representations of $SU(2)_L$ and $SU(2)_R$. We will
deal only with the third generation of quarks and leptons for
which the assignment is
\bea
Q_{L} = \left( \begin{array}{c}
t \\ b \end{array} \right)_L \sim \,(3,\;2,\;1,\;1/3)~,\;\;\;
&Q_{R}=\left( \begin{array}{c}
t \\ b \end{array} \right)_R \sim \,(3,\;1,\;2,\;1/3)~;\nonumber \\
\Psi_{L}= \left( \begin{array}{c}
\nu_\tau \\ \tau \end{array} \right)_L \sim \,(1,\; 2,\;1,\;-1)~,
&\Psi_{R}=\left( \begin{array}{c}
\nu_\tau \\ \tau \end{array} \right)_R \sim \,(1,\; 1,\;2,\;-1)~.
\label{fermdef}
\eea
The usual Dirac masses of the fermions are generated by the VEV of a
bi--doublet Higgs scalar $\phi$:
\beq
\phi =  \left( \begin{array}{cc}
\phi_1^0 & \phi_2^+ \\
\phi_1^-& \phi_2^0 \end{array} \right)\sim\,(1,\; 2,\;2,\;0)~; \;\;\;
\langle \phi \rangle = \left( \begin{array}{cc}
\kappa & 0 \\ 0 & \kappa' \end{array} \right)~.
\label{bidef}
\eeq
However, to arrive at the phenomenologically required symmetry
breaking pattern $SU(3)_c \otimes SU(2)_L \otimes SU(2)_R
\otimes U(1)_{B-L}\rightarrow SU(3)_c \otimes SU(2)_L
\otimes U(1)_{Y}\rightarrow SU(3)_c\otimes U(1)_{e.m.}$, one
needs additional Higgs multiplets. The simplest possibility is
to add two doublets~\cite{LR1,LR2}:
\beq
\chi_{L}=\left( \begin{array}{c}
\chi_L^0 \\ \chi_L^-\end{array}\right)\sim \,(1,\;2,\;1,\;-1)~,\;\;\;
\chi_{R}=\left( \begin{array}{c}
\chi_R^0 \\ \chi_R^-\end{array} \right) \sim \,(1,\;1,\;2,\;-1)~.
\label{chidef}
\eeq
Then $SU(2)_R$ is broken at the right--handed scale $M_R$ by $\langle
\chi_R^0 \rangle = v_R$, and the \EW symmetry is broken by the
VEVs of $\phi$ and possibly of $\chi_L^0$ ($\equiv v_L$). In fact,
$\chi_L$ is not necessary for the above symmetry breakdown pattern;
it is usually introduced to ensure the discrete parity symmetry under
which \footnote{Note that this symmetry also requires the equality
of the $SU(2)_L$ and $SU(2)_R$ gauge coupling constants
\mbox{$g_{2L}=g_{2R}\equiv g_2$}.}
\beq
Q_L \leftrightarrow Q_R~,\;\;\; \Psi_L \leftrightarrow
\Psi_R~,\;\;\; \phi \leftrightarrow \phi^\dagger~,\;\;\;
\chi_L \leftrightarrow \chi_R~,\;\;\;W_L\leftrightarrow W_R~.
\label{discrete}
\eeq
It has been shown~\cite{SPB} that even if the Higgs potential is
exactly symmetric with respect to the discrete parity transformation,
the vacuum of the model may prefer $v_R\gg v_L$. This would give a
very elegant explanation of parity violation at low energies as
being the result of spontaneous symmetry breakdown.

The doublets $\chi_L$ and $\chi_R$ are singlets of either $SU(2)_R$
or $SU(2)_L$ and so cannot interact in a renormalizable way with
the usual quarks and leptons. This means that neutrinos are Dirac
particles and get their masses in the same way as the other fermions.
It is therefore very difficult to understand the smallness of their
masses -- one needs an extreme fine tuning of the corresponding
Yukawa couplings. An attractive solution of this problem was
suggested in Ref.~\cite{LR3} where two Higgs triplets
$\Delta_L \sim (1,\;3,\;1,\;2)$ and $\Delta_R \sim (1,\;1,\;3,\;2)$
were introduced instead of the doublets $\chi_L$ and $\chi_R$.
Leptons can interact with the Higgs triplets through the
Majorana--like Yukawa coupling
\beq
f(\Psi_L^T C \tau_2 \vec{\tau}\vec{\Delta}_L \Psi_L + \Psi_R^T C
\tau_2 \vec{\tau}\vec{\Delta}_R \Psi_R)  + h.c.~,
\label{majorana}
\eeq
where $C$ is the charge conjugation matrix. In this model neutrinos
are Majorana particles, and the seesaw mechanism~\cite{seesaw} is
operative. This mechanism provides a very natural explanation of the
smallness of the usual neutrino masses, relating it to the fact that
the right--handed scale is much higher than the \EW scale. In other
words, parity non--conservation at low energies and the smallness
of neutrino masses have a common origin in this model.

\subsection{Dynamical Symmetry Breaking for the Triplet Model}
\label{sec:triplets}

We now assume, following the approach of Refs.~\cite{BHL,N,Mir,Mar}
to the \SMk that the low--energy ($\mu < \Lambda$) degrees of
freedom of our LR model are just fermions and gauge bosons, with
no fundamental Higgs fields being present. Therefore the Lagrangian
contains only the usual kinetic terms for all gauge fields and
fermions. In addition we postulate in a first step the following
set of gauge--invariant 4-f interactions to be present at low
energies:
\beq
L_{int}=L_{int1}+L_{int2}~,
\label{Lintfirst}
\eeq
\bea
L_{int1}&=&G_1(\bar{Q}_{Li}Q_{Rj})(\bar{Q}_{Rj}Q_{Li})+
[G_2(\bar{Q}_{Li}Q_{Rj})(\tau_2\bar{Q}_{L}Q_{R}\tau_2)_{ij}
+h.c.]\nonumber \\
&&+G_3(\bar{\Psi}_{Li}\Psi_{Rj})(\bar{\Psi}_{Rj}\Psi_{Li})+
[G_4(\bar{\Psi}_{Li}\Psi_{Rj})
(\tau_2\bar{\Psi}_{L}\Psi_{R}\tau_2)_{ij}+h.c.]\nonumber \\
&&+{}[G_5(\bar{Q}_{Li}Q_{Rj})(\bar{\Psi}_{Rj}\Psi_{Li})
+h.c.]+
[G_6(\bar{Q}_{Li}Q_{Rj})(\tau_2\bar{\Psi}_{L}\Psi_{R}\tau_2)_{ij}
+h.c.]\;~,
\label{Lint1first}\\
L_{int2}&=&\tilde{G}_7[(\Psi_L^T C \tau_2 \vec{\tau}\Psi_L)
(\bar{\Psi}_L \vec{\tau} \tau_2 C \bar{\Psi}_L^T)+
(\Psi_R^T C \tau_2 \vec{\tau}\Psi_R)
(\bar{\Psi}_R \vec{\tau} \tau_2 C\bar{\Psi}_R^T)]~.
\label{Lint2}
\eea
In analogy to the BHL model the $G_a$ are dimensionful 4-f couplings
of the order of $\Lambda^{-2}$ motivated by some new physics at the
high energy scale $\Lambda$. The indices $i$ and $j$ refer to isospin
and it is implied that the colour indices of quarks are summed over
within each bracket. Note that the above interactions are not only
gauge invariant, but also (for hermitean $G_2$, $G_4$, $G_5$ and
$G_6$) symmetric with respect to the discrete parity
operation~(\ref{discrete}). However, it should be emphasized that
the 4-f interactions of eqs.~(\ref{Lint1first}) and (\ref{Lint2})
do not constitute the most general set of the gauge-- and
parity--invariant 4-f interactions. We left out undesirable terms
which, if critical, would produce electrically charged or coloured
condensates. We simply assume that the new physics responsible for
the effective 4-f couplings does not produce such terms.

We will use the above Lagrangian and later modifications thereof
to break the LR symmetry dynamically. We assume that the heaviest
(i.e. the third generation) quarks and leptons play a special role
in the symmetry breaking dynamics and so confine ourselves to the
discussion of the third generation. In this limit only the
third--generation fermions are massive while all the light fermions
are considered to be massless. This seems to be a good starting
point from where light fermion masses could, e.g., be generated
radiatively.

One might expect that, just as in the NJL model or in~\cite{BHL},
for strong enough (``critical'') $G_a$ the LR symmetry will be
dynamically broken and the correct pattern of the LR symmetry
breakdown will emerge. However, as we shall see, the situation
with the LR model is more complicated.

One way to explore the symmetry breaking in the model and to study
the composite Higgs scalars is to consider the four--point Green
functions generated by the 4-f couplings of
equations~(\ref{Lint1first}) and (\ref{Lint2}) and study their
poles corresponding to the two--particle bound states. This can be
done analytically in fermion bubble approximation in which the
exact solution can be obtained~\cite{BHL,VL,NJL}. The scale
$\Lambda$ plays again the role of a natural cutoff for the divergent
diagrams. An alternative approach is to rewrite the 4-f structure
in terms of static auxiliary fields (see equations~(\ref{L4F}),
(\ref{auxil}) and Appendix \ref{appLag} for details). These static
auxiliary fields can acquire gauge invariant kinetic terms and
become physical propagating scalar fields.

Loosely speaking, the auxiliary scalars are defined as the square
roots of the original 4-f operators. Therefore the 4-f structure
of eq.~(\ref{Lint1first}) can generate the composite bi--doublet
Higgs field  $\phi$ which (for small lepton 4-f couplings) has the
structure
\beq
\phi_{ij}\sim\alpha (\bar{Q}_{Rj}Q_{Li})+\beta (\tau_2 \bar{Q}_L Q_R
\tau_2)_{ij}~,
\label{phiij}
\eeq
while the 4-f terms of equation~(\ref{Lint2}) give rise to the
composite triplet scalars,
\beq
\vec{\Delta}_L \sim (\Psi_L^T C \tau_2 \vec{\tau}\Psi_L),\;\;\;
\vec{\Delta}_R\sim (\Psi_R^T C  \tau_2 \vec{\tau}\Psi_R)~.
\label{dLR}
\eeq
Note that while the VEVs of $\phi$ are particle--antiparticle
condensates, the VEVs of $\Delta_L$ and $\Delta_R$ are
Majorana--like particle--particle condensates\footnote{A
Majorana--like condensate of right handed neutrinos has been
considered in~\cite{HLP} in the framework of the \SMp}.

The model based on the 4-f couplings of eqs.~(\ref{Lint1first}) and
(\ref{Lint2}) has thus all the necessary ingredients to produce the
composite Higgs fields $\phi$, $\Delta_L$ and $\Delta_R$. However,
as we shall shortly see, it does not lead to the correct pattern of
LR symmetry breakdown.

\begin{figure}[htb]
\centerline{
\rotate[r]{\epsfxsize=31ex \epsffile{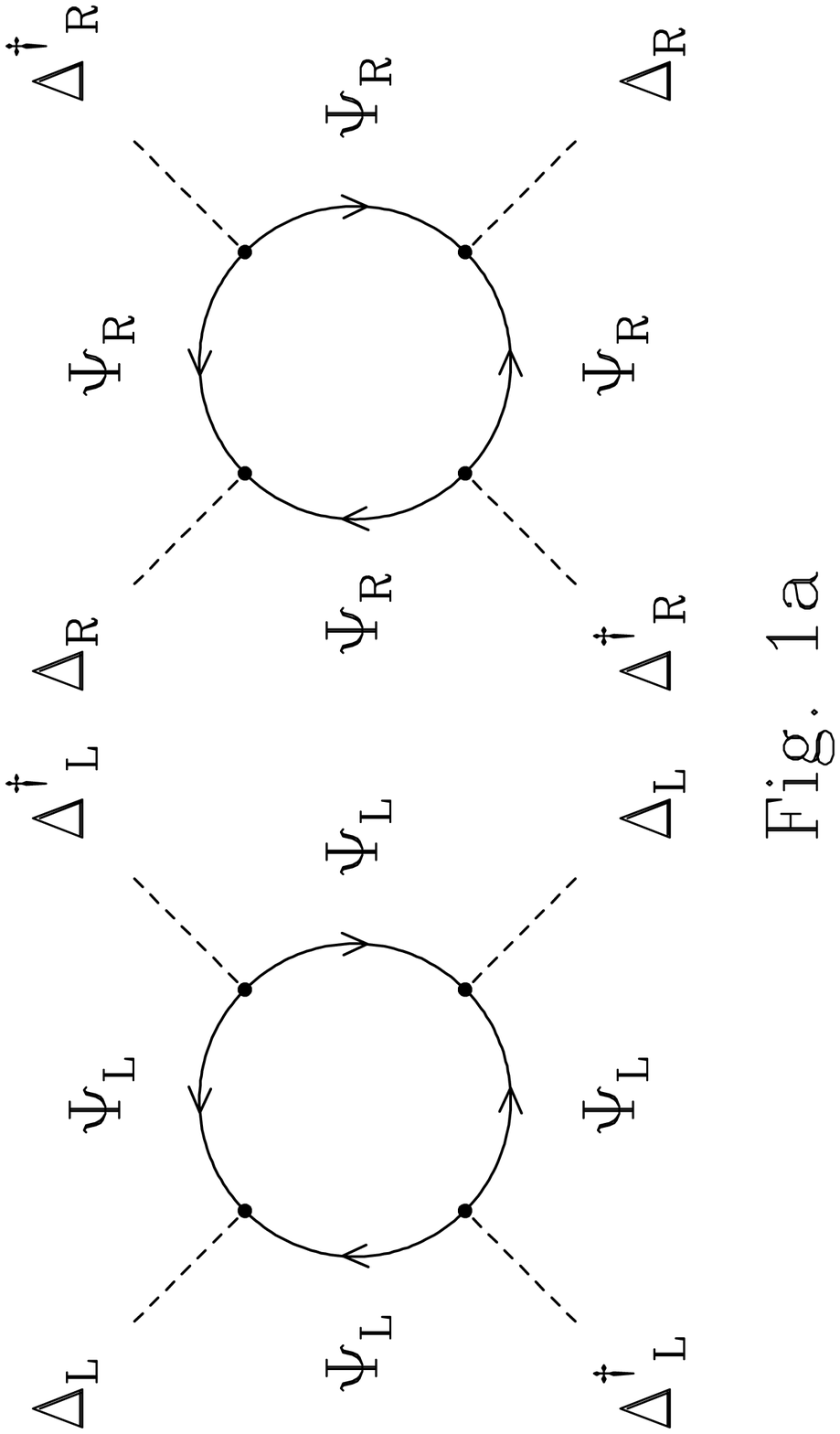} }
\hspace{2em}
\rotate[r]{\epsfxsize=31ex \epsffile{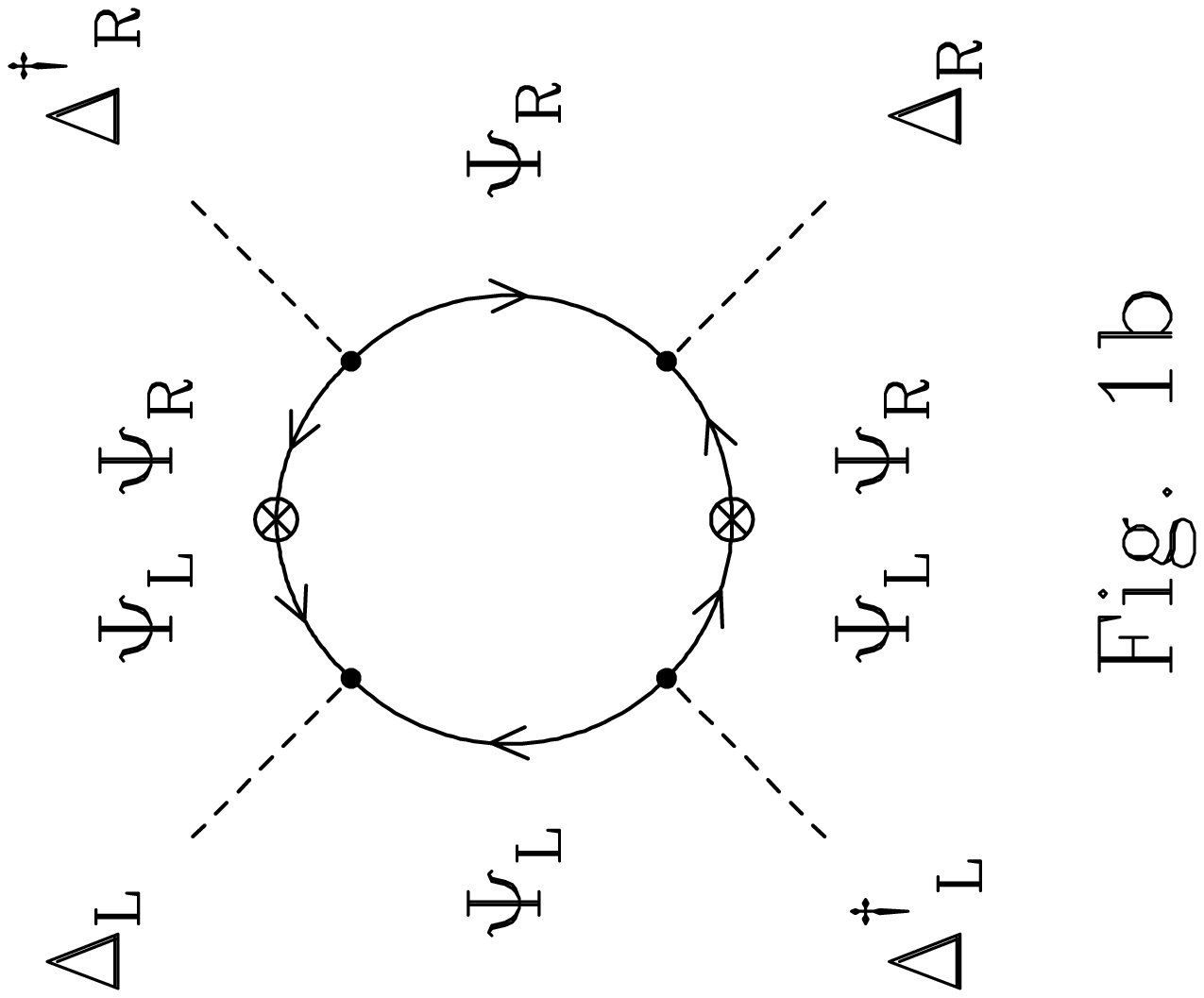} }}
\caption[]{\small\sl
Fermion loop diagrams contributing to the quartic couplings
$\lambda_1$~(Fig.~\ref{fig:delta}a) and
$\lambda_2$~(Fig.~\ref{fig:delta}b) for Higgs triplets.
\label{fig:delta}
}
\end{figure}

For parity to be spontaneously broken, one needs
$\langle \Delta_R \rangle > \langle \Delta_L \rangle$. One can
readily make sure that this can only be satisfied provided
$\lambda_2 > \lambda_1$ where $\lambda_1$ and $\lambda_2$ are the
coefficients of the
$[(\Delta_L^\dagger \Delta_L)^2+(\Delta_R^\dagger \Delta_R)^2]$ and
$2(\Delta_L^\dagger \Delta_L)(\Delta_R^\dagger \Delta_R)$
quartic couplings in the Higgs potential~\cite{SPB}. In the
conventional approach, $\lambda_1$ and $\lambda_2$ are free
parameters and one can always choose $\lambda_2>\lambda_1$.
On the contrary, in the composite Higgs approach based on a
certain set of effective 4-f couplings, the parameters of the
effective Higgs potential are not arbitrary: they are all
calculable in terms of the 4-f couplings $G_a$ and the scale of new
physics $\Lambda$~\cite{BHL}. In particular, in fermion bubble
approximation at one loop level the quartic couplings $\lambda_1$ and
$\lambda_2$ are induced through the Yukawa couplings of
eq.~(\ref{majorana}) and are given by the diagrams of
Fig.~\ref{fig:delta}.
It can be seen from Fig.~\ref{fig:delta}b that to induce the
$\lambda_2$ term one needs the $\Psi_L$--$\Psi_R$ mixing in
the fermion line in the loop, i.e. the lepton Dirac mass term
insertions. However, the Dirac mass terms are generated by the
VEVs of the bi--doublet $\phi$; they are absent at the parity
breaking scale which is supposed to be higher than the \EW scale.
Even if parity and \EW symmetry are broken simultaneously (which
is hardly a phenomenologically viable scenario), this would not
save the situation since the diagram of Fig.~\ref{fig:delta}b is
finite whereas the one of Fig.~\ref{fig:delta}a is logarithmically
divergent and so the inequality $\lambda_2>\lambda_1$ cannot be
satisfied. We also checked that using the full one--loop
Coleman--Weinberg~\cite{CW} effective potential instead of the
truncated one which includes up to the quartic terms in the fields
does not save the situation. Therefore we will consider a model
with a different composite Higgs content. As we have mentioned,
the simplest model includes two doublets, $\chi_L$ and $\chi_R$,
instead of the triplets $\Delta_L$ and $\Delta_R$. The question
is whether it is possible to get the correct pattern of dynamical
symmetry breakdown in this model. As we shall see, this is really
the case.


\section{The Model}
\label{sec:themodel}

\begin{figure}[htb]
\centerline{
\rotate[r]{\epsfxsize=31ex \epsffile{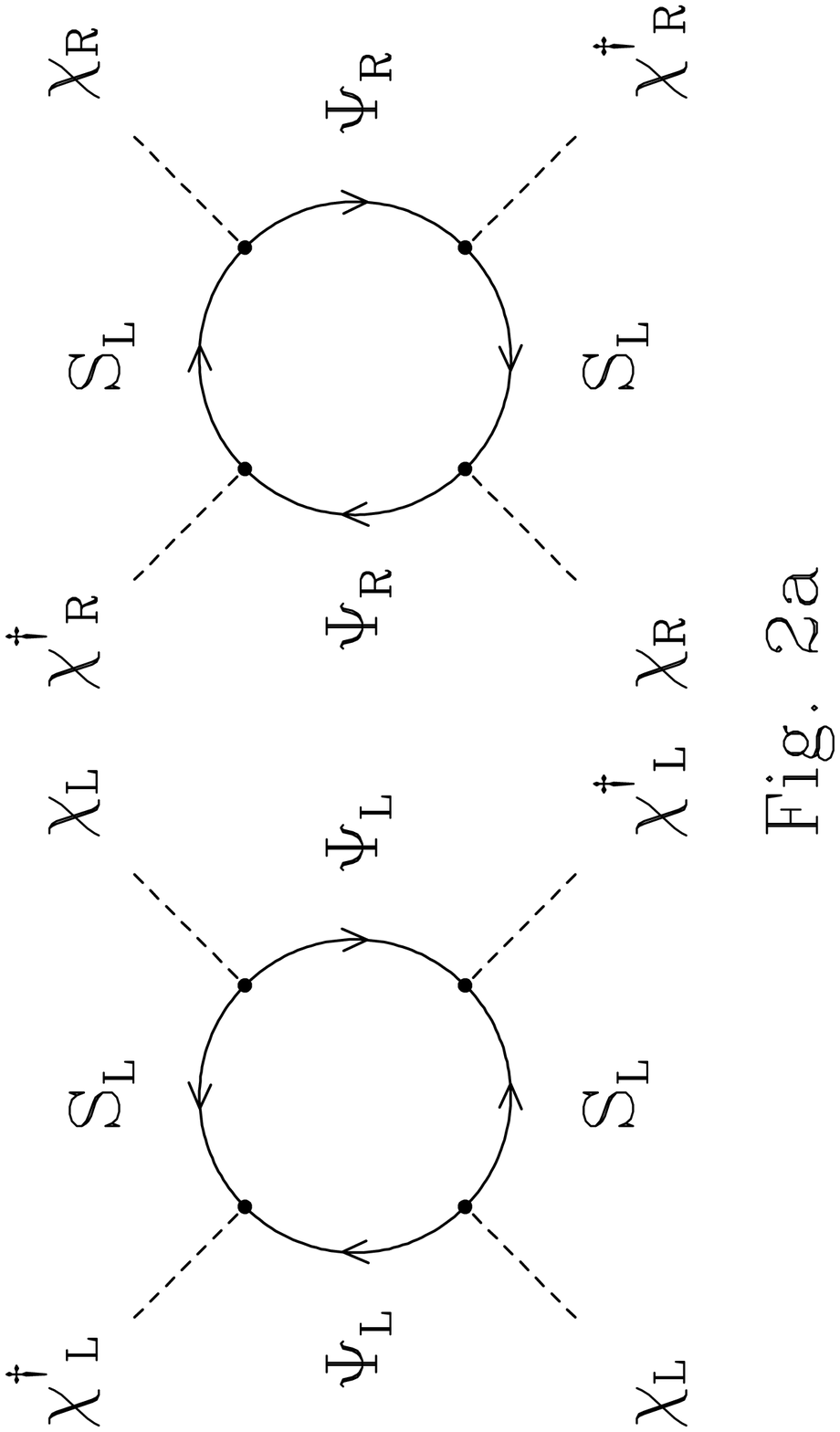} }
\hspace{2em}
\rotate[r]{\epsfxsize=31ex \epsffile{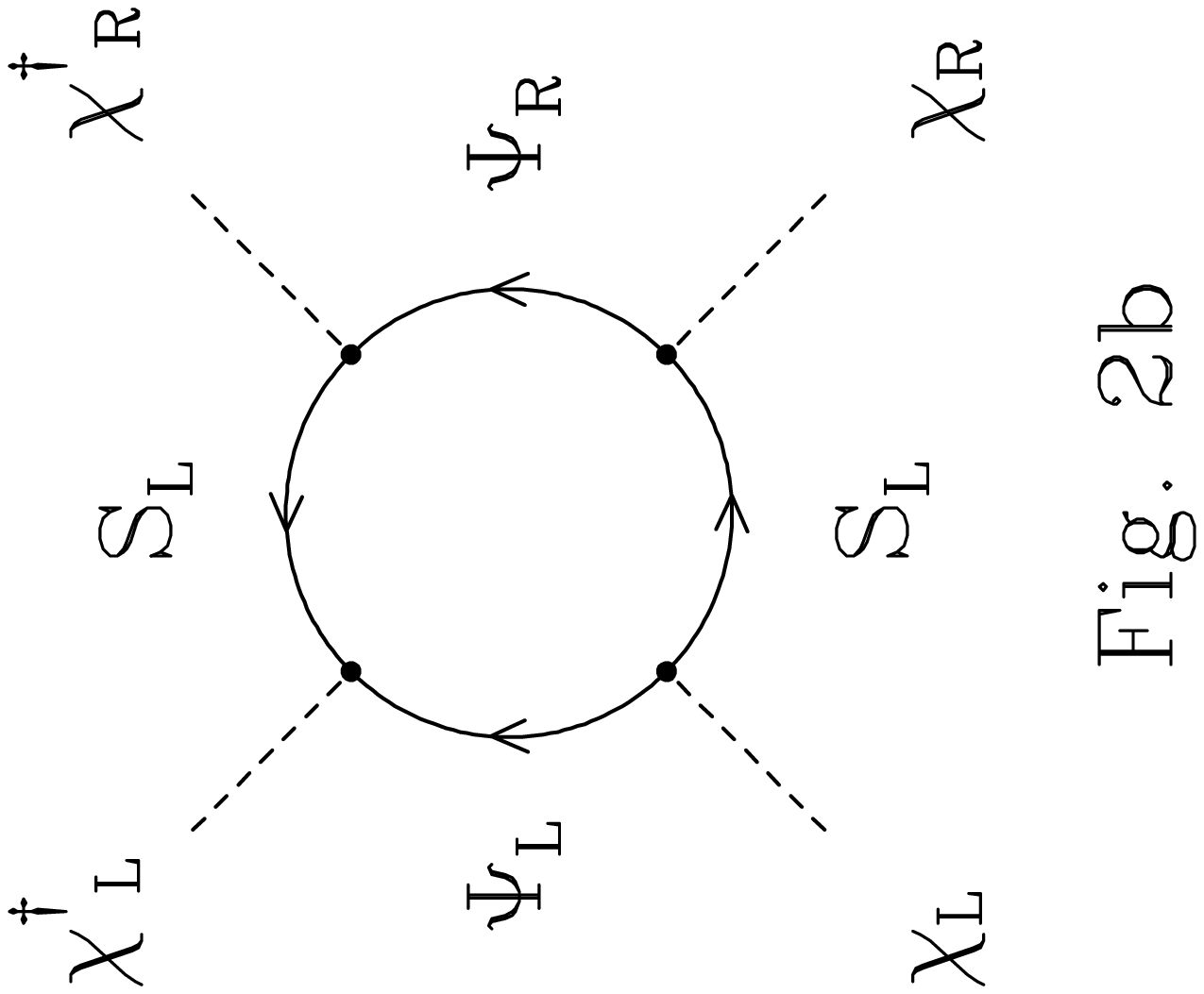} }}
\caption[]{\small\sl
Fermion loop diagrams contributing to the
quartic couplings $\lambda_1$~(Fig.~\ref{fig:doub}a) and
$\lambda_2$~(Fig.~\ref{fig:doub}b)
for the Higgs doublets $\chi_{L/R}$.
\label{fig:doub}
}
\end{figure}

We will now consider the original version of the LR symmetric model
with a bi--doublet $\phi$ and two doublets $\chi_L$ and $\chi_R$,
as defined in eqs.~(\ref{bidef}) and~(\ref{chidef}). In conventional
LR models one can have the doublet Higgs bosons without introducing
any other new particles. In our model, however, $\chi_L$ and
$\chi_R$ are composite and cannot be built from the standard
fermions. We therefore assume that in addition to the usual quark
and lepton doublets of eq.~(\ref{fermdef}) there is a gauge--singlet
fermion
\beq
S_L \sim (1,\;1,\;1,\;0)~.
\label{SL}
\eeq
To maintain discrete parity symmetry one needs a right--handed
counterpart of $S_L$. This can be either another particle, $S_R$,
or the right--handed antiparticle of $S_L$,
$(S_L)^c\equiv C\bar{S}_L^T = S^c_R$. The latter choice is more
economical and, as we shall see, leads to the desired symmetry
breaking pattern. We therefore assume that under parity operation
\beq
S_L \leftrightarrow S^c_R~.
\label{SLC}
\eeq
With the singlet fermion $S_L$ our interaction Lagrangian becomes
\beq
L_{int}=L_{int1}+L_{int3}~,
\label{Lint}
\eeq
and contains now in addition to (\ref{Lint1first}) the following term
\beq
L_{int3}=G_7[(S_L^T C \Psi_L)(\bar{\Psi}_L C \bar{S}_L^T)+
(\bar{S}_L\Psi_R)(\bar{\Psi}_R S_L)]+G_8 (S_L^T C S_L)(\bar{S}_L C
\bar{S}_L^T)~,
\label{Lint3}
\eeq
which is now supposed to substitute for $L_{int2}$ of
eq.~(\ref{Lint2}) since we no longer want the triplet Higgses to
be present in the model. The composite Higgs scalars induced by
the 4-f couplings of eq.~(\ref{Lint3}) are the doublets $\chi_L$
and $\chi_R$ and in addition a singlet $\sigma$:
\beq
\chi_L \sim S_L^T C \Psi_L~, \;\;\;\;  \chi_R \sim \bar{S}_L\Psi_R =
(S^c_R)^T C \Psi_R~, \;\;\;\; \sigma \sim \bar{S}_L C \bar{S}_L^T~.
\label{composite}
\eeq
Using the Yukawa couplings of the doublets $\chi_L$ and $\chi_R$ (see
eq.~(\ref{Laux}) below), one can now calculate the fermion--loop
contributions to the  quartic couplings
$\lambda_1 [(\chi_L^\dagger \chi_L)^2+(\chi_R^\dagger \chi_R)^2]$ and
$2\lambda_2 (\chi_L^\dagger\chi_L)(\chi_R^\dagger \chi_R)$ in the
effective Higgs potential (Fig.~\ref{fig:doub}).
The $\lambda_1$ and $\lambda_2$ terms are now given by similar
diagrams. Since the Yukawa couplings of $\chi_L$ and $\chi_R$
coincide (which is just the consequence of the discrete parity
symmetry), Figs.~\ref{fig:doub}a and \ref{fig:doub}b yield
$\lambda_1=\lambda_2$. Recall that one needs $\lambda_2> \lambda_1$
for spontaneous parity breakdown. As we shall see, taking into
account the gauge--boson loop contributions to $\lambda_1$ and
$\lambda_2$ shown in Fig.~\ref{fig:gauge} will automatically secure
this relation.

We did not include into eq.~(\ref{Lint}) a 4-f term of the kind
[$(S_L^T C S_L)^2 + (\bar{S}_L C \bar{S}_L^T)^2$] which is also
gauge invariant  and parity--symmetric. One possible motivation
for not considering such terms is that if one imagines the new
physics responsible for the 4-f terms as being related to some
new vector boson exchange~\cite{TOPCOLOR,U1,King,Boenisch}, such
terms are never induced. Here we just assume that there is a global
symmetry which precludes these terms. For example, this could be a
global $U(1)$ symmetry under which $S_L$ has the charge +1, the
rest of fermions being neutral [$\chi_L$, $\chi_R$ and $\sigma$
will have the charges +1, $-1$ and $-2$ according to
(\ref{composite})]. This will forbid the
$(S_L^T C S_L)^2 + (\bar{S}_L C \bar{S}_L^T)^2$ terms and also
the bare mass term for $S_L$ which is allowed otherwise by the
gauge symmetry, since $S_L$ is a gauge--singlet fermion.

Now we will switch to the auxiliary field formalism, in which the
scalars $\chi_L$, $\chi_R$, $\phi$ and $\sigma$ have the following
bare mass terms and Yukawa couplings:
\bea
L_{aux}&=&-M_0^2(\chi_L^\dagger \chi_L+\chi_R^\dagger \chi_R)-M_1^2
\tr{(\phi^\dagger \phi)}-\frac{M_2^2}{2}\tr{(\phi^\dagger
\tilde{\phi}+h.c.)}-M_3^2 \sigma^\dagger\sigma \nonumber \\
& & -\left[\bar{Q}_L(Y_1\phi+Y_2\tilde{\phi})Q_R +
\bar{\Psi}_L(Y_3\phi+Y_4\tilde{\phi})\Psi_R + h.c.\right]\nonumber\\
& &-\left[Y_5(\bar{\Psi}_L \chi_L S^c_R+\bar{\Psi}_R \chi_R S_L)
+Y_6 (S_L^T C S_L)\sigma + h.c.\right]~.
\label{Laux}
\eea
Here the field $\tilde{\phi}\equiv \tau_2\phi^*\tau_2$ has the same
quantum numbers as $\phi$: $\tilde{\phi}\sim (1,\;2,\;2,\;0)$. By
integrating out the auxiliary scalar fields one can reproduce the
4-f structures of eqs.~(\ref{Lint}) and express the 4-f couplings
$G_1,...,G_8$ in terms of the Yukawa couplings $Y_1,...Y_6$ and the
mass parameters $M_0^2$, $M_1^2$, $M_2^2$ and $M_3^2$ (the explicit
formulas are given in Appendix~\ref{appLag}). Note that all the mass
terms in $L_{aux}$ are positive. In what follows we use the freedom
of rescaling the scalar fields to choose $M_1=M_3=M_0$. At low
energies ($E < \Lambda)$ the scalar fields will acquire
gauge--invariant kinetic terms, quartic couplings and renormalized
mass terms through radiative corrections. This means that the
auxiliary fields may become physical propagating fields, and their
squared mass terms may become negative at some low energy scale,
leading to non--vanishing VEVs of the composite scalars, i.e. to
fermion condensates~\cite{BHL}.

{}From eq.~(\ref{Laux}) one can readily find the fermion masses.
The masses of the quarks and charged leptons and the Dirac neutrino
mass $m_D$ are given by the VEVs of the bi--doublet (we assume all
the VEVs to be real):
\bea
m_t=Y_1\kappa+Y_2\kappa'~,\;\;\;\;\;m_D=Y_3\kappa+Y_4\kappa'~,
    \nonumber \\
m_b=Y_1\kappa'+Y_2\kappa~,\;\;\;\;\; m_{\tau}=Y_3\kappa'+Y_4\kappa~.
\label{fmass}
\eea
We already mentioned that LR models with only doublet Higgs scalars
usually suffer from the large neutrino mass problem. It turns out
that introducing the singlet fermion $S_L$ not only provides the
spontaneous parity breaking in our model, but also cures the
neutrino mass problem. In fact, as it was first noticed in~\cite{WW},
with an additional singlet neutral fermion $S_L$ the neutrino mass
matrix takes the form (in the basis $(\nu_L, \nu^c_L, S_L)$)
\beq
M_{\nu} = \left(
\begin{array}{c}
\end{array}
\begin{array}{ccc}
0 & m_D & \beta\\
m_D^T & 0 & M^*\\
\beta^{T} & M^{\dagger} & \tilde{\mu}
\end{array} \right)\;,
\label{numass1}
\eeq
where the entries $\beta$, $M$ and $\tilde{\mu}$ can be read off from
eq.~(\ref{Laux}):
\beq
\beta=Y_5 v_L~,\quad M=Y_5 v_R~, \quad\tilde{\mu} = 2Y_6\sigma_0~.
\label{yukawas}
\eeq
Here $\sigma_0\equiv\langle\sigma\rangle$. For $v_R\gg\kappa,\kappa',
v_L$ and $v_R \gta \sigma_0$ one obtains two heavy Majorana
neutrino mass eigenstates  with the masses $\sim M$ and a light
Majorana neutrino with the mass $m_\nu \simeq \tilde{\mu}(m_D^2/M^2)
-2\beta m_D/M$ which vanishes in the limit $M \to \infty$. This is
the modified seesaw mechanism which provides the smallness of the
neutrino mass.


\section{The Effective Potential in Bubble Approximation}
\label{sec:effpot}
In the auxiliary field formalism we can now study the dynamical
symmetry breaking by calculating the effective potential. The
effective mass terms and wave--function renormalization constants
can be obtained from the 2--point scalar Green functions
(Fig.~\ref{fig:twopt}), whereas the quartic terms are given by
the 4--point functions (Fig.~\ref{fig:fourpt}). Higher--order
terms are finite in the limit $\Lambda \to \infty$ and therefore
relatively unimportant. A convenient way to calculate the potential
parameters which automatically takes care of the numerous
combinatorial factors in the diagrams of Figs.~\ref{fig:twopt} and
Fig.~\ref{fig:fourpt} is to calculate the full one--loop
Coleman--Weinberg effective potential of the system and then to
truncate this potential so as to keep only the terms up to and
including quartic ones in the scalar fields. To derive the low
energy effective Lagrangian we introduce an infrared cutoff $\mu$
in the loop integrals. This procedure and its physical meaning are
discussed in Appendix~\ref{appeffpot}.

\subsection{Spontaneous Parity Breaking}
\begin{figure}[htb]
\centerline{
\rotate[r]{\epsfxsize=31ex \epsffile{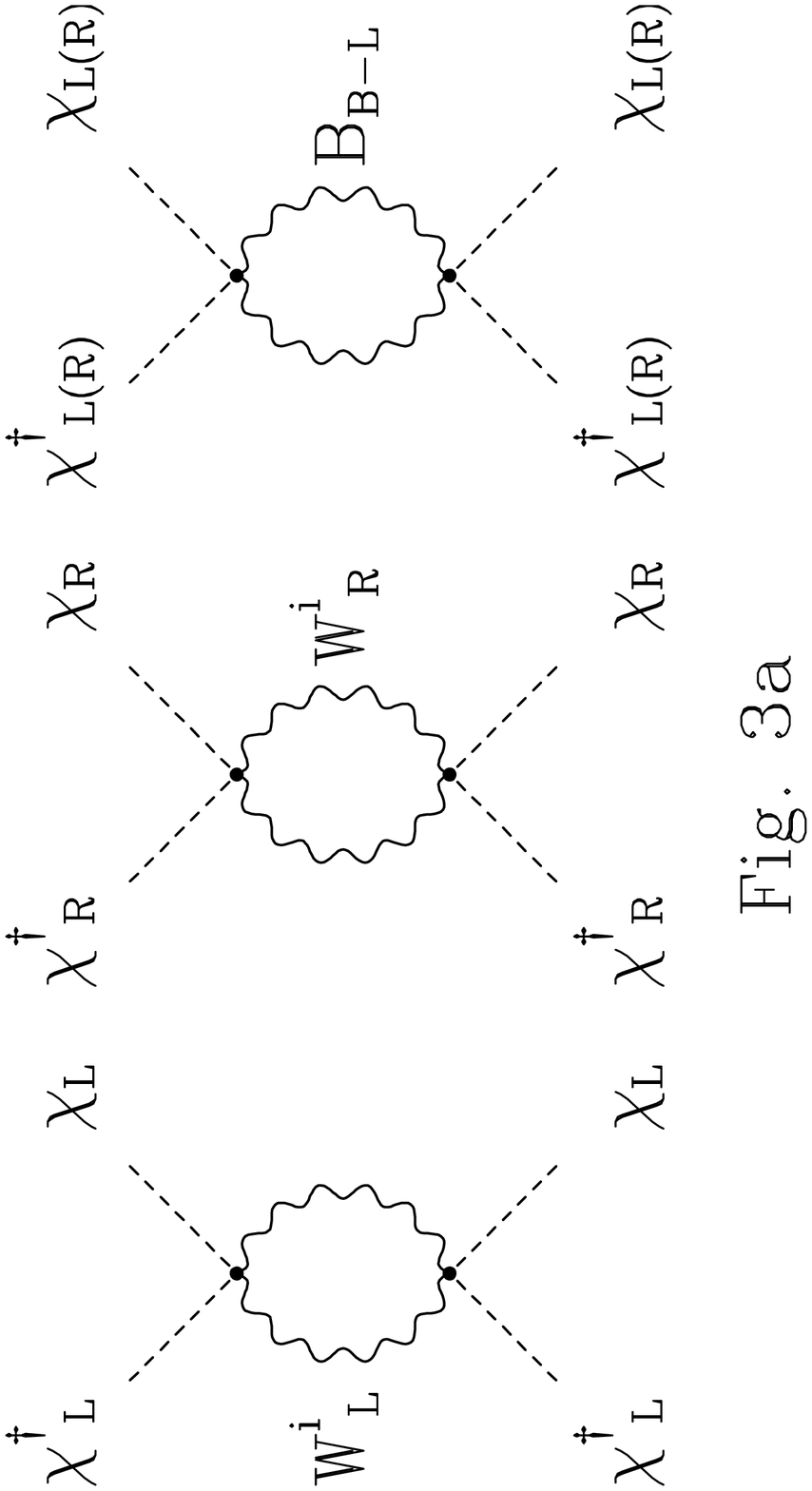} }
\hspace{2em}
\rotate[r]{\epsfxsize=31ex \epsffile{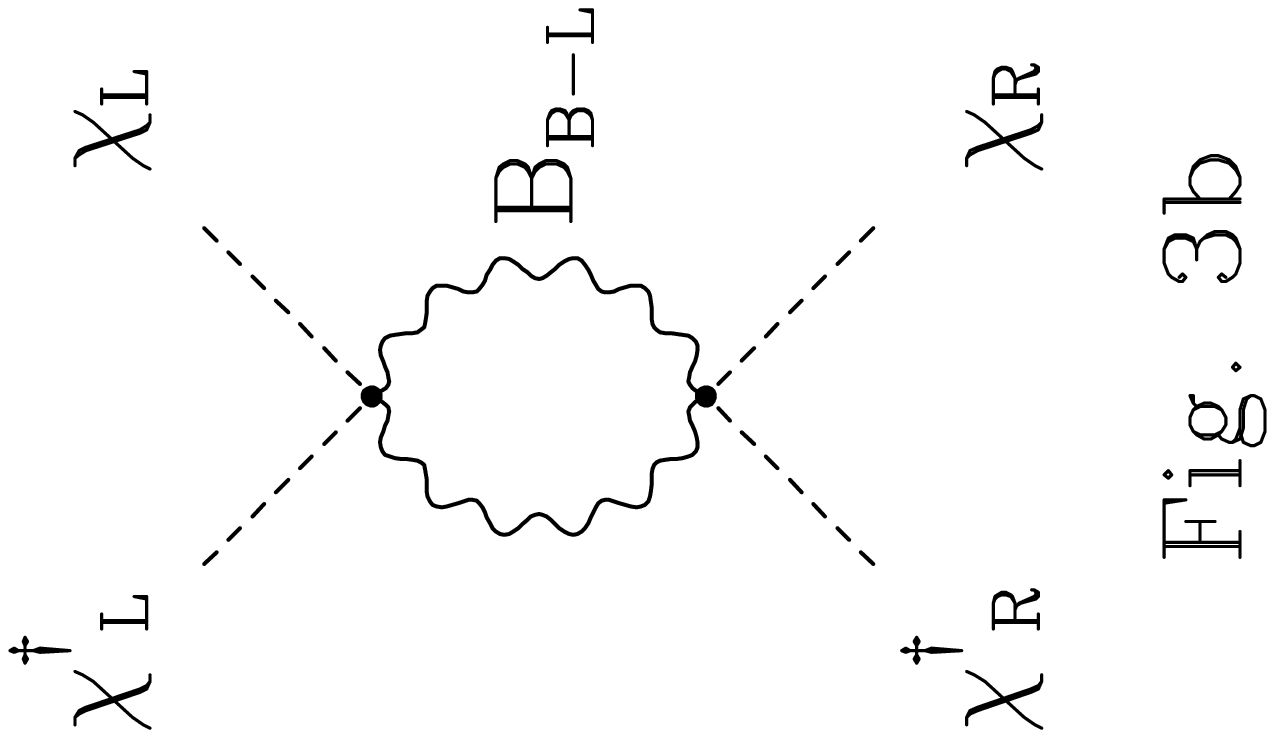} }}
\caption[]{\small\sl
Gauge boson loop diagrams contributing to the quartic couplings
$\lambda_1$~(Fig.~\ref{fig:gauge}a) and
$\lambda_2$~(Fig.~\ref{fig:gauge}b) for the Higgs doublets
$\chi_{L/R}$ in Landau gauge.
\label{fig:gauge}
}
\end{figure}

As we have already emphasized, stability of the vacuum with
$v_L \neq v_R$ requires \mbox{$\lambda_2>\lambda_1$}. While at
fermion loop level we have $\lambda_2=\lambda_1$, we will see that
the gauge boson loop contributions will automatically secure the
relation $\lambda_2>\lambda_1$. Both $\lambda_1$ and $\lambda_2$
obtain corrections from $U(1)_{B-L}$ gauge boson loops (see
Figs.~\ref{fig:gauge}a and~\ref{fig:gauge}b), whereas only for
$\lambda_1$ there are additional diagrams with $W^i_L$ or $W^i_R$
loops (Fig.~\ref{fig:gauge}).
Since all these contributions have a relative minus sign compared
to those from the fermion loops, one finds $\lambda_2>\lambda_1$
irrespective of the values of the Yukawa or gauge couplings or any
other parameter of the model, provided that the $SU(2)$ gauge
coupling $g_2\neq 0$ [compare the expressions for $\lambda_1$ and
$\lambda_2$ in~(\ref{Lambda})]. Thus the condition for spontaneous
parity breaking is automatically satisfied in our model.

We have a very interesting situation here. In a model with composite
triplets $\Delta_L$ and $\Delta_R$ parity is never broken, i.e. the
model is not phenomenologically viable \footnote{At least if one
does not introduce triplet or higher--representation fermions.}.
At the same time, in the model with two composite doublets $\chi_L$
and $\chi_R$ instead of two triplets (which requires introduction
of an additional singlet fermion $S_L$) parity is broken
automatically. This means that, unlike in the conventional LR
models, in the composite Higgs approach {\em whether or not parity
can be spontaneously broken depends on the particle content of the
model rather than on the choice of the parameters of the Higgs
potential}.

\subsection{Effective mass parameters}
Let us now study the symmetry breaking in the model in detail. The
gauge and parity symmetries can only be broken if the relevant
squared mass terms in the effective potential become negative. These
terms are given by the sums of the bare mass terms and the one--loop
corrections coming from the diagrams of Fig.~\ref{fig:twopt}:
\begin{figure}[p]
\centerline{
\rotate[r]{\epsfxsize=35ex \epsffile{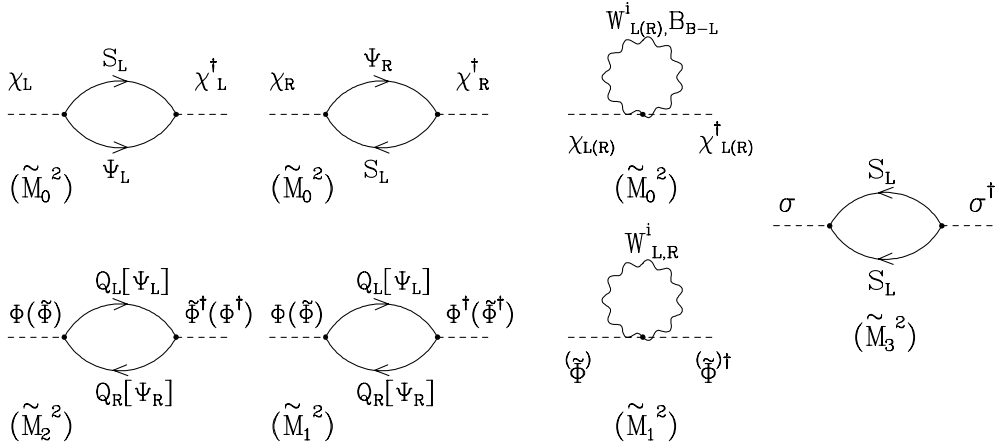} }}
\caption[]{\small\sl
Diagrams contributing to the effective scalar mass terms
$\tilde{M}_0^2 \ldots \tilde{M}_3^2$ in Landau gauge.
\label{fig:twopt}
}
\end{figure}
\begin{figure}[p]
\centerline{
\rotate[r]{\epsfxsize=70ex \epsffile{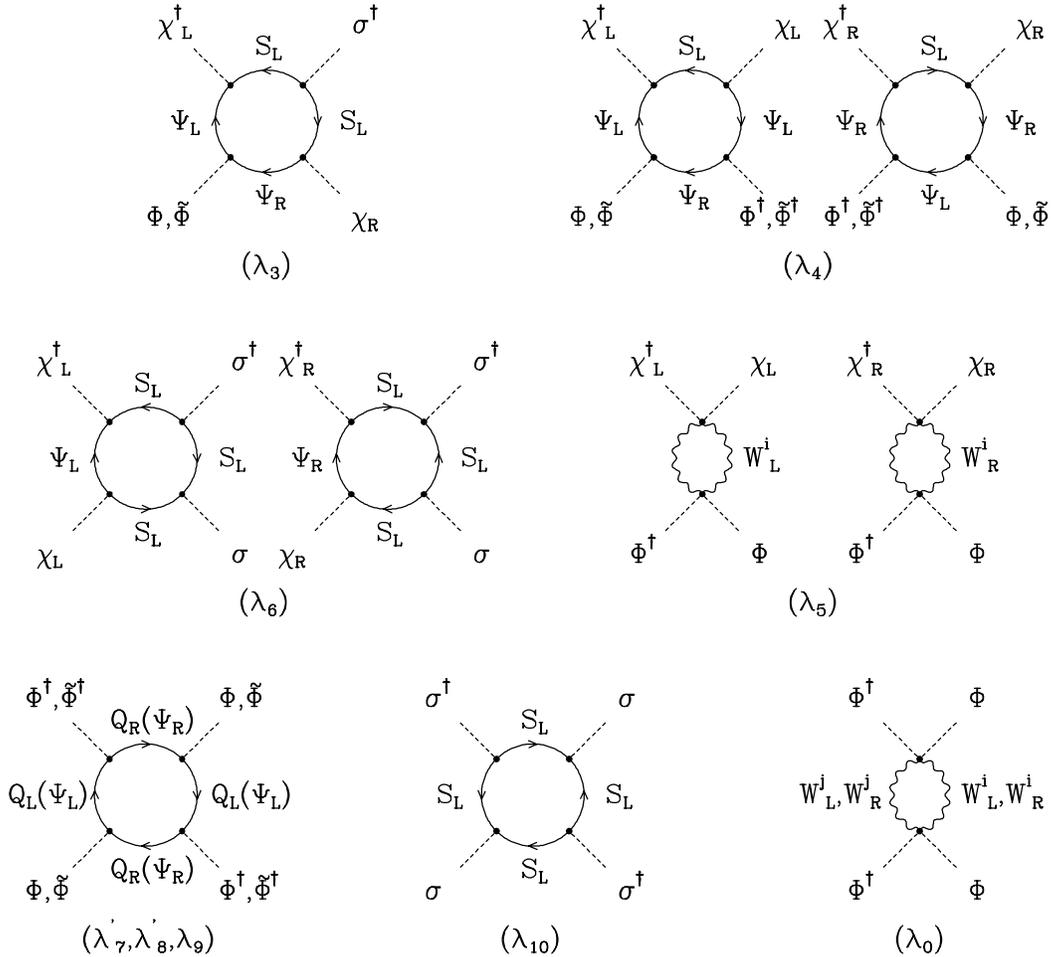} }}
\caption[]{\small\sl
Diagrams contributing to the quartic terms in the Higgs potential in
Landau gauge.
\label{fig:fourpt}
}
\end{figure}
\bea
\tilde{M}_0^2 &=& M_0^2-\frac{1}{8\pi^2}\left [Y_5^2-\frac{3}{8}
Z_{\chi}(3g_2^2+g_1^2)\right ] (\Lambda^2-\mu^2)~,
\label{M02}\\
\tilde{M}_1^2 &=& M_0^2-\frac{1}{8\pi^2}\left \{
\left [N_c(Y_1^2+Y_2^2)+(Y_3^2+Y_4^2)\right ]
-\frac{9}{4}Z_{\phi}g_2^2\right \}(\Lambda^2-\mu^2)~,
\label{M12}\\
\tilde{M}_2^2 &=& M_2^2-\frac{1}{4\pi^2}(N_c Y_1^2 Y_2^2+Y_3^2 Y_4^2)
(\Lambda^2-\mu^2)~,
\label{M22}\\
\tilde{M}_3^2 &=& M_0^2-\frac{1}{4\pi^2}\,Y_6^2\, (\Lambda^2-\mu^2)~.
\label{M32}
\eea
The factors $Z_{\chi}$ and $Z_{\phi}$ multiplying the gauge
contributions in the above formulas for $\tilde{M}^2$ [and also in
the expressions for quartic couplings $\lambda$, see (\ref{Lambda})]
are the prefactors for the gauge invariant kinetic terms of
composite scalars, similar to $Z_\phi$ in eq.~(\ref{Lcomplete}).
They will be given explicitly in eqs.~(\ref{zfactors})
and~(\ref{zfactors2}) below and appear in eqs.~(\ref{M02}),
(\ref{M12}) and (\ref{Lambda}) because the kinetic terms of the
scalars are not yet brought into the canonical form. The running
mass terms coincide with the bare masses at  $\mu=\Lambda$.

While the bare mass parameters $M_i^2$ in eq.~(\ref{Laux}) are
positive, the corresponding running quantities $\tilde{M}_i^2$,
given by eqs.~(\ref{M02}) -- (\ref{M32}), may become negative at
low energy scales provided that the corresponding Yukawa couplings
are large enough. Those values for which this occurs at $\mu=0$ we
shall call the {\em critical } Yukawa couplings. For $\tilde{M}_i^2$
to become negative at some scale $\mu^2 > 0$ the corresponding
Yukawa couplings or combinations of them must be above their
critical values. If this is to happen at scales $\mu\ll\Lambda$
the Yukawa couplings must be fine--tuned very closely to their
critical values to ensure the proper cancelation between the large
bare masses of the scalars and the $\Lambda^2$ corrections in
eqs.~(\ref{M02})--(\ref{M32}). This is equivalent to the usual
fine--tuning problem of gauge theories with elementary Higgs
scalars~\cite{BHL}\footnote{It has been claimed in~\cite{Bl}
that taking into account the loops with the composite Higgs scalars
results in the automatic cancelation of quadratic divergences and
solves the gauge hierarchy problem of the \SM in the BHL approach.
We do not discuss this possibility here.}.

{}From eqs.~(\ref{M02}), (\ref{M12}), (\ref{zfactors}) and
(\ref{zfactors2}) one can find the critical values for the Yukawa
couplings:
\bea
(Y_5^2)_{cr}&\equiv& 8\pi^2
\left(\frac{M_0^2}{\Lambda^2}\right)\left [
1-\frac{3}{8}\,(3g_2^2+g_1^2)\,l_0 \right]^{-1}~,\label{Y5cr} \\
(\tilde{Y}^2)_{cr}&\equiv & 8\pi^2
\left(\frac{M_0^2}{\Lambda^2}\right)
\left [1-\frac{9}{4}g_2^2 \,l_0\right]^{-1}~,    \label{Ycr}
\eea
where $\tilde{Y}^2$ and $l_0$ are defined as
\bea
\tilde{Y}^2 &\equiv& N_c(Y_1^2+Y_2^2)+(Y_3^2+Y_4^2)~,\label{tildeY}\\
l_0 &\equiv & \frac{1}{16\pi^2}\,\ln\left(\frac{\Lambda^2}{\mu^2}
\right)~.
\label{l0}
\eea
For $Y_5^2=(Y_5^2)_{cr}$, $\tilde{M}_0^2$ becomes zero at $\mu^2=0$.
Analogously, for $\tilde{Y}^2=(\tilde{Y}^2)_{cr}$, $\tilde{M}_1^2 =0$
at $\mu^2=0$. Now let us introduce the scales $\mu_{R0}$ and $\mu_1$
through the relations
\bea
\frac{\delta Y_5^2}{(Y_5^2)_{cr}} &\equiv &
\frac{Y_5^2 -(Y_5^2)_{cr}}
{(Y_5^2)_{cr}}\,=\,\frac{\mu_{R0}^2}{ \Lambda^2-
\mu_{R0}^2}\,\simeq \frac{\mu_{R0}^2}{\Lambda^2}~,
\label{deltaY5} \\
\frac{\delta \tilde{Y}^2}{\tilde{Y}^2_{cr}} &\equiv &
\frac{\tilde{Y}^2
-\tilde{Y}^2_{cr}}{\tilde{Y}^2_{cr}}\,=\,\frac{\mu_{1}^2}{ \Lambda^2-
\mu_{1}^2}\,\simeq \frac{\mu_{1}^2}{\Lambda^2}~,
\label{deltaY}
\eea
where the last equalities in (\ref{deltaY5}) and (\ref{deltaY})
hold for $\mu_{R0}^2, \mu_1^2 \ll \Lambda^2$. The meaning of the
scales $\mu_{R0}^2$ and $\mu_1^2$ is very simple, they are the
scales at which $\tilde{M}_0^2$ and $\tilde{M}_1^2$ become zero
for given $\delta Y_5^2>0$ and $\delta\tilde{Y}^2>0$. Consequently,
a negative value of $\mu_1^2$ or $\delta\tilde{Y}^2$ corresponds to
sub--critical $\tilde{Y}^2$, and, as we shall see below, this will
indeed be required in our scenario.

Using eqs.~(\ref{M02}), (\ref{M12}), (\ref{deltaY5}) and
(\ref{deltaY}) one arrives at the following expressions for
$\tilde{M}_0^2(\mu^2)$ and $\tilde{M}_1^2(\mu^2)$:
\bea
\tilde{M}_0^2(\mu^2) &=& M_0^2 \left( \frac{\mu^2-\mu_{R0}^2}
{\Lambda^2-\mu_{R0}^2}\right)\simeq
M_0^2\left( \frac{\mu^2-\mu_{R0}^2}{\Lambda^2}\right)~,
\label{MM02}\\
\tilde{M}_1^2(\mu^2) &=& M_0^2 \left( \frac{\mu^2 - \mu_{1}^2}
{\Lambda^2-\mu_{1}^2}\right)\simeq
M_0^2\left( \frac{\mu^2-\mu_{1}^2}{\Lambda^2}\right)~.
\label{MM12}
\eea
For $\mu^2 \ll \mu_{R0}^2, \mu_1^2$ one finds $\tilde{M}_0^2\simeq -
(M_0^2/\Lambda^2)\,\mu_{R0}^2 \sim -\mu_{R0}^2$ and
$\tilde{M}_1^2\simeq -(M_0^2/\Lambda^2)\,\mu_{1}^2 \sim -\mu_{1}^2$.

\subsection{Pattern of symmetry breaking}
In order to determine the vacuum structure in our model, let us first
consider the extremum conditions for the effective potential $V$,
i.e. the conditions that the first derivatives of $V_{\mbox{eff}}$
with respect to the VEVs of the scalar fields $\sigma_0$, $v_R$,
$v_L$, $\kappa$ and $\kappa'$ vanish:
\beq
\left[\tilde{M}_3^2+\lambda_6(v_L^2+v_R^2)+2\lambda_{10}\sigma_0^2
\right] \sigma_0 + (\lambda_3/2)\, m_D \,v_L\, v_R = 0~;
\label{VS}
\eeq
\beq
\left[\tilde{M}_0^2+2\lambda_1 v_R^2+2\lambda_2 v_L^2 +\lambda_4
m_D^2 +\lambda_5 (\kappa^2+{\kappa'}^2) + \lambda_6\,\sigma_0^2
\right]v_R + (\lambda_3/2)\, m_D\, v_L\, \sigma_0 = 0~;
\label{VVR}
\eeq
\beq
\left[\tilde{M}_0^2+2\lambda_1 v_L^2+2\lambda_2 v_R^2 +\lambda_4
m_D^2 +\lambda_5 (\kappa^2+{\kappa'}^2) + \lambda_6\,\sigma_0^2
\right]v_L + (\lambda_3/2)\, m_D\, v_R\, \sigma_0 = 0~;
\label{VVL}
\eeq
\bea
\left[2\tilde{M}_1^2+2(\lambda_4 Y_3^2 + \lambda_5) (v_L^2+v_R^2)+
4\lambda_7\, \kappa^2+2\lambda_8\,{\kappa'}^2 + 3\lambda_9\,
\kappa \kappa' \right]\kappa + \nonumber \\
2\tilde{M}_2^2 \,\kappa'+\lambda_3 \,Y_3\, v_L \, v_R\,\sigma_0 +
2\lambda_4 \,Y_3\,Y_4\,\kappa' \,(v_L^2+v_R^2)+
\lambda_9\,{\kappa'}^3 = 0~;
\label{VK}
\eea
\bea
\left[2\tilde{M}_1^2+2(\lambda_4 Y_4^2 + \lambda_5) (v_L^2+v_R^2)+
4\lambda_7\, {\kappa'}^2+2\lambda_8\, \kappa^2 + 3\lambda_9\, \kappa
\kappa'  \right]\kappa'  \nonumber + \\
2\tilde{M}_2^2 \,\kappa +\lambda_3 \,Y_4\, v_L \, v_R\,\sigma_0 +
2\lambda_4 \,Y_3\,Y_4\,\kappa \,(v_L^2+v_R^2)+\lambda_9\,\kappa^3=0~.
\label{VK1}
\eea
The quartic couplings $\lambda_i$ are given in
Appendix~\ref{appeffpot} along with the expression for the
effective potential [see eqs.~(\ref{Lambda}) and (\ref{Veff})];
$m_D$ is defined in eq.~(\ref{fmass}).

Multiplying (\ref{VVR}) by $v_L$ and (\ref{VVL}) by $v_R$ and
subtracting we get (for $v_R^2 \neq v_L^2$)
\beq
4\,(\lambda_1-\lambda_2)\,v_L\,v_R = \lambda_3\, m_D\, \sigma_0~,
\label{vevseesaw}
\eeq
which is the analog of the ``VEV seesaw'' relation of
Ref.~\cite{LR3}. Above the scale at which the bi-doublet develops
non--vanishing VEVs $m_D=0$ and hence either $v_L$ or $v_R$ is zero
(or both).

We assume that the scale $\mu_R$ at which parity gets spontaneously
broken (i.e. $\chi_R^0$  develops a VEV) is higher than the \EW scale
$\mu_{EW}\sim 100~GeV$, i.e. that $\tilde{M}_0^2$ changes its sign at
a higher scale than $\tilde{M}_1^2$. This means that the expression
$Y_5^2 - (3/8)Z_\chi(3g_2^2+g_1^2)$ should be bigger than the
combination
\beq
N_c(Y_1^2+Y_2^2)+(Y_3^2+Y_4^2)-\frac{9}{4}Z_{\phi}g_2^2 \equiv
\tilde{Y}^2 - \frac{9}{4}Z_{\phi}g_2^2~.
\label{ineq}
\eeq
At scales $\mu > \mu_{EW}$ the VEVs of the bi--doublet are zero and
it is sufficient to consider eqs.~(\ref{VS}) and (\ref{VVR}). It
follows from eqs.~(\ref{VVL}) and (\ref{vevseesaw}) that $v_L$ can
be consistently set equal to zero in this energy range.

Let us now consider eqs.~(\ref{VS}) and (\ref{VVR}) with
$\kappa=\kappa'=m_D=0$. We have now essentially two possibilities.
Either
\beq
Y_5^2-\frac{3}{8}Z_\chi(3g_2^2+g_1^2)>2\,Y_6^2~,
\label{condit}
\eeq
or vice versa. If eq.~(\ref{condit}) is satisfied, with decreasing
scale $\mu$, $\tilde{M}_0^2$ will become negative earlier than
$\tilde{M}_3^2$, i.e. $\chi_R$ will develop a VEV earlier than
$\sigma$. It is easy to see that in fact $\sigma$ will never
develop a VEV in this case. Indeed, solving (\ref{VVR}) for $v_R$,
substituting it into (\ref{VS}) and using eqs.~(\ref{M02}) and
(\ref{M32}) one can make sure that the combination $\tilde{M}_3^2
+\lambda_6\,v_R^2$, which is the effective ``driving term'' for
the VEV of $\sigma$, never gets negative even if $\tilde{M}_3^2$
does. This means that $\sigma_0$ never appears. Conversely, if the
condition opposite to the one of eq.~(\ref{condit}) holds,
$\tilde{M}_3^2$ will become negative first and the sum
$\tilde{M}_0^2+\lambda_6\,\sigma_0^2$, which is the effective driving
term for $v_R$ in eq.~(\ref{VVR}), never gets negative, i.e. $v_R=0$.
Clearly this situation is phenomenologically unacceptable, and from
now on we therefore assume that eq.~(\ref{condit}) is satisfied,
which means that we choose the 4-f couplings $G_7$ and $G_8$
accordingly.

We have demonstrated that above the \EW scale the VEVs of $\chi_R$
and $\sigma$ never coexist. The question remains whether below the
\EW scale, when the VEVs of the bi--doublet $\phi$ come into play,
$\sigma$ and $\chi_L$ can acquire non--vanishing VEVs. This cannot
be answered by just studying the first derivative conditions
(\ref{VS}) -- (\ref{VK1}). However, as we will show in
Appendix~\ref{apphiggs}, the condition that the matrices of second
derivatives of the effective potential with respect to the fields
be positive definite (i.e. the vacuum stability condition) results
in $\sigma_0=0=v_L$ even when $\kappa,\kappa' \neq 0$. The only
exception may be the situation when the inequality in
eq.~(\ref{condit}) becomes equality. We do not consider such a
possibility since it requires an extreme fine--tuning of the
Yukawa couplings.

Let us now consider the vacuum structure below the \EW breaking
scale. The non--vanishing VEVs are $v_R$, $\kappa$ and $\kappa'$.
Since $m_t \gg m_b$, it follows from  eq.~(\ref{fmass}) that
$\kappa$ should be much larger than $\kappa'$ or vice versa
provided no significant cancelation between $Y_1\kappa'$ and
$Y_2\kappa$ occurs. This is also welcome because of the stringent
upper limit on the $W_L-W_R$ mixing~\cite{LS}:
$\xi \approx 2\kappa\kappa'/(v_R^2-v_L^2) < 0.0025$. To further
simplify the discussion at this point, we shall therefore make the
frequently used assumption~\cite{zero} $\kappa'=0$ which does not
change any symmetry properties of the model. The general case
$\kappa,\kappa' \ne 0$ is discussed in section~\ref{sec:fixp} and
Appendix~\ref{apphiggs}.

For $\kappa'=0$ the relation $m_t \gg m_b$ translates into
$Y_1 \gg Y_2$. In the conventional approach this assumption does
not lead to any contradiction with phenomenology. However, as we
shall see, in our case the condition $\kappa'=0$ cannot be exact.
Consistency of eqs.~(\ref{VK}) and (\ref{VK1}) with $\kappa'=0$
requires $Y_1 Y_2=0$, $Y_3 Y_4=0$ and $M_2^2=0$ (this gives
$\tilde{M}_2^2=\lambda_9 = 0$; as follows from eq.~(\ref{Lambda}),
all the terms in the effective potential which are linear in
$\kappa'$ become zero in this limit, as they should). The condition
$Y_1 Y_2=0$ along with $\kappa'=0$ implies that either $Y_1=0$,
$m_t=0$ or $Y_2=0$, $m_b=0$. The first possibility is obviously
phenomenologically unacceptable whereas the second one can be
considered as a reasonable first approximation; we therefore assume
$Y_2=0$, $Y_1 \neq 0$.

The situation is less clear for the lepton Yukawa couplings $Y_3$ and
$Y_4$. Since $m_\tau \ll m_t$ and the Dirac mass $m_D$ of $\nu_\tau$
is unknown, one can choose either $Y_3 \neq 0$, $Y_4=0$ or $Y_3=0$,
$Y_4 \neq 0$. In Appendix~\ref{apphiggs} we will show that the vacuum
stability condition in our model requires $m_\tau^2-m_D^2>0$, which
implies $Y_4^2 >Y_3^2$, therefore we choose $Y_3=0$ and $Y_4\neq 0$.

\noindent Now with $\sigma_0=v_L=\kappa'=Y_2=Y_3=0$ one can readily
obtain the solutions of eqs.~(\ref{VVR}) and (\ref{VK}):
\bea
v_R^2=\frac{\tilde{M}_1^2\,\lambda_5-2\tilde{M}_0^2\,\lambda_7}
{4\lambda_1\lambda_7-\lambda_5^2}~,
\label{vr2}\\
\kappa^2=\frac{\tilde{M}_0^2\,\lambda_5-2\tilde{M}_1^2\,\lambda_1}
{4\lambda_1\lambda_7-\lambda_5^2}~.
\label{k2}
\eea
To estimate the magnitudes of these VEVs we rewrite them
using eqs.~(\ref{MM02}) and (\ref{MM12}):
\beq
v_R^2(\mu^2)=\frac{1}{4\lambda_1\lambda_7-\lambda_5^2}
\left(\frac{M_0^2}
{\Lambda^2}\right)\left\{\left[2\lambda_7\,\mu_{R0}^2+|\lambda_5|\,
\mu_1^2 \right]-\left[2\lambda_7+|\lambda_5|\right]\mu^2\right\}~,
\label{nvr2}
\eeq
\beq
\kappa^2(\mu^2)=\frac{1}{4\lambda_1\lambda_7-\lambda_5^2}
\left(\frac{M_0^2}
{\Lambda^2}\right)\left\{\left[|\lambda_5|\,\mu_{R0}^2+2\lambda_1\,
\mu_1^2 \right]-\left[2\lambda_1+|\lambda_5|\right]\mu^2\right\}~,
\label{nk2}
\eeq
where we have taken into account that $\lambda_5<0$
[see~(\ref{Lambda})]. From these equations one can find the scales
$\mu_R^2$ and $\mu_{EW}^2$ at which parity and \EW symmetry get
broken, i.e. the VEVs $v_R^2$ and $\kappa^2$ become non--zero:
\bea
\mu_R^2 &=& \frac{2\lambda_7\,\mu_{R0}^2+|\lambda_5|\,\mu_1^2}
{2\lambda_7+|\lambda_5|}
\simeq\mu_{R0}^2+\frac{|\lambda_5|}{2\lambda_7}
\,\mu_1^2\simeq \mu_{R0}^2~,
\label{mur2}\\
\mu_{EW}^2 &=& \frac{|\lambda_5|\,\mu_{R0}^2+2\lambda_1\,\mu_1^2}
{2\lambda_1+|\lambda_5|}\simeq \frac{|\lambda_5|}{2\lambda_1}
\,\mu_{R0}^2+\mu_1^2~,
\label{muew2}
\eea
where in the last equalities in eqs.~(\ref{mur2}) and (\ref{muew2})
we have taken into account $|\lambda_5|\ll \lambda_1,\,\lambda_7$.
Note that eqs.~(\ref{nvr2}) and (\ref{nk2}), as well as
eqs.~(\ref{vr2}) and (\ref{k2}) above, are only valid for
$\mu \leq \mu_R$ and $\mu \leq \mu_{EW}$, respectively.

Now one can rewrite eqs.~(\ref{nvr2}) and (\ref{nk2}) as
\bea
v_R^2(\mu^2) &=&\left(\frac{M_0^2}{\Lambda^2}\right)
\frac{2\lambda_7+|\lambda_5|}{4\lambda_1\lambda_7-\lambda_5^2}
\left( \mu_{R}^2-\mu^2 \right)\simeq \left(\frac{M_0^2}{\Lambda^2}
\right)\frac{ \mu_{R}^2-\mu^2 }{2\lambda_1}~,
\label{newvr2}\\
\kappa^2(\mu^2) &=& \left(\frac{M_0^2}{\Lambda^2}\right)
\frac{2\lambda_1+|\lambda_5|}{4\lambda_1\lambda_7-\lambda_5^2}
\left( \mu_{EW}^2-\mu^2 \right)\simeq \left(\frac{M_0^2}{\Lambda^2}
\right) \frac{\mu_{EW}^2-\mu^2 }{2\lambda_7}~.
\label{newk2}
\eea
We are interested in $v_R^2(0)\equiv v_R^2$ and $\kappa^2(0)\equiv
\kappa^2$ which determine the masses of all the fermions and gauge
bosons in the model:
\beq
v_R^2 \simeq \left(\frac{M_0^2}{\Lambda^2}\right)
\frac{\mu_{R}^2}{2\lambda_1}~,
\label{vr20}
\eeq
\beq
\kappa^2\simeq \left( \frac{M_0^2}{\Lambda^2}\right)
\frac{\mu_{EW}^2}{2\lambda_7}~.
\label{k20}
\eeq
This gives
\beq
\frac{\kappa^2}{v_R^2}\simeq \left(\frac{\lambda_1}{\lambda_7}\right)
\frac{\mu_{EW}^2}{\mu_R^2}\sim \frac{\mu_{EW}^2}{\mu_R^2}\simeq
\frac{|\lambda_5|}{2\lambda_1}+\frac{\mu_1^2}{\mu_R^2}~.
\label{ratio}
\eeq
Recall now that in conventional LR models with
$\mu_{EW}\ll\mu_R \ll\Lambda_{GUT}$ (or $\Lambda_{\rm Planck}$)
one has to fine--tune two gauge hierarchies:
$\Lambda_{GUT}\dash\mu_R$ and $\mu_R\dash\mu_{EW}$.
We have a similar situation here: to achieve $\mu_{EW} \ll \mu_R
\ll \Lambda$ one has to fine--tune two Yukawa couplings, $Y_5^2$ and
$\tilde{Y}^2$ [see (\ref{deltaY5}) and (\ref{deltaY})]. Tuning of
$Y_5^2$ allows for the hierarchy $\mu_R^2 \ll \Lambda^2$; one then
needs to adjust $\tilde{Y}^2$ (or $\mu_1^2$) to achieve
$\mu_{EW}^2 \ll \mu_R^2$ through eq.~(\ref{ratio}).

Since $\lambda_5$ only contains relatively small gauge couplings
while $Y_5 \sim \cal{O}$$(1)$, we typically have
$|\lambda_5|/2\lambda_1\sim 10^{-2}$. Thus, if there is no
significant cancellation between the two terms in (\ref{ratio}),
one gets the right--handed scale of the order of a few $TeV$.
Unfortunately, such a low LR scale scenario is not viable.
As we shall show in Sec.~\ref{sec:bubblepred}, the squared masses of
two Higgs bosons in our model become negative (i.e. the vacuum
becomes unstable) unless $v_R\,\gtap \,20~TeV$. This requires
some cancelation in eq.~(\ref{ratio})\footnote{Note that this
does not increase the number of the parameters to be tuned but
just shifts the value to which one of them should be adjusted.},
and then the right--handed scale $v_R\sim \mu_R$ can in principle
lie anywhere between a few tens of $TeV$ and $\Lambda$. However,
if one prefers ``minimal cancelation'' in eq.~(\ref{ratio}),
by about two orders of magnitude or so, one would arrive at a value
of $v_R$ around $20~TeV$. It is interesting that the partial
cancelation of the two terms in (\ref{ratio}) implies $\mu_1^2<0$,
i.e. $\tilde{Y}^2$ must be below its critical value [see
eq.~(\ref{deltaY})]. This means that $\tilde{M}_1^2$ never
becomes negative. In fact it is the $\tilde{M}_0^2$ term,
responsible for parity breakdown, that also drives the VEV of the
bi--doublet. One can see from eq.~(\ref{VK}) that the effective
driving term for $\kappa$ is $\tilde{M}_1^2+\lambda_5 v_R^2$;
it may become negative for large enough $v_R^2$ even if
$\tilde{M}_1^2$ is positive since $\lambda_5<0$ [see
eq.~(\ref{Lambda})]. Thus we have a tumbling scenario where the
breaking of parity and $SU(2)_R$ occurring at the scale $\mu_R$
causes the breaking of the \EW symmetry at a lower scale $\mu_{EW}$.


\section{Predictions in Bubble Approximation}
\label{sec:bubblepred}
We can now obtain the predictions for fermion and Higgs boson
masses in bubble approximation. As we have mentioned before, the
static Higgs fields acquire gauge--invariant kinetic terms at low
energies through radiative corrections, and the corresponding wave
function renormalization constants can be derived from the 2--point
Green functions. Direct calculation yields
\bea
Z_{\phi} & = & \sixtpi\, \[N_c (Y_{1}^{2}+Y_{2}^{2})+
Y_{3}^{2}+Y_{4}^{2}\]\, \lnlambda~,
\label{zfactors}\\
Z_{\chi} & = & \sixtpi\, Y_{5}^{2}\, \lnlambda~,
\label{zfactors2}\\
Z_{\sigma} & = & \sixtpi\, 2 Y_6^2\, \lnlambda~. \label{zfactors3}
\eea
To extract physical observables one should first rescale the Higgs
fields so as to absorb the relevant $Z$ factors into the definitions
of the scalar fields and bring their kinetic terms into the
canonical form. This amounts to dividing the squared mass terms by
the corresponding $Z$ factors, Yukawa couplings by $\sqrt{Z}$ and
multiplying the scalar fields and their VEVs by $\sqrt{Z}$.
Renormalization factors of the quartic couplings depend on the
scalar fields involved and can be readily read off from the
effective potential (see Appendix \ref{apprenorm} for more details).

As we have already pointed out, the minimization of the effective
Higgs potential gives $\sigma_0=0=v_L$. This reduces the neutrino
mass matrix (\ref{numass1}) to
\beq
M_{\nu} = \left(
\begin{array}{c}
\end{array}
\begin{array}{ccc}
0 & m_D & 0\\
m_D & 0 & M\\
0 & M & 0
\end{array} \right)\ \; .
\label{numass2}
\eeq
Diagonalization of this matrix gives the following neutrino mass
eigenstates and eigenvalues:
\beq
\begin{array}{lclrrlcc}
\nu_{1} & = & \cos\theta~\nu_{L}-\sin\theta~S_{L} & & & m_{1} & =
            & 0  \\
\nu_{2} & = & \frac{1}{\sqrt{2}}\(\sin\theta~\nu_{L}+\nu_{L}^{c}+\cos
           \theta~S_{L}\) & & & m_{2} & = & \sqrt{M^{2}+m_{D}^{2}} \\
\nu_{3} & = & \frac{1}{\sqrt{2}}\(\sin\theta~\nu_{L}-\nu_{L}^{c}+\cos
           \theta~S_{L}\) & & & m_{2} & = & -\sqrt{M^{2}+m_{D}^{2}}
\end{array}
\label{neutrinomasses}
\eeq
with the mixing angle
\beq
\sin\theta = \frac{m_{D}}{\sqrt{M^{2}+m_{D}^{2}}}~.
\label{numix}
\eeq
Thus, we have one massless left--handed neutrino $\nu_{1}$ and two
heavy Majorana neutrinos with degenerate masses
$\sqrt{M^{2}+m_{D}^{2}}$ and opposite $CP$--parities which combine
to form a heavy Dirac neutrino. Since $m_D \ll M$ the \EW eigenstate
$\nu_L\equiv \nu_{\tau}$ is predominantly the massless eigenstate
whereas the right--handed neutrino $\nu_R$ and the singlet fermion
$S_L$ consist predominantly of the heavy eigenstates.

The gauge boson masses for our symmetry breaking scenario can be
found in Appendix~\ref{appeffpot}. For $v_R \gg \kappa,\kappa'
\gg v_L$, which implies strong parity violation at low energies
and small LR--mixing, the masses reduce to
\beq
\begin{array}{ll}
m^2_{W_L} \approx \frac{g_2^2}{2}(\hat{\kappa}^2+\hat{\kappa}'^2)~, &
m^2_{W_R} \approx \frac{g_2^2}{2} \hat{v}_R^2 ~,\\
 & \\
m_{Z_L}^2 \approx \frac{g_2^2}{2} \sec\theta_W^2
(\hat{\kappa}^2~+\hat{\kappa}'^2)~, &
m_{Z_R}^2 \approx \frac{g_2^2+g_1^2}{2} \hat{v}_R^2~, \end{array}
\eeq
where the ``hats'' denote renormalized quantities. The usual VEV of
the \SM should be identified with
$(\hat{\kappa}^2+\hat{\kappa}'^2)^{1/2}$.

\subsection{Relations between the scalar VEVs and fermion masses for
$\kappa'=0$}

We will now assume that $\kappa'=0$ which will simplify the
discussion considerably. The general case $\kappa,\kappa' \ne 0$
will be considered at the end of this section and in
Sec.~\ref{sec:fixp}. It was shown above that $\kappa'=0$ requires
$Y_2=0=Y_3$, which yields
\beq
m_t=Y_1\kappa = \hat{Y}_1\hat{\kappa}~, \quad m_{\tau}=Y_4\kappa =
\hat{Y}_4\hat{\kappa}~, \quad m_b=m_D=0~.
\label{fermmasses}
\eeq
Vanishing Dirac neutrino mass $m_D$ implies $\sin\theta=0$, and the
heavy neutrino mass is now
\beq
M=Y_{5}v_{R}= \hat{Y}_5 \hat{v}_R~.
\eeq
{}From eqs.~(\ref{fermmasses}) and (\ref{zfactors}) and the
definition of the  renormalized Yukawa couplings one can readily
find
\beq
\hkap2=(174~GeV)^{2}
= N_c m_{t}^{2}\(1+\frac{Y_{4}^{2}}{N_{c}Y_{1}^{2}}\)\sixtpi\lnlambda
\approx  m_{t}^{2}
\frac{N_c}{16\pi^2}\lnlambda~.
\label{mtopbubble}
\eeq
This expression coincides with the one derived in bubble
approximation by BHL~\cite{BHL} and we will see in
Sec.~\ref{sec:fixp} that this already corresponds to the \RG
analysis in bubble approximation. Eq.~(\ref{mtopbubble}) gives
the top quark mass in terms of the known \EW VEV and the scale of
new physics $\Lambda$. For example, for $\Lambda=10^{15}~GeV$ one
finds $m_t=165~GeV$. However, this result is limited to the bubble
approximation, and the \RG improved result will be substantially
higher. Note that $m_t \approx 180~GeV$, which is the central value
of the Fermilab results~\cite{CDF,D0}, would mean
$\hat{Y}_1 \approx 1$, or
\beq
\l_0 \equiv \sixtpi \lnlambda \approx \frac{1}{3}~.
\label{ln}
\eeq
Similar considerations lead to the following relation between the
right--handed VEV $\hat{v}_R$, heavy neutrino mass $M$ and the scale
$\Lambda$:
\beq
\hvr2 =  M^{2}\sixtpi\lnlambda~.
\label{Vr2}
\eeq
Note that $\mu \approx m_t$ is understood in eqs.~(\ref{mtopbubble})
and (\ref{ln}), whereas $\mu \approx M$ in eq.~(\ref{Vr2}). However,
we assume $m_t,M\ll \Lambda$ and $M/m_t \ll \Lambda/M$  throughout
this paper, therefore $\ln\frac{\Lambda^2}{m_t^2} \approx
\ln\frac{\Lambda^2}{M^2}$, i.e. the logarithms are universal.
{}From eqs.~(\ref{mtopbubble}) and (\ref{Vr2}) one thus finds
\beq
\frac{\hvr2}{M^2}\approx \frac{1}{3}\frac{\hkap2}{m_{t}^2}~.
\label{VrtoM}
\eeq
The mass of the $\tau$ lepton is not predicted in our model since
it is only weakly coupled to the bi--doublet; it is given by
\beq
m_{\tau} = \frac{Y_4}{Y_1} m_t~,
\label{mtau}
\eeq
and can be adjusted to a desirable value by choosing the proper
magnitude of the ratio $Y_4/Y_1$ (i.e. of $G_3/G_1$, see
eq.~(\ref{G4f}) of Appendix A).

\subsection{Higgs Boson Masses}
The Higgs boson masses can be obtained either as the poles of the
corresponding propagators or, in the auxiliary field approach, by
diagonalizing the matrices of second derivatives of the effective
Higgs potential. In either case it is essential to use the
(non--trivially satisfied) gap equation to cancel the large bare
mass term in order to obtain a small composite Higgs mass. For
example in the first approach, the inverse propagators of the
composite scalar fields have a generic form
\beq
[iD(p^2)]^{-1} = M_0^2 +\Pi(p^2)~,
\label{prop}
\eeq
where $M_0^2 \sim \Lambda^2$ is the bare mass term and $\Pi(p^2)$
is the polarization operator. Using the gap equation one obtains
$\Pi(p^2)=-M_0^2-Z[p^2-4 m_f^2+{\cal O}(m_f^2/\ln(\Lambda^2/m_f^2)]$.
Here $m_f$ is the mass of the fermion whose bound state forms the
composite Higgs boson. Thus the $M_0^2$ term in (\ref{prop}) is
exactly canceled, resulting in a light composite scalar with a mass
of about $2m_f \ll \Lambda$. Without this cancelation the Higgs
boson mass would have been of order $\Lambda$. Then the scalar
field could either correspond to a real physical propagating state,
or to a broad resonance, or the composite state might not exist at
all, since at energy scales of ${\cal{O}}({\Lambda})$ the effective
4-f operators are not sufficient to reliably describe the theory.
In any case this field would be decoupled from the low--energy
spectrum of the model. In our model this is the case for the
$\sigma$ field since we assume the 4-f coupling $G_8$ to be
sub--critical (see eq.~(\ref{condit}) and the following discussion).

The situation is, however, quite different for the $\chi_L$ fields
even though the corresponding gap equation (\ref{VVL}) is only
trivially satisfied $(v_L=0)$. First, one observes that $v_R\neq 0$
and so the gap equation for $v_R$ $is$ non--trivially satisfied.
At the same time, the gap equations for $v_R$ and $v_L$
[eqs.~(\ref{VVR}) and (\ref{VVL})] have very similar structure;
direct inspection shows that they differ from each other just by
the interchange of $\lambda_1$ and $\lambda_2$, which is a
consequence of the discrete parity symmetry. One can therefore use
the information contained in the gap equation for $v_R$ to cancel
the large bare mass term in the propagator of $\chi_L$. As a result,
$\chi_L$ turns out to be a light physical propagating state.
Direct calculation shows that its mass vanishes in the limit
$(\lambda_2-\lambda_1)=0$, i.e. $g_2 = 0$, which corresponds to
the fermion bubble approximation (see eq.~(\ref{MH40}) below and
eqs.~(\ref{imag}) and (D60) in Appendix~\ref{apphiggs}).

It should be possible to understand these vanishing scalar masses
as a signal of some symmetry. Indeed, in the limit
$\lambda_2=\lambda_1$ the ($\chi_L,\chi_R$) sector of the
effective Higgs potential [eq.~(\ref{Veff})] depends on $\chi_L$
and $\chi_R$ only through the combination
$(\chi_L^\dagger \chi_L+\chi_R^\dagger \chi_R)$. This means that the
potential has a global $SU(4)$ symmetry which is larger than the
initial $SU(2)_L\times SU(2)_R\times U(1)_{B-L}$ symmetry
[$\{\chi_L^0,\chi_L^-,\chi_R^0,\chi_R^-\}$ forms a fundamental
representation of $SU(4)$]. After $\chi_R^0$ gets a non--vanishing
VEV $v_R$, the symmetry is broken down to $SU(3)$, resulting in
$15 - 8=7$ \GBsp Three of them ($\chi_R^\pm$ and $\Im \chi_R^0$)
are eaten by the $SU(2)_R$ gauge bosons $W_R^\pm$ and $Z_R$, and
the remaining four ($\chi_L^\pm$, $\Re \chi_L^0$ and $\Im \chi_L^0$)
are physical massless \GBsp The $SU(4)$ symmetry is broken by the
$\phi$--dependent terms in the effective potential and by $SU(2)$
gauge interactions. As a result, $\chi_L^\pm$, $\Re \chi_L^0$ and
$\Im \chi_L^0$ acquire small masses and become pseudo--\GBsp
In fact, the origin of this approximate $SU(4)$ symmetry can be
traced back to the 4-f operators of eq.~(\ref{Lint3}). It is an
accidental symmetry resulting from the gauge invariance and parity
symmetry of the $G_7$ term. Note that no such symmetry occurs in
conventional LR models.

We now present our results for the composite Higgs bosons for the
case $\sigma_0=v_L=\kappa'=0$. Further information and more general
results are contained in Appendix \ref{apphiggs}. We have the
following \GBs in our model\footnote{Hereafter we omit the ``hats''
over the renormalized quantities.}:
\beq
\begin{array}{lll}
G_1^\pm =\frac{1}{\sqrt{v_R^2+\kappa^2}}\(\kappa \phi_2^\pm +v_R
\chi_R^\pm \) \approx \chi_R^\pm~, &\;\;& G_1^0 = \chi_{Ri}^0~,
\label{GB1} \\
G_2^\pm = \phi_1^\pm~,    &\;\;& G_2^0 = \phi_{1i}^0~, \label{GB2}
\end{array}
\eeq
where $G_1^\pm$,$G_1^0$ are eaten by $W_R^\pm$, $Z_R$  and
$G_2^\pm$,$G_2^0$ by $W_L^\pm$, $Z_L$,
respectively.
\noindent
The physical Higgs boson sector of the model contains two $CP$--even
neutral scalars
\bea
H_1^0 & \approx &\(1-\frac{\lambda_5^2}{8\lambda_1^2}\frac{\kappa^2}
{v_R^2}\)\chi_{Rr}^0+\frac{\lambda_5}{2\lambda_1}\frac{\kappa}{v_R}
\phi_{1r}^0\approx \chi_{Rr}^0~\;, \label{H10}  \\
H_2^0 & \approx & -\frac{\lambda_5}{2\lambda_1}\frac{\kappa}{v_R}
\chi_{Rr}^0
+\(1-\frac{\lambda_5^2}{8\lambda_1^2}\frac{\kappa^2}
{v_R^2}\)\phi_{1r}^0 \approx \phi_{1r}^0\;, \label{H20}
\eea
with the masses
\bea
M_{H_1^0}^2 & \simeq & 4M^{2}\[1-\frac{3}{16}
 \(3g_2^{4}+2g_2^{2}g_1^{2}+g_1^{4}\)l_0^2\]\approx 4M^2 \;,
\label{MH10}\\
M_{H_2^0}^2 & \simeq & 4 m_t^2 \(1-\frac{m_{\tau}^{2}}{3 m_{t}^{2}} -
\frac{9}{4}g_2^4 l_0^2 \) \approx 4 m_{t}^{2}\;, \label{MH20}
\eea
which are directly related to the two steps of symmetry breaking,
$SU(2)_R\times U(1)_{B-L}\rightarrow U(1)_Y$ and
$SU(2)_L\times U(1)_Y    \rightarrow U(1)_{em}$.
Further, there are the charged Higgs boson $H_3^\pm$ with its neutral
$CP$--even and $CP$--odd partners $H_{3r}^0$ and $H_{3i}^0$,
\bea
H_3^\pm &=&\frac{1}{\sqrt{v_R^2+\kappa^2}}\(-\kappa \chi_R^\pm +v_R
\phi_2^\pm \) \approx \phi_2^\pm \;, \label{H3pm} \\
H_{3r}^0 & = & \phi_{2r}^0 \;, \quad \quad H_{3i}^0 = \phi_{2i}^0 \;,
\label{H30}
\eea
and the previously mentioned pseudo--\GBs $\chi_L$:
\bea
H_4^\pm = \chi_L^\pm\;, \quad \quad
H_{4r}^0=\chi_{Lr}^0\;, \quad \quad H_{4i}^0=\chi_{Li}^0~,
\eea
with the masses
\bea
M_{H_3^\pm}^2 &\approx&  \frac{2}{3}M^{2}
\frac{m_{\tau}^{2}}{m_{t}^{2}}~,
\label{MH3pm} \\
M_{H_{3r}^0}^2 = M_{H_{3i}^0}^2 & \approx & \frac{2}{3}M^{2}
\frac{m_{\tau}^{2}}{m_{t}^{2}} - \frac{1}{2} M_{H_2^0}^2~,
\label{MH30}\\
M_{H_4^\pm}^2 &=& \frac{3}{8}\(3g_2^4+2g_2^2g_1^2\)l_0^2\,M^2+
2 m_{\tau}^2~,\label{MH4pm} \\
M_{H_{4r}^0}^2 = M_{H_{4i}^0}^2 &=&
\frac{3}{8}\(3g_2^4+2g_2^2g_1^2\)l_0^2\,M^2~. \label{MH40}
\eea
Note that the last expression for the mass of the neutral
pseudo--\GBs is proportional to $\lambda_2-\lambda_1$.

Altogether we have 4 physical charged scalars, 4 $CP$--even and
2 $CP$--odd physical neutral scalars. The mass of the scalar $H_2^0$,
which is the analog of the \SM Higgs boson [eq.~(\ref{MH20})],
essentially coincides with the one obtained in bubble approximation
by BHL~\cite{BHL}. This just reflects the fact that this boson is
the $t\bar{t}$ bound state with a mass of  $\approx 2 m_t$.
Analogously, the mass of the heavy $CP$--even scalar
$H_1^0 \approx \chi_{Rr}^0$ is approximately $2M$ since it is a
bound state of heavy neutrinos, $\chi_{R}^{0}\sim\bar{S}_{L}\nu_R=
[\bar{\nu}_{2L}(\nu_{2L})^c-\bar{\nu}_{3L}(\nu_{3L})^c]/2$.

In conventional LR models only one scalar, which is the analog of
the \SM Higgs boson, is light (at the \EW scale), all the others
have their masses of the order of the right--handed scale
$M$~\cite{LR2,LR3,LRdesh}. In our case, the masses of those
scalars are also proportional to $M$, but all of them except the
mass of $H_1^0$ have some suppression factors. The masses
of $H_4 = \chi_L$ are suppressed because of their pseudo--Goldstone
nature and vanish in the limit $g_2\rightarrow 0$, $m_{\tau}\to 0$.
In fact, though the $SU(2)$ gauge coupling constant $g_2$ is
smaller than the typical Yukawa constants in our model, it is not
too small; taking the estimate of $\l_0$ from eq.~(\ref{ln}), one
arrives at $M_{\chi_L} \sim 10^{-1} M$.

The mass of the charged scalars $H_3^\pm \approx \phi_2^\pm$ is
suppressed by the factor $m_{\tau}/m_t$ and is therefore of the
order $10^{-2} M$. The masses of the neutral $H_{3r}^0$ and
$H_{3i}^0$ are even smaller; they are related to the masses of
charged $H_3^\pm$ and the \SM Higgs $H_2^0$ by eq.~(\ref{MH30}).
In fact this equation imposes an upper limit on the \SM Higgs boson
mass $M_{H_2^0}$ (for a given $M$) or a lower limit on the
right--handed mass $M$ (for a given $M_{H_2^0}$). These limits follow
from the requirement that $M_{H_3^0}^2$ be positive, i.e. from the
vacuum stability condition. For example, for $M_{H_2^0} \approx
200~GeV$ we find $M\gtap 17~TeV$. This is the lower bound on the
right--handed scale that we mentioned\footnote{The current
experimental lower bound on the \SM Higgs boson mass is only
$60~GeV$, which would yield $M\gtap 5~TeV$. It is interesting to
notice that the same lower limit $M\gtap 5~TeV$ would result from
eq.~(\ref{MH3pm}) and the experimental lower bound on the charged
Higgs boson mass $M^\pm > 45~GeV$ which follows from the LEP data.
However, in our model the mass of the standard--model--like
(composite) neutral Higgs boson must be at least of the order of
the top quark mass, which yields the above--mentioned estimate
$M\gtap 17~TeV$.} in Sec.~\ref{sec:effpot}.

Thus, we have a number of intermediate scale Higgs bosons with
relations between masses of various scalars
[eqs.~(\ref{MH3pm})--(\ref{MH40})] and between fermion and Higgs
boson masses [eqs.~(\ref{MH10}) and (\ref{MH20})] which are in
principle testable. If the right--handed scale $v_R$ is of the
order of a few tens of $TeV$, the neutral $CP$--even and $CP$--odd
scalars $H_{3r}^0$ and $H_{3i}^0$ can be even lighter than the
\EW Higgs boson. In fact, they can be as light as $\sim 60~GeV$
and so might be observable at LEP2.

Finally, we would like to comment on the approximation $\kappa'=0$
which we used so far. Clearly, this is an oversimplification: In a
realistic model with non--vanishing $m_b$ and $m_D$ both $\kappa$
and $\kappa'$ must be non--zero (note that the Yukawa couplings
$Y_2$ and $Y_3$ will also be non--zero in this case). However,
these masses are not predicted in our model and can merely be
adjusted to desirable values. The Dirac neutrino mass $m_D$ is
unknown and so remains a free parameter; however, it must be
smaller than $m_{\tau}$ in our model in order to satisfy the vacuum
stability condition $(Y_4^2-Y_3^2)(\kappa^2-\kappa'^2)>0$ [see
eqs.~(\ref{AppMH3pm}) and (\ref{AppMH3i})] which is equivalent
to $m_{\tau}^2-m_D^2>0$. The Higgs boson masses and mass
eigenstates for the general case $\kappa,\kappa' \ne 0$ are given
in Appendix~\ref{apphiggs}. In the next section we will obtain \RG
improved predictions for the fermion masses and show that some
interesting results (including a viable top quark mass) emerge for
sizeable values of $\kappa'$, $\kappa' \sim \kappa$.


\section{Renormalization Group Improved Predictions}
\label{sec:fixp}

In the preceding sections we studied our model and derived its
predictions in bubble approximation, taking into account only
fermion and $SU(2)_L\times SU(2)_R \times U(1)_{B-L}$ gauge boson
loops. However, important corrections arise from QCD effects and
loops with composite Higgs scalars. Following the approach of
BHL~\cite{BHL}, we will incorporate these effects by solving the
full one--loop renormalization group equations of the low energy
effective LR--model with boundary conditions corresponding to
compositeness.

These boundary conditions follow from the vanishing of the
radiatively induced kinetic terms for the Higgs scalars at the
scale $\Lambda$, where the composite particles break up into their
constituents:
\beq
Z_{\phi}(\mu^2 \to \Lambda^2)=Z_{\chi}(\mu^2 \to \Lambda^2)=
Z_{\sigma}(\mu^2 \to \Lambda^2)=0~.
\label{zboundary}
\eeq
After rescaling the scalar fields so as to bring their kinetic terms
into the canonical form (see Sec.~\ref{sec:bubblepred} and
Appendix~\ref{apprenorm}), eqs.~(\ref{zboundary}) result in boundary
conditions for the Yukawa and quartic couplings of the low energy
effective Lagrangian. These conditions are similar to those
obtained by BHL [eq.~(\ref{compcon})] and have the following
generic form:
\beq
\hat{Y}^2 = \frac{Y^2}{Z} \stackrel{\mu^2\rightarrow\Lambda^2}
{\longrightarrow}\infty~,\;\;
\hat{\lambda}=\frac{\lambda}{Z^2}\stackrel{\mu^2\rightarrow\Lambda^2}
{\longrightarrow} \infty~,\;\;
\frac{\hat{\lambda}}{\hat{Y}^4}=\frac{\lambda}{Y^4}
\stackrel{\mu^2\rightarrow\Lambda^2}{\longrightarrow} 0~.
\label{comp3}
\eeq
The renormalized parameters of our model derived in bubble
approximation already satisfy these compositeness conditions; for
example, the renormalized Yukawa couplings are
\bea
\hat{Y}_1^2(\mu) & = & \frac{Y_1^2}{Z_\phi}
                 \approx \[\frac{3}{16\pi^2} \lnlambda \]^{-1}~;
                \;\;\;\;(\hat{Y}_4 \ll \hat{Y}_1)~,
\label{boundcon1} \\
\hat{Y}_4^2(\mu) & \approx & \frac{Y_4^2}{Y_1^2}\; \hat{Y}_1^2(\mu)
                         =   \frac{G_3}{G_1}    \; \hat{Y}_1^2(\mu)~,
\label{boundcon2}  \\
\hat{Y}_5^2(\mu) & = & \frac{Y_5^2}{Z_\chi} =\[
                \frac{1}{16\pi^2} \lnlambda \]^{-1}~,
\label{boundcon3} \\
\hat{Y}_6^2(\mu) & = & \frac{Y_6^2}{Z_\sigma} =\[
                \frac{2}{16\pi^2} \lnlambda \]^{-1}
\label{boundcon6}~.
\eea
Obviously they diverge as $\mu\rightarrow\Lambda$. Furthermore, their
running coincides exactly with that described by the fermion loop
contributions to one--loop $\beta$--functions of the corresponding
LR--theory. These fermion loop contributions can be read off from
the trace terms in eqs.~(\ref{beta1})--(\ref{beta6}) in
Appendix~\ref{appbeta} where the full set of gauge and Yukawa
$\beta$--functions for our model is given.

The idea is now to identify the Landau poles in the {\em full}
one--loop \RG evolution of couplings with the compositeness scale
$\Lambda$ and run the couplings down to low energy scales.

\subsection{The case $\kappa'=0$}

We will first consider the simplified scenario with $\kappa'=0$.
The case $\kappa'\neq 0$ which leads to phenomenologically
acceptable fermion masses will be discussed in the next subsection.
\renewcommand{\thefootnote}{\fnsymbol{footnote}}
In our one--generation scenario the \RG equations for the Yukawa
couplings in the limit $\kappa'=0$ (which requires $Y_2=Y_3=0$, see
Sec.~\ref{sec:effpot}) reduce to\footnote[2]{In the following we
will omit the hats over the renormalized quantities.}
\renewcommand{\thefootnote}{\arabic{footnote}}
\bea
\SIXTPI \frac{dY_1}{dt} & = &
5Y_1^3+Y_1 Y_4^2 - \(8 g_3^2+\frac{9}{2}g_2^2+\frac{1}{6}g_1^2\)Y_1~,
\label{betatrun1}\\
\SIXTPI \frac{dY_4}{dt} & = &
3Y_4^3+3 Y_4 Y_1^2+Y_4 Y_5^2 -
\(\frac{9}{2}g_2^2+\frac{3}{2}g_1^2\)Y_4~,
\label{betatrun2}\\
\SIXTPI \frac{dY_5}{dt} & = &
 \frac{7}{2}Y_5^3 + Y_5\(Y_4^2+[2Y_6^2]\)
              -\(\frac{9}{4}g_2^2+\frac{3}{4}g_1^2\)Y_5~,
\label{betatrun3}\\
\SIXTPI \frac{dY_6}{dt} & = &
 [6]Y_6^3 + 4Y_6 Y_5^2~.
\label{betatrun4}
\eea
For large values of $Y_i$ the $Y_i^3$--contributions in the
$\beta$--functions are dominant; they quickly drive the couplings
down to values of order one as the scale $\mu$ decreases. In this
regime gauge and other Yukawa coupling contributions become
important, and the interplay of these contributions and $Y_i^3$
terms result in so--called infrared quasi--fixed
points~\cite{quasiIR}. Thus a large range of initial values of
the Yukawa couplings at the cutoff is focused into a small range
at low energies. The masses of the fermions will then be given
implicitly by conditions of the kind $Y(m)\cdot VEV = m$.

Strictly speaking it is not legitimate to evolve the Yukawa
couplings with one--loop $\beta$--functions to their Landau poles,
i.e. in the non--perturbative regime. However, it has been argued
in~\cite{BHL} that this should not result in any significant errors
since (1) the running time $t=\ln\mu$ in the non--perturbative
domain is only a few percent of the total running time and, more
importantly, (2) the infrared quasi--fixed point structure of the
\RG equations makes the predictions fairly insensitive to the
detailed behavior of the solutions in the large Yukawa coupling
domain. Lattice gauge theory has generally confirmed the
reliability of perturbation theory in this fixed point
analysis~\cite{latticeFP}.

Except for switching to the \SM \RG equations below the parity
breaking scale the only relevant threshold effects in the evolution
are due to the masses of the $\sigma$ scalars. We have shown in
bubble approximation that the VEVs of $\sigma$ and $\chi_R$ do not
coexist; phenomenology then dictates the choice $\sigma_0=0$,
$v_R \neq 0$ which requires the 4-f coupling $G_8$ to be
sub--critical, or at least satisfying the condition (\ref{condit}).
We will assume that the same holds true beyond the bubble
approximation (although we were unable to prove that) and consider
$\sigma$ to be non--propagating, or at least decoupled from the
low--energy spectrum of the model. We will therefore switch off
effects from propagating $\sigma$ scalars directly at the cutoff
by neglecting the contributions in square brackets of
eqs.~(\ref{betatrun3}) and (\ref{betatrun4}). In this limit the
running of $Y_6$ does not influence the running of the other
Yukawa couplings.

In numerical calculations a large number $Y_i(\Lambda)$ must be
used as boundary condition for Yukawa couplings instead of infinity.
Fortunately, the infrared quasi--fixed point structure of the \RG
equation makes the solutions fairly insensitive to the actual
values of $Y_i(\Lambda)$ provided that they are large
enough~\cite{BHL,quasiIR}. In fact, the infrared quasi--fixed--point
behavior sets in already for $Y_i(\Lambda)\gtap 5$. In our
calculations we have chosen $Y_5(\Lambda)=10$; taking e.g. $10^3$
instead of 10 results only in a correction of about $0.4\%$ in the
low--energy value of $Y_5$. The fermion--loop results of
eqs.~(\ref{boundcon1})--(\ref{boundcon6}) imply a fixed ratio
between the running coupling constants which one could, as a first
approximation, also impose as the boundary condition at the cutoff
for the full \RG evolution, e.g. $Y_1(\Lambda)=3\,Y_5(\Lambda)=30$.
Fortunately, once again, the numerical results depend very weakly
on this scaling factor, and for this purpose it could just as well
be taken to be unity.
\begin{figure}[htb]
\centerline{
\epsfysize=84ex
\rotate[r]
{\epsffile{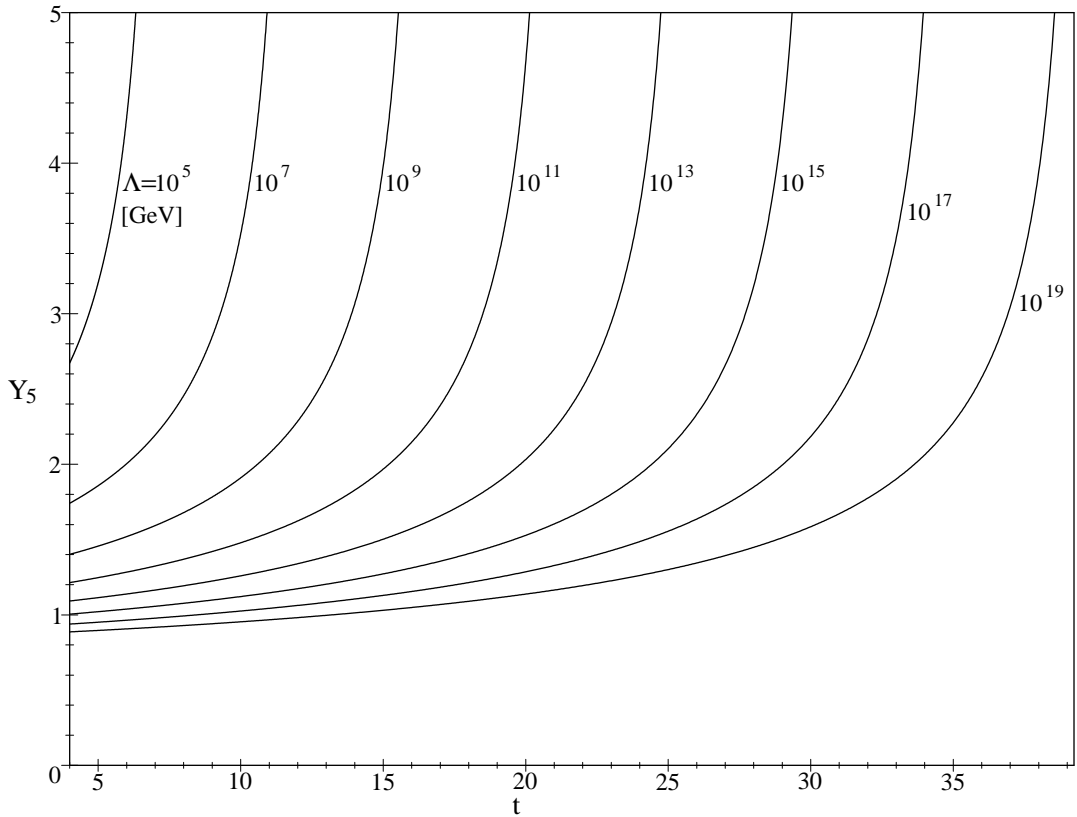}}}
\caption[]{\small\sl
Renormalization Group evolution of the $Y_5$ Yukawa coupling for
various compositeness scales $\Lambda,\; t=\ln(\mu /m_Z)$.
\label{fig:Y5fixed}}
\end{figure}
Fig.~\ref{fig:Y5fixed} shows $Y_5(\mu)$ obtained by numerically
solving eqs.~(\ref{betatrun1})--(\ref{betatrun3}) for various
values of the cutoff $\Lambda$. The heavy neutrino mass $M$ is
determined by the equation
\beq
M = Y_5(M)\cdot \hat{v}_R~,
\eeq
and for $v_R\sim \mu_R \ll \Lambda$ one finds values of
$Y_5(M)\simeq Y_5(\mu_R)$ roughly between 1 and 2.

The evolution of $Y_1$ and $Y_4$ below the parity breaking scale is
determined by the usual \SM $\beta$--functions~\cite{oneloopb}. It
turns out that the numerically most important difference between
the LR and \SM $\beta$--functions for $Y_1$ is a contribution of
$1/2\; Y_1^3$ coming from the self energy diagram with a $\phi_2^+$
scalar exchange. Since the mass of $\phi_2^+$ is not of the order
of the right--handed scale but is suppressed by a factor
$\approx 10^{-2}$ we switch to the \SM $\beta$--functions two
orders of magnitude below the parity breaking scale $\mu_R$.
\begin{figure}[htb]
\centerline{
\epsfysize=84ex
 \rotate[r]{ \epsffile{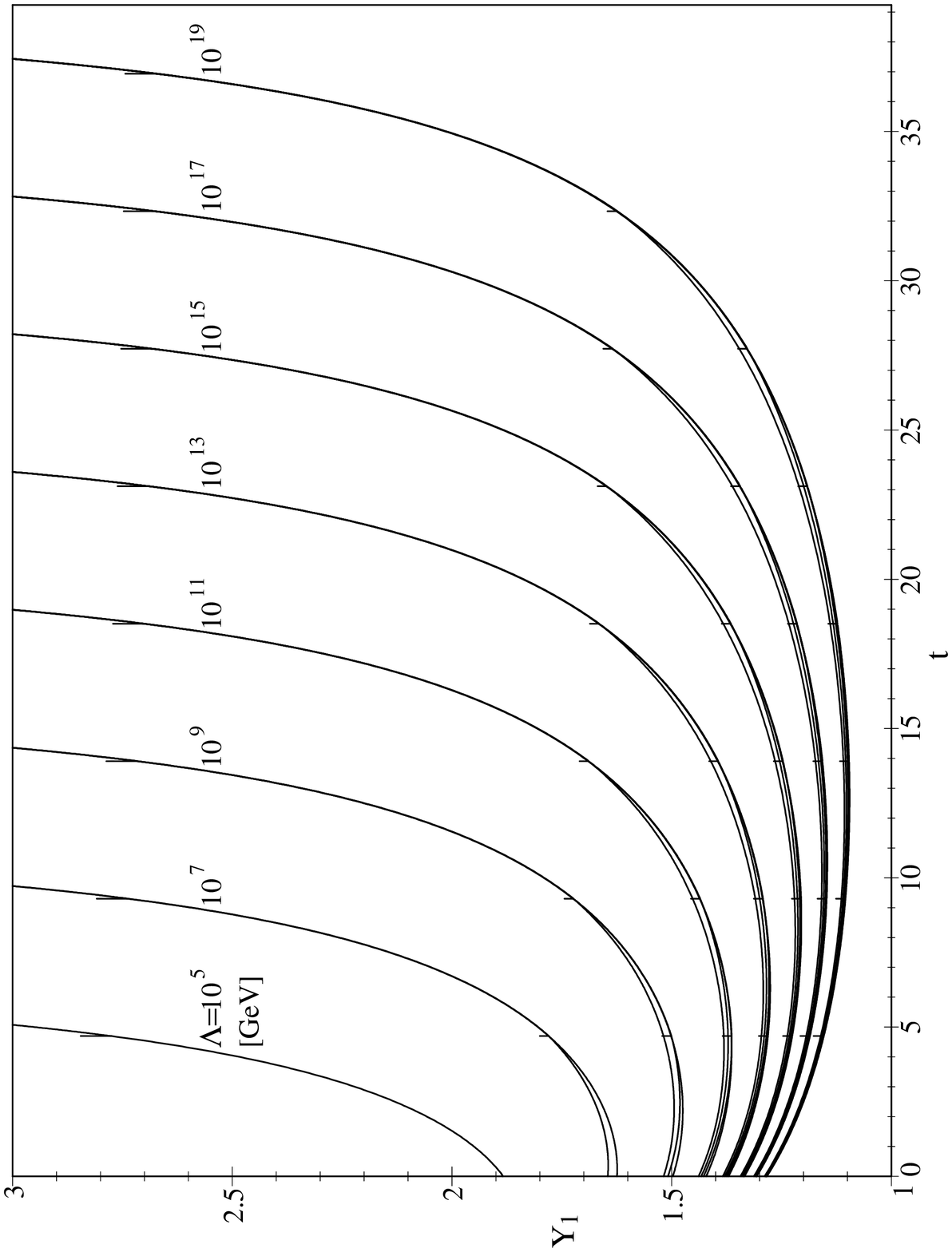} }}
\caption[]{\small\sl
Renormalization Group evolution of the $Y_1$ Yukawa coupling for
various parity breaking scales $\mu_R$ (indicated by little ticks)
and compositeness scales $\Lambda;\; t=\ln(\mu /m_Z)$.
\label{fig:Y1fixed}}
\end{figure}

In Fig.~\ref{fig:Y1fixed} we present our numerical solutions for
$Y_1(\mu)$ for various values of $\Lambda$ and $\mu_R$. One can
clearly see the infrared quasi--fixed point structure of the
solutions. The values of $Y_1$ at $t=0$ $(\mu=m_Z)$ are to some
extent sensitive to the magnitude of the cutoff but fairly
insensitive to the scale where parity breaks. This is because in
fact the previously mentioned contribution to the $Y_1$
$\beta$--function from the $\phi_2^+$ exchange makes only a
relatively small difference between $(9/2)\,Y_1^3$ in the \SM and
$5\,Y_1^3$ in the LR symmetric model with a bi--doublet. For the
same reason our top quark mass prediction in the $\kappa'=0$ case
is very similar to the one of the BHL model~\cite{BHL}, which is
too high compared to the recent experimental results~\cite{CDF,D0}.
We also conclude that the \RG analysis for the minimal left--right
symmetric model in Ref.~\cite{rothstein}, which predicted a lower
top quark mass, is wrong due to the incorrect trace terms in the
$\beta$--functions. Even for a rather high cutoff
$\Lambda\approx 10^{17}~GeV$ one obtains a top quark mass about
$229~GeV$.

\begin{figure}[htb]
\centerline{
\epsfysize=84ex
\rotate[r]
{\epsffile{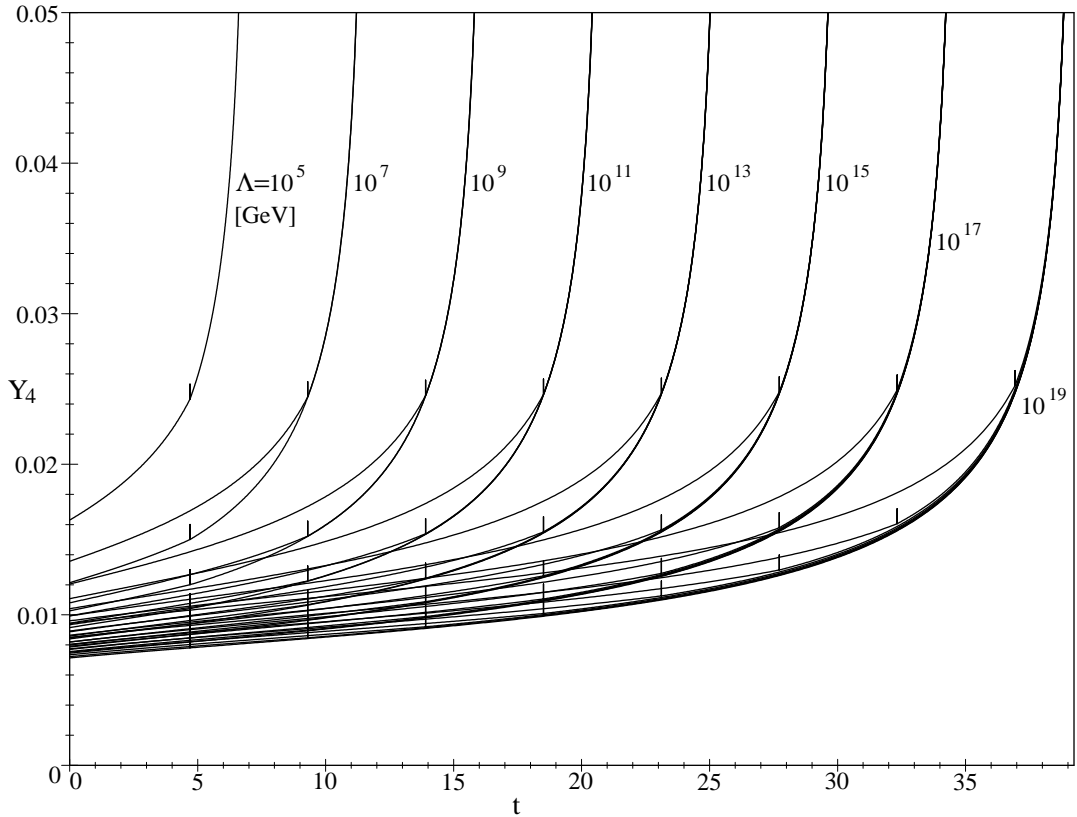}}}
\caption[]{\small\sl
Renormalization Group evolution of the $Y_4$ Yukawa coupling with the
boundary condition $Y_4 = 0.005 Y_1$ at $\Lambda$.
\label{fig:Y4fixed}}
\end{figure}

The evolution of the Yukawa coupling $Y_4$ which determines
$m_\tau$ is not governed by a fixed point; the reason for this is
that the $\tau$ lepton is too light and so does not contribute much
to the composite Higgs bi--doublet, which is driven by large $Y_1$
and not by $Y_4$. In other words, our model exhibits a top
condensate (along with a heavy--neutrino condensate) rather than
a tau condensate. One can readily obtain a suitable low--energy value
of $Y_4$ by choosing a proper value of $Y_4(\Lambda)$ (see
Fig.~\ref{fig:Y4fixed}). Thus, as we already mentioned in
Sec.~\ref{sec:bubblepred}, although $m_\tau$ is not predicted in
our model, it can be easily adjusted to the correct value.

\subsection{The general case $\kappa,\kappa' \ne 0$}

In the preceding discussion we assumed for simplicity that only one
of the two neutral components of the bi--doublet $\phi$ acquires a
VEV ($\kappa\ne 0$, $\kappa'=0$). Apparently the resulting
fixed--point value of the top quark Yukawa coupling is in this
limit outside the phenomenologically acceptable region~\cite{CDF,D0},
and moreover the assumption $\kappa'=0$ implies a zero bottom quark
mass. Evidently a more realistic scenario with $m_b \ne 0$ would
require either $\kappa' \ne 0 $ or $Y_2\ne 0$. In conventional LR
models these two conditions are unrelated and can be satisfied
separately, but in our model $Y_2\ne 0$ automatically means
$\kappa'\ne 0$ and vice versa. In this subsection we will show that
it is indeed possible to obtain viable top and bottom quark masses
for a range of values of $\kappa' \sim \kappa$.

{}From the minimization condition of the effective
potential~(\ref{VS}--\ref{VK1}) one finds two approximate solutions,
$\kappa'/\kappa\approx - Y_3/Y_4$ and
$\kappa'/\kappa\approx Y_4/Y_3$.
Without loss of generality we assume $|\kappa'| < |\kappa|$; then the
latter solution is excluded by the vacuum stability condition [see
eq.~(\ref{charged})], and only the solution
\beq
\frac{\kappa'}{\kappa} = - \frac{Y_3}{Y_4}
          + {\cal O}\(\frac{\kappa^2}{\vr2}\)
\label{y34cond}
\eeq
remains. On the other hand, to obtain $m_t \gg m_b$ for
$\kappa \sim \kappa'$ one requires the condition
\beq
 \frac{Y_2}{Y_1} \approx -\frac{\kappa'}{\kappa}
\label{y12cond}
\eeq
to be satisfied. From eq.~(\ref{y34cond}) it is seen that the ratio
of $\kappa'/\kappa$ is determined by the low energy values of $Y_3$
and $Y_4$. Since these values are not governed by infrared
quasi--fixed points and depend on the boundary conditions at the
cutoff, the ratio
\beq
\tan \beta \equiv \frac{\kappa}{\kappa'}
\eeq
is a free parameter in our model. Unfortunately, the \RG evolution
of the Yukawa couplings does not automatically yield an infrared
value of $Y_2/Y_1$ satisfying eq.~(\ref{y12cond}) as a result of,
e.g., an attractive fixed point. However, this ratio runs fairly
slowly and by choosing its initial value appropriately one can
always obtain the desirable value at low energies.

One can now study the \RG evolution of the full set of Yukawa
couplings. It turns out that the combination $\sqrt{Y_1^2+Y_2^2}$
exhibits a fixed point behavior which is similar to the one of
$Y_1$ discussed in the previous subsection. Below the right--handed
scale one should switch to the \SM $\beta$--functions of $Y_t$ and
$Y_b$ which are obvious linear combinations of $Y_1$ and $Y_2$.
Imposing the relation~(\ref{y12cond}) we find those values of
$\tan \beta$ which result in viable top quark masses, depending
on the values of the cutoff $\Lambda$ and the right--handed
scale $\mu_R$. This is in fact similar to supersymmetric or 2--Higgs
doublet models in which the top mass depends on $\tan\beta$ which
is essentially a free parameter.
\begin{figure}[htb]
\centerline{
\epsfysize=88ex
\rotate[r]
{\epsffile{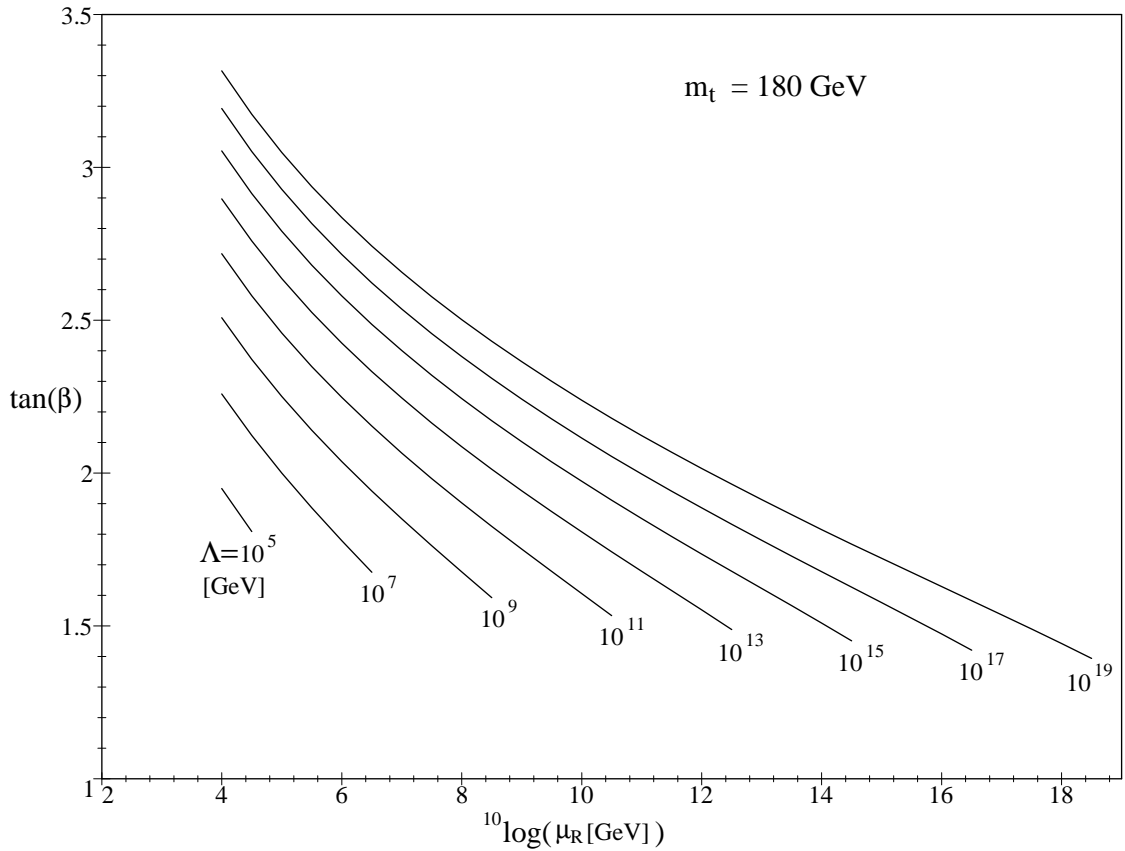}}}
\caption[]{\small\sl
Values of $\tan\beta$ for $m_t = 180~GeV$ and various magnitudes of
the cutoff $\Lambda$ and right--handed scale $\mu_R$.
\label{fig:figy12}}
\end{figure}

In Fig.~\ref{fig:figy12} we show our results for $\tan\beta$
assuming a top mass of $m_t=180~GeV$. One observes that a viable
top mass can be obtained for a large range of possible values of
the cutoff and for various parity breaking scales. This is in
contrast with the top condensate approach to the \SM\cite{BHL}
where the lowest possible (but still too high) top mass arises for
the largest possible cutoff. As can be seen from the figure, in our
model we can have a viable top quark mass for values of the
cut--off as low as $200~TeV$ and parity breaking scale about
$20~TeV$. This means that there is only a minimal amount of
fine--tuning involved and the gauge hierarchy problem gets
significantly ameliorated. If one considers different values
for the top mass, the whole set of curves in Fig.~\ref{fig:figy12}
is slightly shifted vertically, e.g., for
$m_t = (168 - 192)~GeV$~\cite{CDF,D0}
one finds an overall range for $\tan\beta$ of $(1.3 - 4.0)$.

\section{Phenomenological Considerations}
\label{sec:phen}

A detailed study of all phenomenological aspects of our model is
outside the scope of the present paper. We therefore discuss here
only a few major points.

In the general version of our model with non--vanishing $\kappa'$
we were able to obtain viable top and bottom quark masses. The
renormalization group fixed point analysis leads to fermion mass
predictions which are less constrained than in the BHL
model~\cite{BHL} since the fermions acquire their masses through
more than one VEV. The resulting masses depend on the fixed points
and on the ratio of those VEVs, in our case
$\tan\beta=\kappa/\kappa'$, which is a free parameter. The
Fermilab Tevatron results for $m_t$ can be reproduced for
$\tan\beta\simeq$1.3--4. In the limiting case of $\kappa'=0$
($\tan\beta\to \infty$) one arrives at a practically unique value
of $m_t$, which turns out to be very similar to that of BHL, i.e.
too high as compared to experiment.

In addition to the usual quarks and leptons, we have a neutral
gauge singlet fermion $S_L$. Its existence along with the
conditions $v_L=0=\sigma_0$ which follow from the minimization of
the Higgs potential result in the tau neutrino $\nu_{\tau}$ being
massless in our model. This also applies to $\nu_e$ and $\nu_{\mu}$
if the model is directly generalized to include the first two
generations of fermions\footnote{We assume here that there is one
singlet fermion $S_{Li}$ for each quark--lepton family. One can
also consider the situation when there is a unique $S_L$ for all
three families. In this case only one active neutrino would be
massless and the remaining two would have masses of the order of
$m_D$.}. The reason for this is that the model possesses a global
$U(1)$ symmetry $S_L\to e^{i\alpha}S_L$,
$\chi_L\to e^{i\alpha}\chi_L$, $\chi_R\to e^{-i\alpha}\chi_R$,
$\sigma\to e^{-2i\alpha}\sigma$. After $\chi_R^0$ develops a VEV this
symmetry gets redefined, but since $v_L=\sigma_0=0$, there still
exists an unbroken global $U(1)$ symmetry such that $\nu_{\tau}$
remains massless. The situation when this symmetry is spontaneously
broken and the properties of  the resulting Majoron were considered
(within the standard elementary Higgs mechanism of symmetry
breaking) in~\cite{AJRV}.

In the limiting case $\kappa'\to 0$ the Dirac neutrino mass term
$m_D$ vanishes and there is no neutrino mixing. However, as we
already mentioned, realistic fermion masses require
$\kappa, \kappa'\neq 0$. In this case $m_D\neq 0$ and one could
expect interesting effects of lepton flavour violation mediated
by $S_L$~\cite{BER,3E}. The reason for this is that one can in
principle have quite a sizeable Dirac neutrino mass and hence
mixing in the lepton sector without generating inadmissible heavy
physical neutrino states (in fact, the neutrinos taking part in
the usual \EW interactions remain exactly massless however large
$m_D$). Another interesting effect of $S_L$ is the possibility
of having $CP$ non--conservation in the lepton sector even though the
physical neutrinos are massless~\cite{CP}. However, as follows from
eqs.~(\ref{fmass}) and (\ref{y34cond}), due to the dynamical nature
of the LR symmetry breaking, in our model $m_D$ is suppressed by a
factor ${\cal O}(\kappa^2/v_R^2)$ as compared to the charged lepton
mass. This means that even for a right--handed scale as low as
$\sim 20~TeV$ the flavour violating and $CP$ non--conservation
effects in the lepton sector caused by the mixing of light and
heavy neutrinos may not be observable. In this respect our
dynamical model is more restrictive than the version of the same
model based on the conventional Higgs mechanism~\cite{AJRV} which
allows for a large Dirac neutrino mass term $m_D$. However, as we
shall discuss shortly, there still may exist observable effects of
lepton flavour violation mediated by $\chi_L$ and $\chi_R$ scalar
boson exchanges.

We will now discuss the implications of the Higgs sector of our
model. Since $\chi_L$ does not mix with other Higgs multiplets,
and the components of $\chi_R$ are either eaten by the right--handed
gauge bosons or heavy, our low energy Higgs sector is very similar
to that of the two--Higgs--doublet \SM (2HDSM). In our case there
are however very specific relations between the masses of scalars
(see Sec.~\ref{sec:bubblepred}) which would be a distinctive
signature for our model.

Extra Higgs and gauge boson multiplets contribute to radiative
corrections like e.g. in the $\Delta \rho$ parameter and so might
be constrained by \EW precision data. The contribution of the
mixing with $SU(2)_R$ gauge bosons is however negligible for
$v_R\gtap 20~TeV$~\cite{Altarelli}. So is the $\chi_L$ contribution
to $\Delta \rho$; this follows from the fact that the charged
and neutral components of $\chi_L$ are nearly degenerate [see
eqs.~(\ref{MH4pm}) and (\ref{MH40})]. The bi--doublet contribution to
$\Delta\rho$ coincides with the Higgs boson contributions in the
2HDSM. In general it depends on the masses of the relevant scalars
and two mixing angles, $\alpha$ and $\beta$, where $\beta$ coincides
with our $\beta\equiv \tan^{-1}(\kappa/\kappa')$ and $\alpha$ is
the mixing angle of neutral $CP$--even Higgs bosons~\cite{2HDSM}.
In our case $\alpha\simeq\beta$, and the mass of the lightest
$CP$--even scalar coincides with the mass of its $CP$--odd partner
(see eq.~(\ref{MH30}) and Appendix~\ref{apphiggs}). This reduces
the Higgs boson contributions to $\Delta\rho$ to a simple expression
\beq
\Delta\rho_{H}\approx \frac{G_F}{8\sqrt{2}\pi^2}\(M_{H_3^\pm}^2 +
M_{H_{3i}^0}^2-2\,\frac{M_{H_3^\pm}^2 M_{H_{3i}^0}^2}{M_{H_3^\pm}^2 -
M_{H_{3i}^0}^2}\ln\frac{M_{H_3^\pm}^2 }
{M_{H_{3i}^0}^2}\)\equiv \frac{G_F}{8\sqrt{2}\pi^2}\,f(M_{H_3^\pm}^2,
M_{H_{3i}^0}^2)~,
\label{drho}
\eeq
where $M_{H_3^\pm}^2$ and $M_{H_{3i}^0}^2$ are the masses of the
charged and neutral $CP$--odd Higgs bosons given by
eqs.~(\ref{MH3pm}) and (\ref{MH30}) respectively. Electro--weak
precision measurements require $f(M_{H_3^\pm}^2, M_{H_{3i}^0}^2)$
to be $\ltap 3\times (100~GeV)^2$~\cite{LE-PDG}. This imposes an
upper bound on the mass splitting between $M_{H_3^\pm}^2$ and
$M_{H_{3i}^0}^2$. In our model $M_{H_3^\pm}^2-M_{H_{3i}^0}^2=
M_{H_2^0}^2/2$ where $M_{H_2^0}$ is the \SM Higgs boson mass,
which gives
\beq
f(M_{H_3^\pm}^2,M_{H_{3i}^0}^2)\approx \left \{
\begin{array}{l}
M_{H_2^0}^4/12 M_{H_3^\pm}^2,\quad M_{H_2^0}^2\ll 2 M_{H_{3i}^0}^2\\
M_{H_2^0}^2/2,\quad \quad \quad M_{H_2^0}^2 \gg 2 M_{H_{3i}^0}^2
\end{array}\right.~.
\label{f}
\eeq
For $M_{H_2^0}^2 \ll 2 M_{H_{3i}^0}^2$ (i.e. for $M$ not too close
to its lower bound of about $20~TeV$) one then obtains the following
upper limit on the \SM Higgs boson mass: $M_{H_2^0}\ltap
[36 M_{H_3^\pm}^2\,(100~GeV)^2] ^{1/4}\approx 70\sqrt{M/TeV}$.
For example, for $M\approx 100~TeV$ this gives $M_{H_2^0} \ltap
700~GeV$. For $M_{H_2^0}^2 \gg 2 M_{H_{3i}^0}^2 $ (which means
that $M$ is close to its lower bound) one finds $M_{H_2^0}\ltap
245~GeV$. These constraints are less restrictive than those
following from the vacuum stability condition in our model.

In conventional LR models only one scalar particle, the neutral
$CP$--even Higgs boson $H_2^0$ which is the analog of the \SM one,
is light; all the others are at the right--handed scale
$v_R$~\cite{LR2,LR3,LRdesh}. In our model, only one Higgs boson
($H_1^0$) is at the right--handed scale. In addition to $H_2^0$
which is at the \EW scale we have $\chi_L^\pm$, $\chi_{Lr}^0$ and
$\chi_{Li}^0$ whose masses are typically one order of magnitude
below the right--handed scale, and $\phi_2^\pm$ whose mass is
about two orders of magnitude smaller than $v_R$. The masses of
neutral $\phi_{2r}^0$ and $\phi_{2i}^0$ are even smaller [see
eqs.~(\ref{MH3pm}) and (\ref{MH30})].

As mentioned, the vacuum stability conditions require the
right--handed scale $v_R$ to lie above $\sim 20~TeV$ in our model.
This is an order of magnitude more stringent than the bounds which
follow from phenomenological considerations (see~\cite{LS} for a
recent analysis). For $v_R$ close to its lower bound one can expect
$\phi_2^{\pm}$ to be detectable at LHC whereas $\phi_{2r}^0$ and
$\phi_{2i}^0$ may have a mass as small as $60-100~GeV$ which might
also be accessible at LEP2.

It is interesting that the Higgs sector of our model can naturally
lead to a new positive contribution to
$R_b\equiv\Gamma(Z\to b\bar{b})/\Gamma (Z\to hadrons)$ which may be
desirable in order to reconcile the experimental result
$R_b=0.2205\pm 0.0016$~\cite{Rb} with theoretical predictions. It
has been shown in~\cite{Denner1} that this can be achieved in the
two--Higgs--doublet \SM (2HDSM) with $\tan\beta\sim 50$ and
degenerate or almost degenerate neutral $CP$--odd and $CP$--even
scalars with mass $\sim 50~GeV$. In our case the role of these
scalars is naturally played by $\phi_{2r}^0$ and $\phi_{2i}^0$,
and for $Y_2 \ll Y_1$ one has $\tan \beta \approx m_t/m_b \sim 50$.
Note, however, that for such high values of $\tan\beta$ the top
quark mass turns out to be too high as compared to the CDF/D0
results.

Low--lying neutral scalars usually pose a problem in LR models since
they can mediate strong flavour--changing neutral currents. Typical
lower bounds for the masses of such scalars, which come mainly from
the $K_L-K_S$ mass difference, are of the order of a few
$TeV$~\cite{LS}. In the limit when the fermions of the first two
generations are massless which we were studying in the present
paper, there are however no flavour--changing neutral currents.
The situation in a more realistic version of the model will depend
very much on how the fermions of the first and the second
generations acquire their masses. One possibility is a radiative
mechanism induced by some horizontal interactions\footnote{A
scenario in which the masses of the bottom quark and $\tau$ lepton
are generated radiatively in the top--condensate approach was
suggested in~\cite{BM}.}. It has been argued recently that the
smallness of the light quark masses might lead naturally to
smallness of flavour--violating couplings of extra scalars to
light quarks; as a result, the bounds of the masses of these
scalars get significantly relaxed, and these masses may well be
at the \EW scale~\cite{FCNC}. Note that this does not require
introducing any discrete symmetry to suppress flavour--changing
neutral currents.

Light scalars can result in sizeable lepton--flavour violating
decays such as $\mu\to e\gamma$. For example, the contribution of
the one--loop diagrams with $S_L$ and charged $\chi_L$ or $\chi_R$
in the loop to the branching ratio $BR(\mu\to e\gamma)$ is of the
order
\beq
\vert (Y_5)_{\mu a}^\dagger (Y_5)_{ae}\vert^2\;
\frac{3\alpha}{8\pi}\frac{1}
{G_F^2 M^4}~.
\label{muegamma}
 \eeq
Assuming that the flavour--off--diagonal $Y_5$ couplings are of the
order of the diagonal ones (i.e. $\sim 1$), one finds that the
contribution of eq.~(\ref{muegamma}) becomes comparable with the
present experimental upper bound $BR(\mu\to e\gamma)<4.9\times
10^{-11}$ for $M\simeq 20~TeV$ which is just above the lower bound
on $M$ following from the vacuum stability condition. Contributions
to $BR(\mu\to e\gamma)$ from the loops with $\phi_2$ become
important if the masses of its components are at the \EW
scale~\cite{Chang}. Thus we may have observable $\mu\to e\gamma$
decay in our model.

In our phenomenological considerations we were mainly assuming that
the right--handed scale $v_R$ is not far from its lower bound. In
this case one can expect several low or intermediate scale scalars
in addition to the \SM Higgs boson to be present in the model, and
also some interesting observable effects of lepton flavour violation
mediated by the singlet fermion $S_L$ and doublet Higgs scalars
$\chi_L$ and $\chi_R$. However, as was already mentioned in
Sec.~\ref{sec:effpot}, the right--handed scale can in principle
be anywhere between $\sim 20~TeV$ and $\Lambda$ in our model.
The low values of $v_R$ would result if one requires the ``minimal
cancelation'' in eq.~(\ref{ratio}). Still, an unknown dynamics
leading to our low--energy 4-f interactions may prefer a higher
degree of cancelation, and so the right--handed scale may be very
high. In this case, unfortunately, our model will be practically
untestable.


\section{Discussion and Conclusions}
\label{sec:concl}

The model we have presented here is to our knowledge the first
successful attempt to break left--right symmetry dynamically. It
is consistent with the currently available experimental data. A
striking feature of the dynamical approach turns out to be the
fact that whether or not parity can be spontaneously broken at
low energies depends on the particle content of the model and not
on the choice of the quartic couplings in the Higgs potential,
as it is the case in the conventional approach. Our model exhibits
a tumbling scenario where the breaking of parity and $SU(2)_R$
at a right handed scale $\mu_R$ eventually drives the breaking of
the \EW symmetry at a lower scale $\mu_{EW}\sim 100~GeV$.

Our model has 9 input parameters (eight four--fermion couplings
$G_1,...,G_8$ and the scale of new physics $\Lambda$) in terms of
which we calculate 16 physical observables (5 fermion masses,
8 Higgs boson masses and 3 VEVs $\kappa$, $\kappa'$ and $v_R$),
so there are $16-9=7$ predictions. First, the symmetry breaking
is studied and the resulting predictions are derived in bubble
approximation. The predictions for the fermion masses are then
renormalization group improved thus including all the \EW and QCD
corrections to one--loop order. Unlike in the minimal BHL model,
the top quark is not predicted as an infrared quasi--fixed point
value of the Yukawa coupling times the known \EW VEV. The mass
formula is more complex here and in addition only the sum of
squares of the bi--doublet VEV's is fixed to be
$\kappa^2+\kappa'^2\simeq (174~GeV)^2$ while
$\tan\beta=\kappa/\kappa'$ is essentially a free parameter. Our
model gives a viable top quark mass value for
$\tan\beta\simeq $1.3--4 and exhibits a number of low and
intermediate scale Higgs bosons. Furthermore it predicts relations
between masses of various scalars and between fermion and Higgs
boson masses which are in principle testable. If the right--handed
scale $\mu_R$ is of the order of a few tens of $TeV$, the neutral
$CP$--even and $CP$--odd scalars $\phi_{2r}^0$ and $\phi_{2i}^0$
can be even lighter than the \EW Higgs boson. In fact, they could
be as light as $\sim 60-100~GeV$ and so might be observable at LEP2.
Such light $\phi_{2r}^0$ and $\phi_{2i}^0$ might also provide a
positive contribution to
$R_b=\Gamma(Z\to b\bar{b})/\Gamma(Z\to hadrons)$ which could account
for the discrepancy between the LEP observations and the \SM
predictions.

Our model is formulated in terms of attractive 4-f interactions
which are allowed by the symmetries and which trigger condensation
if the couplings are strong enough. Since 4-f interactions are not
renormalizable in the usual sense they should be regarded as
effective low energy approximations of heavy degrees of freedom
which have been integrated out. One could try to generate the
interactions from renormalizable gauge theories as it has been
done for the case of the BHL model. This goes however beyond the
scope of the present paper and will be addressed elsewhere.

\ \vskip .5cm

We are grateful to Zurab~Berezhiani, Darwin~Chang, Richard~Dawid,
Rabi~Mohapatra, Ulrich~Nierste, Graham~Ross, Goran~Senjanovi\'{c},
Alexei~Smirnov and Mikhail~Voloshin for useful discussions. Thanks
are also due to Stefano~Ranfone, who participated in the earlier
stage of the work. This work was supported in part by the Spanish
DGICYT under grants PB92-0084 and SAB93--0090 (E.A. and J.V.) and
by the German DFG under contract number Li519/2--1 (M.L. and E.S.).

\newpage

\section*{Appendix}
\appendix
\section{Auxiliary fields and 4-f terms} \label{appLag}
In the auxiliary field formalism the 4--fermion interaction terms of
eqs.~(\ref{Lint1first}) and (\ref{Lint3}) are rewritten in terms of
the static auxiliary scalar fields $\phi$, $\chi_L$, $\chi_R$ and
$\sigma$ with mass terms and Yukawa couplings but no kinetic terms
(\ref{Laux}). Since the Lagrangian is quadratic in the newly
defined scalar fields they can always be integrated out in the
functional integral. Equivalently, one can use the equations of
motion of the auxiliary scalar fields to express them in terms of
the fermionic degrees of freedom. After substituting the result
into the auxiliary Lagrangian (\ref{Laux}) one recovers the original
4-f structure. Direct comparison gives the following expressions
for the 4-f couplings $G_a$ in terms of Yukawa couplings and bare
mass terms of the auxiliary scalar fields:

\bea
G_1 & = &
\frac{1}{M_1^4-M_2^4}\[M_1^2\,(Y_1^2+Y_2^2)-2M_2^2\,Y_1Y_2 \]
      \nonumber\\
G_2 & = &
\frac{1}{M_1^4-M_2^4}\[M_1^2\,Y_1Y_2-\frac{1}{2}M_2^2\,(Y_1^2+Y_2^2)
   \] \nonumber\\
G_3 & = &
\frac{1}{M_1^4-M_2^4}\[M_1^2\,(Y_3^2+Y_4^2)-2M_2^2\,Y_3Y_4 \]
      \nonumber\\
G_4 & = &
\frac{1}{M_1^4-M_2^4}\[M_1^2\,Y_3Y_4-\frac{1}{2}M_2^2\,(Y_3^2+Y_4^2)
   \] \nonumber\\
G_5 & = &
\frac{1}{M_1^4-M_2^4}\[M_1^2\,(Y_1Y_3+Y_2Y_4)-M_2^2\,(Y_1Y_4+Y_2Y_3)
   \] \nonumber\\
G_6&= & \frac{1}
{M_1^4-M_2^4}\[M_1^2\,(Y_1Y_4+Y_2Y_3)-M_2^2\,(Y_1Y_3+Y_2Y_4)\]
      \nonumber\\
G_7 & = & \frac{Y_5^2}{M_0^2}~, \quad \quad G_8 = \frac{Y_6^2}{M_3^2}
\label{G4f}
\eea
Of 10 parameters $Y_1,...,Y_6$, $M_0^2,...,M_3^2$, three are
redundant: they can be eliminated by rescaling the fields $\phi$,
$\sigma$ and $(\chi_L, \chi_R)$. We use this freedom to choose
$M_1=M_3=M_0$, and in addition we could, for example, fix $M_0$ or
one of the bare Yukawa couplings. Therefore we have only seven
physical parameters to describe eight 4-f couplings $G_1,...,G_8$.
This means that the set of auxiliary fields which we introduced is
insufficient to describe the most general set of 4-f terms in
eqs.~(\ref{Lint1first}) and (\ref{Lint3}) (note however that new
physics responsible for these 4-f terms need not produce the most
general case). It is easy to see that the number of physical
parameters in the $(\chi_L,\chi_R)$ and $\sigma$ sectors corresponds
to the number of the 4-f terms, so the problem lies in the $\phi$
sector. In fact, with one bi--doublet only 4 of 6 couplings
$G_1,...,G_6$ are independent (using eqs.~(\ref{G4f}) one can make
sure that, e.g., $G_5$ and $G_6$ can be expressed through
$G_1,...G_4$). To describe the most general case one needs at
least two bi--doublets. However, in the limit $\kappa'=0$ we have
$Y_2=Y_3=0$ and $M_2=0$, therefore $G_2=G_4=G_5=0$,
$G_6=\sqrt{G_1 G_3}$, and the number of physical parameters in
$L_{aux}$ exactly coincides with the number of the independent 4-f
couplings.

\section{The effective potential}  \label{appeffpot}
At energies below the cutoff $\Lambda$ the auxiliary fields will
acquire gauge invariant kinetic terms, quartic interactions and
mass corrections via fermion and gauge boson loops. The kinetic
terms and mass renormalization can be derived from the 2--point
scalar Green functions, whereas the quartic couplings are given by
the 4--point functions (see Figs.~\ref{fig:twopt} and
\ref{fig:fourpt}).

The effective Higgs potential $V_{\mbox{eff}}$ of the model can be
calculated directly from Figs.~\ref{fig:twopt} and \ref{fig:fourpt}.
However, a more convenient way to calculate  $V_{\mbox{eff}}$ is to
consider the full one--loop Coleman--Weinberg effective
potential~\cite{CW}. If one substitutes the VEVs for the scalar
fields and expresses the masses of all fermions and gauge bosons
in terms of these VEVs, one arrives at a very simple expression for
the Coleman--Weinberg  potential,
\beq
V_{CW}=
\sum_p \eta_p \cdot \frac{1}{32\pi^2}\int\limits_{0}^{\Lambda^2}
dk^2\,k^2\ln\[1+\frac{m_p^2}{k^2}\]~,
\label{V3}
\eeq
where $m_p$ is the mass of the p$th$ particle and $\eta_p$ is related
to the number of degrees of freedom:
\beq
\eta_p = \left\{\begin{array}{rl}
 1 & \mbox{for scalar particles} \\
-4 & \mbox{for Dirac fermions, multiplied by $N_c$ for quarks} \\
-2 & \mbox{for Majorana fermions} \\
 3 & \mbox{for neutral gauge bosons} \\
 6 & \mbox{for charged gauge bosons} \end{array} \right.
\label{V4}
\eeq
In our calculation of $V_{\mbox{eff}}$ no contributions with Higgs
bosons in the loops will be included. The reason for this is that
the scalars are composite particles and their very existence is (to
the leading order) the result of the one--loop effective potential
or, in 4-f language, due to the infinite bubble--sum (infinite loop
order) of the constituent fermions. The loops with propagating
Higgs scalars therefore correspond to a mixed loop order, and due
to double counting problems it is difficult to self--consistently
take into account the feedback of propagating Higgs bosons into
the effective potential. However, the propagating Higgs effects
can be consistently incorporated in the \RG approach which is
discussed in Sec.~\ref{sec:fixp}.

Instead of using the full one--loop Coleman--Weinberg potential one
can truncate it so as to keep the terms up to and including quartic
terms in the fields, since higher order contributions are finite in
the limit $\Lambda\to\infty$ and so relatively unimportant. While
the full Coleman--Weinberg potential of eq.~(\ref{V3}) is
infrared--finite, the truncated one is infrared--divergent, and
so one has to introduce an infrared cutoff $\mu$. Integrating over
the momenta $\mu^2\leq k^2\leq\Lambda^2$ in (\ref{V3}) is equivalent
to integrating out the field degrees of freedom with high momenta
and so results in an effective Lagrangian at the scale
$\mu$~\cite{BHL,HLP,Luty,Bando} in the sense of Wilson's
renormalization group approach. For energy scales $\mu$
lower than the masses of the particles in the loops the scale $\mu$
should be replaced by the relevant particle masses.

Using this approach one arrives at the following formula for the
effective potential:
\bea
V_{\mbox{eff}} \Big|_{VEV}& =
 & M_0^2(v_L^2+v_R^2) + M_1^2(\kap2+\kapp2)
 +2M_2^2\kappa\kappa' + M_3^2 \sigma_0^2 \nonumber\\
 & - & \frac{1}{32\pi^2}\bigg[ 2(m_1^2+m_2^2+m_3^2)+4(N_c m_t^2
    + N_c m_b^2 + m_{\tau}^2) \nonumber \\
 & & \hspace{5em}-6(m_W^2+m_{W'}^2)-3(m_Z^2+m_{Z'}^2) \bigg]
      (\Lambda^2-\mu^2) \nonumber \\
 & + & \frac{1}{32\pi^2}\bigg[m_1^4+m_2^4+m_3^4+
2(N_c m_t^4+N_c m_b^4+m_{\tau}^4) \nonumber \\
 & &\hspace{5em}-3(m_W^4+m_{W'}^4)-\frac{3}{2}(m_Z^4+m_{Z'}^4)\bigg]
    \lnlambda
\label{V5}
\eea
Here $m_{1}$, $m_2$ and $m_3$ are the eigenvalues of the neutrino
mass matrix $M_{\nu}$ of eq.~(\ref{numass1}). The fermion masses
can be expressed in terms of the scalar VEVs using
eqs.~(\ref{fmass}) and (\ref{yukawas}). The sums of the squares and
fourth powers of neutrino masses can be obtained by taking the
traces of $M_{\nu}^2$ and $M_{\nu}^4$ respectively. This gives
\bea
m_1^2+m_2^2+m_3^2 & = & 4Y_6^2\sigma_0^2+2(Y_3\kappa+Y_4\kappa')^2
                        +2Y_5^2(v_L^2+v_R^2) \nonumber \\
m_1^4+m_2^4+m_3^4 & = & 16\[Y_6^4\sigma_0^4+
                        Y_6^2 Y_5^2\sigma_0^2(v_L^2+v_R^2)
  +Y_6 Y_5^2(Y_3\kappa+Y_4\kappa')\sigma_0 v_L v_R \] \nonumber \\
 & & +2 \[ (Y_3\kappa+Y_4\kappa')^2 +Y_5^2 (v_L^2+v_R^2) \]^2
\label{sums}
\eea
The gauge boson masses in the LR model with a bi--doublet $\phi$
and doublets $\chi_L$ and $\chi_R$ are
\bea
m^2_{W'/W}& = &\frac{g_2^2}{2}(\kap2+\kapp2)+\frac{g_2^2}{4}
 (v_L^2+v_R^2)
 \pm \frac{g_2^2}{4}\sqrt{(v_L^2-v_R^2)^2+16(\kappa\kappa')^2}~, \\
\label{mWfull}
m^2_{Z'/Z} &=& \frac{g_2^2}{2}(\kap2+\kapp2)+\frac{g_2^2+g_1^2}{4}
(v_L^2+v_R^2)
\pm \frac{1}{4}\bigg[ (g_2^2+g_1^2)^2(v_L^2+v_R^2)^2 \nonumber \\
& & +4g_2^4(\kap2+\kapp2)^2-4g_2^2g_1^2(\kap2+\kapp2)(v_L^2+v_R^2)
-4g_2^2(g_2^2+2g_1^2) v_L^2 v_R^2 \bigg]^\frac{1}{2} \;\;.
\label{mgaugebos}
\eea
Note that the $U(1)_{B-L}$ gauge coupling $g_1$ is different from the
standard--model $U(1)$ coupling $g'$:
\beq
g_1^2 = {g'}^{2}\frac{1-s_W^2}{1-2s_W^2}~, \quad \quad
s_W^2 \equiv \sin^2\theta_W = \frac{g_1^2}{g_2^2+2g_1^2}~,
\label{relation}
\eeq
where $\theta_W$ is the Weinberg angle.

The procedure outlined above yields the one--loop effective
potential in terms of the scalar VEVs, from which the
potential in terms of the fields can be recovered. This gives
\bea
V_{\mbox{eff}}&= &\tilde{M}_0^2(\chi_L^\dagger\chi_L
          +\chi_R^\dagger\chi_R)
          +\tilde{M}_1^2 \tr{(\phi^\dagger \phi)}
          +\frac{\tilde{M}_2^2}{2}
           \tr{(\phi^\dagger \tilde{\phi}+h.c.)}+\tilde{M}_3^2
           \sigma^\dagger\sigma \nonumber \\
& & +\lambda_1[(\chi_L^\dagger \chi_L)^2+(\chi_R^\dagger \chi_R)^2]
    +2\lambda_2 (\chi_L^\dagger \chi_L)(\chi_R^\dagger \chi_R)
\nonumber\\
& &  + \frac{1}{2}\lambda_3[\chi_L^\dagger(Y_3\phi+
                            Y_4 \tilde{\phi})\chi_R
       \sigma^\dagger + h.c.] \nonumber \\
& &  +\lambda_4[\chi_L^\dagger(Y_3\phi+Y_4 \tilde{\phi})
                    (Y_3\phi^\dagger+Y_4\tilde{\phi}^\dagger)\chi_L
           +\chi_R^\dagger(Y_3\phi^\dagger+Y_4\tilde{\phi}^\dagger)
                      (Y_3\phi+Y_4 \tilde{\phi})\chi_R] \nonumber \\
& & +\lambda_5(\chi_L^\dagger \chi_L
    +\chi_R^\dagger \chi_R)\tr(\phi^\dagger\phi)
+\lambda_6(\chi_L^\dagger\chi_L+\chi_R^\dagger \chi_R)\sigma^\dagger
      \sigma\nonumber \\
& & +\lambda_7' \tr(\phi^\dagger\phi\phi^\dagger\phi)
    +\frac{1}{3}\lambda_8'\tr(\phi^\dagger\tilde{\phi}
      \tilde{\phi}^\dagger\phi) \nonumber \\
& & +\frac{1}{12}\lambda_8'
        [\tr(\phi^\dagger\tilde{\phi}\phi^\dagger\tilde{\phi})+h.c.]
    +\frac{1}{2} \lambda_9
     [\tr(\phi^\dagger\phi\phi^\dagger\tilde{\phi})+h.c.]\nonumber \\
& & +\lambda_0
     [\tr(\phi^\dagger\phi)]^2 + \lambda_{10}(\sigma^\dagger\sigma)^2
\label{Veff}
\eea
with the quartic couplings
\bea
\lambda_0 & = &
\sixtpi\[-\frac{3}{2}g_2^4Z_\phi^2\] \lnlambda \nonumber \\
\lambda_1 & = & \sixtpi\[ Y_5^4-\frac{3}{16}(3g_2^4+2g_2^2g_1^2+
g_1^4)Z_\chi^2\]\lnlambda \nonumber \\
\lambda_2 & = &
\sixtpi\[ Y_5^4-\frac{3}{16}g_1^4 Z_\chi^2 \] \lnlambda
 \nonumber \\
\lambda_3 & = & \sixtpi\[ 8Y_5^2 Y_6\] \lnlambda
 \nonumber \\
\lambda_4 & = & \sixtpi\[ 2Y_5^2\] \lnlambda
 \nonumber \\
\lambda_5 & = &
\sixtpi\[ -\frac{9}{8} g_2^4 Z_\phi Z_\chi \] \lnlambda
 \nonumber \\
\lambda_6 & = & \sixtpi\[ 8Y_5^2 Y_6^2 \] \lnlambda
 \nonumber \\
\lambda_7' & =&\sixtpi\[ N_c
(Y_1^4+Y_2^4)+(Y_3^4+Y_4^4)\]\lnlambda \nonumber \\
\lambda_8' & = & \sixtpi\[ 12(N_cY_1^2Y_2^2+Y_3^2Y_4^2)\] \lnlambda
\nonumber \\
\lambda_9 & = & \sixtpi\[ 4[N_cY_1Y_2(Y_1^2+Y_2^2)+
                Y_3Y_4(Y_3^2+Y_4^2)] \]
 \lnlambda\nonumber \\
\lambda_{10} & = & \sixtpi\[ 8 Y_6^4 \] \lnlambda\;,
\label{Lambda}
\eea
and the effective mass terms are given by
eqs.~(\ref{M02})--(\ref{M32}). Note that all the quartic couplings
except $\lambda_3$ and $\lambda_4$ are proportional to the fourth
power of Yukawa and/or gauge couplings. The reason for the
different structure of $\lambda_3$ and $\lambda_4$ is that they
enter into eq.~(\ref{Veff}) being multiplied by Yukawa couplings.

To minimize the effective potential and find the vacuum of our
model we will also need the effective potential in terms of the
VEVs of the scalar fields:
\bea
V_{\mbox{eff}} \Big|_{VEV}& = &
\tilde{M}_0^2(v_L^2+v_R^2)+\tilde{M}_1^2(\kap2+\kapp2)
+2 \tilde{M}_2^2\kappa \kappa'+ \tilde{M}_3^2
   \sigma_0^2 \nonumber \\
& & +\lambda_1(v_L^4+v_R^4)+\lambda_2 \, 2v_L^2 \vr2
+\lambda_3 v_L v_R \sigma_0 (Y_3 \kappa+Y_4 \kappa') \nonumber\\
& &+\lambda_4 (Y_3 \kappa+Y_4 \kappa')^2(v_L^2+v_R^2)
   +\lambda_5 (\kap2+\kapp2)(v_L^2+v_R^2)
   +\lambda_6 (v_L^2+v_R^2)\sigma_0^2 \nonumber \\
& &+ \lambda_7 (\kappa^4+\kappa'^4)
+\lambda_8 \kap2 \kapp2
+\lambda_9 \kappa \kappa' (\kap2+\kapp2)
+\lambda_{10} \sigma_0^4~.
\label{Vvevs}
\eea
Here
\beq
\lambda_7=\lambda_7'+\lambda_0~,\quad \quad
\lambda_8=\lambda_8'+2\lambda_0~.
\label{newlambdas}
\eeq
Differentiating (\ref{Vvevs}) with respect to the VEVs $\sigma_0$,
$v_R$, $v_L$, $\kappa$ and $\kappa'$ gives the extremum conditions
(\ref{VS})--(\ref{VK1}) for $V_{\mbox{eff}}$. They constitute the gap
equations in our model in the auxiliary field approach. In order
for an extremum of the potential to be a true minimum, the matrices
of second derivatives of $V_{\mbox{eff}}$ with respect to the fields
must be positive definite. These matrices are studied in
Appendix~\ref{apphiggs}.

\section{Renormalization} \label{apprenorm}
The wave function renormalization constants of the composite scalar
fields in bubble approximation can be directly obtained from the
2--point Green functions given by the diagrams of
Fig.~\ref{fig:twopt}. The results are presented in
eqs.~(\ref{zfactors})--(\ref{zfactors3}). To derive the physical
predictions, it is convenient to absorb the $Z$--factors into the
definitions of the scalar fields so as to bring their kinetic
terms into the canonical form. This amounts to the following
re--definition of the fields and the parameters of the effective
Lagrangian:
\bea
\hat{\phi} \equiv \sqrt{Z_{\phi}}\phi~, &
 \hat{\tilde{\phi}} \equiv \sqrt{Z_{\phi}}\tilde{\phi}~, &
 \hat{\kappa} \equiv \sqrt{Z_{\phi}}\kappa~,\quad
 \hat{\kappa'} \equiv \sqrt{Z_{\phi}}\kappa' ~,\quad
\label{rescal1}\\
\hat{\chi} \equiv \sqrt{Z_{\chi}}\chi~, &
\hat{v}_{L,R} \equiv \sqrt{Z_{\chi}} v_{L,R}~, &
\hat{\sigma} \equiv \sqrt{Z_{\sigma}}\sigma~, \quad
\hat{\sigma}_0 \equiv {\displaystyle\sqrt{Z_{\sigma}} }\sigma_0~,
\quad
\label{rescal2}
\eea
\beq
\hat{\tilde{M}_0}^{2} \equiv\frac{1}{Z_\chi} \tilde{M}_{0}^{2}~,\quad
\hat{\tilde{M}}_{1,2}^2 \equiv\frac{1}{Z_\phi} \tilde{M}_{1,2}^{2}~,
\quad
\hat{\tilde{M}_3}^{2} \equiv\displaystyle{\frac{1}{Z_\sigma}}
\tilde{M}_{3}^{2}~,  \label{rescal3}
\eeq
\beq
\hat{Y_{i}} \equiv \frac{1}{\sqrt{Z_\phi}} Y_{i}\;\; (i=1,\ldots,4)~,
\quad
\hat{Y_{5}} \equiv \frac{1}{\sqrt{Z_\chi}} Y_{5}~, \quad
\hat{Y_{6}} \equiv \frac{1}{\displaystyle{\sqrt{Z_\sigma}} } Y_{6}
\label{rescal4}
\eeq
\beq
\hat{\lambda}_{1,2} \equiv {\displaystyle
    \frac{1}{Z_\chi^2}}\lambda_{1,2}~, \quad
\hat{\lambda}_3 \equiv {\displaystyle
    \frac{1}{Z_\chi \sqrt{Z_\sigma}} }\lambda_3~, \quad
\hat{\lambda}_{4} \equiv {\displaystyle
   \frac{1}{Z_\chi }}\lambda_{4}~, \quad
\hat{\lambda}_{5} \equiv {\displaystyle
   \frac{1}{Z_\chi Z_\phi}}\lambda_{5}~,
\label{rescal5}
\eeq
\beq
 \hat{\lambda}_6   \equiv {\displaystyle
   \frac{1}{Z_\chi Z_\sigma}}\lambda_6~, \quad
\hat{\lambda}_{0,7,8,9} \equiv {\displaystyle
   \frac{1}{Z_\phi^2}}\lambda_{0,7,8,9}~, \quad
\hat{\lambda}_{10}   \equiv {\displaystyle
   \frac{1}{Z_\sigma^2}}\lambda_{10}~.
\label{rescal6}
\eeq
\section{Higgs boson mass matrices}
\label{apphiggs}
The symmetric mass matrices of the scalar fields can be found as the
matrices of second derivatives of $V_{\mbox{eff}}$ with respect to
the field components. We quote here the unrenormalized mass
matrices; the renormalized ones can be obtained by applying the
same rule as in eq.~(\ref{rescal3}).

The mass matrix of the charged fields in the basis
($\phi^{+}_1,\;\chi^{+}_L,\;\phi^{+}_2,\;\chi^{+}_R$) is
\bea
\(M^2_\pm\)_{11} & = & \tilde{M}^2_{{1}}+{{\it
v_{R}}}^{2}{Y_{{3}}}^{2}\lambda_{{
4}}+\lambda_{{4}}{{\it v_{L}}}^{2}{Y_{{4}}}^{2}+\lambda_{{5}}{{
\it v_{L}}}^{2}+{{\it v_{R}}}^{2}\lambda_{{5}}+2\,\lambda_{{7}}
{\kappa'}^{2} \nonumber \\
 & & +2\,\lambda_{{7}}{\kappa}^{2}+\kappa\,\kappa'\,\lambda_{{9}}\\
\(M^2_\pm\)_{12}& = & {\frac {{\it v_{R}}\,\sigma_{0}\,Y_{{3}}
\lambda_{{3}
}}{2}}+\lambda_{{4}}{\it v_{L}}\,{Y_{{3}}}^{2}\kappa-\lambda_{{4}
}{Y_{{4}}}^{2}\kappa\,{\it v_{L}}\\
\(M^2_\pm\)_{13}& = & -\tilde{M}^2_{{2}}-\lambda_{{4}}{{\it
v_{L}}}^{2}Y_{{3}}Y_
{{4}}-{{\it v_{R}}}^{2}Y_{{3}}Y_{{4}}\lambda_{{4}}+2\,\lambda_{{7
}}\kappa\,\kappa'-\lambda_{{8}}\kappa\,\kappa' \nonumber \\
 & & -{\frac {\lambda_{{9}}{
\kappa'}^{2}}{2}}-{\frac {{\kappa}^{2}\lambda_{{9}}}{2}}\\
\(M^2_\pm\)_{14}& = & -{\frac {{\it v_{L}}\,\sigma_{0}\,Y_{{4}}
\lambda_{{3
}}}{2}}-\lambda_{{4}}{\it v_{R}}\,{Y_{{4}}}^{2}\kappa'+\lambda_{{4}}{
Y_{{3}}}^{2}\kappa'\,{\it v_{R}}\\
\(M^2_\pm\)_{22}& = & \tilde{M}^2_{{0}}+2\,\lambda_{{1}}{{\it
v_{L}}}^{2}+2\,
\lambda_{{2}}{{\it v_{R}}}^{2}+2\,\lambda_{{4}}Y_{{3}}\kappa\,Y_{
{4}}\kappa'+\lambda_{{4}}{Y_{{3}}}^{2}{\kappa'}^{2} \nonumber \\
 & & +\lambda_{{4}}{Y_{{
4}}}^{2}{\kappa}^{2}+\lambda_{{5}}{\kappa}^{2}+\lambda_{{5}}{
\kappa'}^{2}+\lambda_{{6}}{\sigma_{0}}^{2}\\
\(M^2_\pm\)_{23}& = & -{\frac {{\it v_{R}}\,\sigma_{0}\,Y_{{4}}
\lambda_{{3
}}}{2}}+\lambda_{{4}}{Y_{{3}}}^{2}\kappa'\,{\it v_{L}}-\lambda_{{4}}{
\it v_{L}}\,{Y_{{4}}}^{2}\kappa'\\
\(M^2_\pm\)_{24}& = & {\frac{\sigma_{0}\,\left (Y_{{4}}\kappa+Y_{{3}}
\kappa'\right )\lambda_{{3}}}{2}}\\
\(M^2_\pm\)_{33}& = &\tilde{M}^2_{{1}}+\lambda_{{4}}{{\it v_{L}}}^{2}
{Y_{{3}}}^
{2}+\lambda_{{4}}{{\it v_{R}}}^{2}{Y_{{4}}}^{2}+\lambda_{{5}}{{
\it v_{L}}}^{2}+{{\it v_{R}}}^{2}\lambda_{{5}} \nonumber \\
 & & +2\,\lambda_{{7}}{\kappa'}
^{2}+2\,\lambda_{{7}}{\kappa}^{2}+\kappa\,\kappa'\,\lambda_{{9}}\\
\(M^2_\pm\)_{34}& = & {\frac {{\it v_{L}}\,\sigma_{0}\,Y_{{3}}
\lambda_{{3}
}}{2}}-\lambda_{{4}}{Y_{{4}}}^{2}\kappa\,{\it v_{R}}+\lambda_{{4}
}{\it v_{R}}\,{Y_{{3}}}^{2}\kappa\\
\(M^2_\pm\)_{44}& = & \tilde{M}^2_{{0}}+2\,
\lambda_{{1}}{{\it v_{R}}}^{2}+2\,
\lambda_{{2}}{{\it v_{L}}}^{2}+2\,\lambda_{{4}}Y_{{3}}\kappa\,Y_{
{4}}\kappa'+\lambda_{{4}}{Y_{{3}}}^{2}{\kappa'}^{2} \nonumber \\
 & &+\lambda_{{4}}{Y_{{4}}}^{2}{\kappa}^{2}+\lambda_{{5}}{\kappa}^{2}
+\lambda_{{5}}{\kappa'}^{2}+\lambda_{{6}}{\sigma_{0}}^{2}\;\;.
\eea
The mass matrix of the neutral $CP$--odd Higgs bosons (imaginary
components of the fields) in the basis ($\phi^0_{1i},\;\phi^0_{2i},\;
\chi^0_{Li},\;\chi^0_{Ri},\;\sigma_i$) is
\bea
\(M^2_{0i}\)_{11}& = & \tilde{M}^2_{{1}}+\lambda_{{4}}\left ({{\it
v_{L}}}^{2}{Y_
{{3}}}^{2}+{{\it v_{R}}}^{2}{Y_{{3}}}^{2}\right )+\left ({{\it v_{L}
}}^{2}+{{\it v_{R}}}^{2}\right )\lambda_{{5}}+2\,\lambda_{{7}}{
\kappa}^{2} \nonumber\\
 & & +{\frac {\lambda_{{8}}{\kappa'}^{2}}{3}}+\kappa\,\kappa'\,
\lambda_{{9}}\\
\(M^2_{0i}\)_{12}& = & -\tilde{M}^2_{{2}}+\lambda_{{4}}\left (-{{\it
v_{L}}}^{2}Y
_{{3}}Y_{{4}}-{{\it v_{R}}}^{2}Y_{{3}}Y_{{4}}\right )-{\frac {2\,
\lambda_{{8}}\kappa\,\kappa'}{3}}-\frac{ {\kappa}^{2}+
{\kappa'}^{2} }{2} \lambda_{{9}}\\
\(M^2_{0i}\)_{13}& = & {\frac {{\it v_{R}}\,\sigma_{0}\,Y_{{3}}
\lambda_{{3}}}{2}}\\
\(M^2_{0i}\)_{14}& = & -{\frac {{\it v_{L}}\,\sigma_{0}\,Y_{{3}}
\lambda_{{3}}}{2}}\\
\(M^2_{0i}\)_{15}& = & {\frac {{\it v_{L}}\,{\it v_{R}}\,Y_{{3}}
\lambda_{{3}}}{2}}\\
\(M^2_{0i}\)_{22}& = & \tilde{M}^2_{{1}}+\lambda_{{4}}\left ({{\it
v_{L}}}^{2}{Y_
{{4}}}^{2}+{{\it v_{R}}}^{2}{Y_{{4}}}^{2}\right )+\left ({{\it v_{L}
}}^{2}+{{\it v_{R}}}^{2}\right )\lambda_{{5}}+2\,\lambda_{{7}}{
\kappa'}^{2} \nonumber \\
 & & +{\frac {\lambda_{{8}}{\kappa}^{2}}{3}}+\kappa\,\kappa'\,
\lambda_{{9}}\\
\(M^2_{0i}\)_{23}& = & -{\frac {{\it v_{R}}\,\sigma_{0}\,Y_{{4}}
\lambda_{{3}}}{2}}\\
\(M^2_{0i}\)_{24}& = & {\frac {{\it v_{L}}\,\sigma_{0}\,Y_{{4}}
\lambda_{{3}}}{2}}\\
\(M^2_{0i}\)_{25}& = & -{\frac {{\it v_{L}}\,{\it v_{R}}\,Y_{{4}}
\lambda_{{3}}}{2}}\\
\(M^2_{0i}\)_{33}& = & \tilde{M}^2_{{0}}+2\,\lambda_{{1}}{{\it
v_{L}}}^{2}+2\,
\lambda_{{2}}{{\it v_{R}}}^{2}+\lambda_{{4}}\left ({Y_{{3}}}^{2}{
\kappa}^{2}+2\,Y_{{3}}\kappa\,Y_{{4}}\kappa'
+{Y_{{4}}}^{2}{\kappa'}^{2}\right ) \nonumber \\
& & +\left ({\kappa}^{2}+{\kappa'}^{2}\right )\lambda_{{5}}+
\lambda_{{6}}{\sigma_{0}}^{2}\\
\(M^2_{0i}\)_{34}& = & \left({\frac{\sigma_{0}\,Y_{{4}}\kappa'}{2}}+{
\frac {\sigma_{0}\,Y_{{3}}\kappa}{2}}\right )\lambda_{{3}}\\
\(M^2_{0i}\)_{35}& = & -\left (\frac{ {\it v_{R}}\,Y_{{4}}\kappa'}{2}
+\frac{ {\it v_{R}} \,Y_{{3}}\kappa}{2}\right )\lambda_{{3}}\\
\(M^2_{0i}\)_{44}& = & \tilde{M}^2_{{0}}+2\,\lambda_{{1}}{{\it
v_{R}}}^{2}+2\,
\lambda_{{2}}{{\it v_{L}}}^{2}+\lambda_{{4}}\left ({Y_{{3}}}^{2}{
\kappa}^{2}+2\,Y_{{3}}\kappa\,Y_{{4}}\kappa'
+{Y_{{4}}}^{2}{\kappa'}^{2}\right ) \nonumber \\
 & & +\left ({\kappa}^{2}+{\kappa'}^{2}\right )\lambda_{{5}}+
\lambda_{{6}}{\sigma_{0}}^{2}\\
\(M^2_{0i}\)_{45}& = &
\left ({\frac {{\it v_{L}}\,Y_{{4}}\kappa'}{2}}+{
\frac {{\it v_{L}}\,Y_{{3}}\kappa}{2}}\right )\lambda_{{3}}\\
\(M^2_{0i}\)_{55}& = & \tilde{M}^2_{{3}}+\left ({{\it v_{L}}}^{2}+
{{\it v_{R}}}^{2} \right )\lambda_{{6}}+2\,
\lambda_{{10}}{\sigma_{0}}^{2}
\eea
The mass matrix of the neutral $CP$--even Higgs bosons (real
components of the fields) in the basis
($\sigma_r,\;\chi^0_{Lr},\;\phi^0_{2r},\;\phi^0_{1r},\;\chi^0_{Rr}$)
is
\bea
\(M^2_{0r}\)_{11}& = & \tilde{M}^2_{{3}}+\left ({{\it v_{L}}}^{2}
+{{\it v_{R}}}^{2}\right )\lambda_{{6}}+6\,\lambda_{{10}}
{\sigma_{0}}^{2} \\
\(M^2_{0r}\)_{12}& = &
\left ({\frac {{\it v_{R}}\,Y_{{4}}\kappa'}{2}}+{\frac
{{\it v_{R}}\,Y_{{3}}\kappa}{2}}\right )\lambda_{{3}}
+2\,\sigma_{0}\,{\it v_{L}}\,\lambda_{{6}}\\
\(M^2_{0r}\)_{13}& = & {\frac {{\it v_{L}}\,{\it v_{R}}\,Y_{{4}}
\lambda_{{3}}}{2}}\\
\(M^2_{0r}\)_{14}& = & {\frac {{\it v_{L}}\,{\it v_{R}}\,Y_{{3}}
\lambda_{{3}}}{2}}\\
\(M^2_{0r}\)_{15}& = &
\left ({\frac {{\it v_{L}}\,Y_{{4}}\kappa'}{2}}+{\frac
{{\it v_{L}}\,Y_{{3}}\kappa}{2}}\right )\lambda_{{3}}+2\,\sigma_{0}\,
{\it v_{R}}\,\lambda_{{6}}\\
\(M^2_{0r}\)_{22}& = & \tilde{M}^2_{{0}}+6\,\lambda_{{1}}{{\it
v_{L}}}^{2}+2\,
\lambda_{{2}}{{\it v_{R}}}^{2}
+\lambda_{{4}}\left ({Y_{{3}}}^{2}{\kappa}^{2}
+2\,Y_{{3}}\kappa\,Y_{{4}}\kappa'+{Y_{{4}}}^{2}{\kappa'}^{2}\right )
\nonumber \\
 & & +\left ({
\kappa}^{2}+{\kappa'}^{2}\right )\lambda_{{5}}
+\lambda_{{6}}{\sigma_{0}}^{2}\\
\(M^2_{0r}\)_{23}& = & {\frac {{\it v_{R}}\,\sigma_{0}\,Y_{{4}}
\lambda_{{3}}}
{2}}+\lambda_{{4}}\left (2\,{\it v_{L}}\,Y_{{3}}\kappa\,Y_{{4}}
+2\,{\it v_{L}}
\,{Y_{{4}}}^{2}\kappa'\right )+2\,{\it v_{L}}\,\kappa'\,
\lambda_{{5}}\\
\(M^2_{0r}\)_{24}& = & {\frac {{\it v_{R}}\,\sigma_{0}\,Y_{{3}}
\lambda_{{3}}}
{2}}+\lambda_{{4}}\left (2\,{\it v_{L}}\,Y_{{3}}Y_{{4}}\kappa'
+2\,{\it v_{L}}\,
{Y_{{3}}}^{2}\kappa\right )+2\,{\it v_{L}}\,\kappa\,\lambda_{{5}}\\
\(M^2_{0r}\)_{25}& = & 4\,\lambda_{{2}}{\it v_{L}}\,{\it v_{R}}
+\left ({\frac
{\sigma_{0}\,Y_{{4}}\kappa'}{2}}+{\frac {\sigma_{0}\,Y_{{3}}\kappa}
{2}}\right)\lambda_{{3}}\\
\(M^2_{0r}\)_{33}& = & \tilde{M}^2_{{1}}+\lambda_{{4}}\left ({{\it
v_{L}}}^{2}{Y_{{4}}}^
{2}+{{\it v_{R}}}^{2}{Y_{{4}}}^{2}\right )+\left ({{\it v_{L}}}^{2}
+{{\it v_{R}}}^{2}\right )\lambda_{{5}}+6\,\lambda_{{7}}{\kappa'}^{2}
\nonumber \\
 & & +\lambda_{{8}}{\kappa}^{2}+3\,\kappa\,\kappa'\,\lambda_{{9}}\\
\(M^2_{0r}\)_{34}& = & \tilde{M}^2_{{2}}+\lambda_{{4}}\left ({{\it
v_{R}}}^{2}Y_{{3}}
Y_{{4}}+{{\it v_{L}}}^{2}Y_{{3}}Y_{{4}}\right )+2\,
\lambda_{{8}}\kappa\,
\kappa'+\left ({\frac {3\,{\kappa}^{2}}{2}}
+{\frac {3\,{\kappa'}^{2}}{2}}\right )\lambda_{{9}}\\
\(M^2_{0r}\)_{35}& = & {\frac {{\it v_{L}}\,\sigma_{0}\,Y_{{4}}
\lambda_{{3}}}
{2}}+\lambda_{{4}}\left (2\,{\it v_{R}}\,Y_{{3}}\kappa\,Y_{{4}}+2\,
{\it v_{R}}\,{Y_{{4}}}^{2}\kappa'\right )+2\,{\it v_{R}}\,\kappa'\,
\lambda_{{5}}\\
\(M^2_{0r}\)_{44}& = & \tilde{M}^2_{{1}}+\lambda_{{4}}\left ({{\it
v_{L}}}^{2}
{Y_{{3}}}^{2}+{{\it v_{R}}}^{2}{Y_{{3}}}^{2}\right )
+\left ({{\it v_{L}}}^{2}
+{{\it v_{R}}}^{2}\right )\lambda_{{5}}+6\,\lambda_{{7}}{\kappa}^{2}
\nonumber \\
 & & +\lambda_{{8}}{\kappa'}^{2}+3\,\kappa\,\kappa'\,\lambda_{{9}}\\
\(M^2_{0r}\)_{45}& = & {\frac {{\it v_{L}}\,\sigma_{0}\,Y_{{3}}
\lambda_{{3}}}
{2}}+\lambda_{{4}}\left (2\,{\it v_{R}}\,{Y_{{3}}}^{2}\kappa
+2\,{\it v_{R}}\,
Y_{{3}}Y_{{4}}\kappa'\right )+2\,{\it v_{R}}\,\kappa\,\lambda_{{5}}\\
\(M^2_{0r}\)_{55}& = & \tilde{M}^2_{{0}}+6\,\lambda_{{1}}{{\it
v_{R}}}^{2}+2\,
\lambda_{{2}}{{\it v_{L}}}^{2}+\lambda_{{4}}
\left ({Y_{{3}}}^{2}{\kappa}^{2}
+2\,Y_{{3}}\kappa\,Y_{{4}}\kappa'+{Y_{{4}}}^{2}{\kappa'}^{2}\right )
\nonumber \\
 & & +\left ({\kappa}^{2}+{\kappa'}^{2}\right )\lambda_{{5}}
+\lambda_{{6}}{\sigma_{0}}^{2}\;\;,
\eea

In the minimum of the effective potential all the Higgs boson mass
matrices must be positive semi--definite, i.e. must have only
positive or zero eigenvalues (vacuum stability conditions). This
is equivalent to the requirement that all the physical Higgs boson
masses be non--negative. Zero eigenvalues correspond to the
Goldstone bosons; if all of them are eaten by gauge bosons, no
zero--mass scalars will be present in the physical spectrum of the
model.

A matrix is positive semi--definite if and only if all its principal
minors are non--negative; this, in particular, means that all the
diagonal elements of the scalar mass matrices must be non--negative.
Let us now consider the principal minor $\Delta$ corresponding to the
sub--matrix of $M^2_{0r}$ acting in the basis
$(\chi_{Rr}^0,\;\sigma_r)$. Assuming $\sigma_0\neq 0$, $v_R\neq 0$
and using the first derivative conditions (\ref{VS})--(\ref{VK1})
one can rewrite the elements of this sub--matrix in the following
form:
\bea
\(M^2_{0r}\)_{55} & = &
4\lambda_1 v_R^2+2(\lambda_2-\lambda_1)v_L^2 \approx
4\lambda_1 v_R^2~,
\label{M55}\\
\(M^2_{0r}\)_{11} & = & 4\lambda_{10}\sigma_0^2
+2(\lambda_2-\lambda_1)\frac{v_R^2}{\sigma_0^2}v_L^2~,
\label{M11}\\
\(M^2_{0r}\)_{15} &=&
2\lambda_6 v_R\sigma_0-2(\lambda_2-\lambda_1)\frac{v_R}
{\sigma_0} v_L~.
\label{M15}
\eea
Using eq.~(\ref{vevseesaw}) one can show that the second term in
(\ref{M15}) is always small as compared to the first one (note that
eq.~(\ref{vevseesaw}) implies $v_L \ll \sigma_0$ when both $v_L$ and
$\sigma_0$ are non--zero). One therefore finds
\beq
\Delta \equiv \(M^2_{0r}\)_{55} \(M^2_{0r}\)_{11}-\(M^2_{0r}\)_{15}^2
\approx 4\frac{v_R^2}{\sigma_0^2}\[2\lambda_1(\lambda_2-\lambda_1)
v_R^2v_L^2+(4\lambda_1\lambda_{10}-\lambda_6^2)\sigma_0^4\]~.
\label{Delta}
\eeq
Since $(4\lambda_1\lambda_{10}-\lambda_6^2)\approx ({}-32Y_5^4 Y_6^4)
[\ln(\Lambda^2/\mu^2)/16\pi^2]^2 <0$, the minor $\Delta$ can only be
positive for small enough $\sigma_0$. Using again
eq.~(\ref{vevseesaw}) one finds the following condition:
\beq
\sigma_0^2<
\frac{\lambda_3^2}{32(\lambda_2-\lambda_1)\lambda_{10}}m_D^2~.
\label{sigmacond}
\eeq
Solving eqs.~(\ref{VS})--(\ref{VK1}) for $\sigma_0$ assuming
non--vanishing $v_L$ and $\sigma_0$ gives
\beq
\sigma_0^2=\frac{\(2\tilde{M}_3^2\lambda_1-\tilde{M}_0^2\lambda_6\)
-\[\frac{\lambda_3^2\lambda_1}{4(\lambda_2-\lambda_1)}+\lambda_4
\lambda_6\] m_D^2-\lambda_5\lambda_6(\kappa^2+\kappa'^2)}
{\lambda_6^2-4\lambda_1\lambda_{10}}~.
\label{sigma0}
\eeq
Analysis of this expression shows that the condition
(\ref{sigmacond}) can only be satisfied if the inequality
(\ref{condit}) is replaced by equality. Since this requires an
extreme fine--tuning of the Yukawa couplings, we do not pursue
such a possibility. Thus, for the solutions with $v_L\neq 0$,
$\sigma_0\neq 0$ we find $\Delta<0$ which means that this
solution is not a minimum of the effective potential. As we shall
shortly see, the solution of eqs.~(\ref{VS})--(\ref{VK1}) with
$v_L=0=\sigma_0$ leads to non--negative masses of all the Higgs
bosons, i.e. is a minimum of $V_{\mbox{eff}}$. Since this solution
is unique\footnote{The solution is unique for given values of the
parameters of the model, in particular, of Yukawa couplings related
to the 4-f couplings by the formulas of Appendix A. Here we discuss
the non--trivial solution which requires the conditions
(\ref{condit}) and $Y_5^2>(Y_5^2)_{cr}$ to be satisfied. The
trivial solution corresponds to sub--critical values of the
Yukawa couplings and so does not coexist with the non--trivial one.}
it describes the true vacuum of the model in bubble approximation.

Let us now rewrite the scalar mass matrices substituting the
first--derivative conditions (i.e. the solutions of
eqs.~(\ref{VS})--(\ref{VK1}) with $v_L=0=\sigma_0$). We start with
the charged scalar fields. The fields $\chi_L^\pm$ are no longer
mixed with the rest of the charged Higgs bosons, i.e. they are
mass eigenstates. Their mass is given by
\bea
M_{\chi_L^\pm}^2 &=&
2(\lambda_2-\lambda_1)v_R^2+\lambda_4 (Y_4^2-Y_3^2)
(\kappa^2-\kappa'^2) \nonumber \\
&=& 2(\lambda_2-\lambda_1)v_R^2+\lambda_4 (m_{\tau}^2-m_D^2)~.
\label{mass1}
\eea
The rest of the charged scalar mass matrix takes the form [in the
basis $(\phi_2^+,\;\phi_1^+,\;\chi_R^+)$]
\beq
M_{\pm}^2 = \lambda_4 (Y_4^2-Y_3^2)
\(
\begin{array}{ccc}
\frac{\kappa^2}{\kappa^2-\kappa'^2}v_R^2 &
\frac{\kappa\kappa'}{\kappa^2-\kappa'^2}v_R^2 & {}-v_R\kappa\\
\frac{\kappa\kappa'}{\kappa^2-\kappa'^2}v_R^2 &
\frac{\kappa'^2}{\kappa^2-\kappa'^2}v_R^2 & {}-v_R\kappa'\\
{}-v_R\kappa & {}-v_R\kappa' & \kappa^2-\kappa'^2
\end{array}
\right)
\label{charged}
\eeq
This matrix has two zero eigenvalues corresponding to the
Goldstone bosons eaten by $W_{1,2}^\pm$:
\bea
G_1^\pm &=& \[\(\frac{\kappa^2+\kappa'^2}{\kappa^2-\kappa'^2}\)^2
+\frac{\kappa^2+\kappa'^2}{v_R^2}\]^{{}-1/2}\left\{
\frac{\kappa}{v_R}\phi_2^\pm+\frac{\kappa'}{v_R}\phi_1^\pm
+\(\frac{\kappa^2+\kappa'^2}{\kappa^2-\kappa'^2}\)\chi_R^\pm
\right\}~, \label{AppG1pm} \\
G_2^\pm &=& \frac{\kappa}{\sqrt{\kappa^2+\kappa'^2}}\phi_1^\pm -
\frac{\kappa'}{\sqrt{\kappa^2+\kappa'^2}}\phi_2^\pm ~.
\label{AppG2pm}
\eea
The massive eigenstate of the matrix (\ref{charged}) is
\beq
H_3^\pm=\[1+\frac{\kappa'^2}{\kappa^2}+\frac{(\kappa^2-\kappa'^2)^2}
{v_R^2\kappa^2}\]^{{}-1/2}\(\phi_2^\pm
+\frac{\kappa'}{\kappa}\phi_1^\pm
-\frac{\kappa^2-\kappa'^2}{v_R \kappa}\chi_R^\pm \)
\label{AppH3pm}
\eeq
with the mass
\beq
M_{H_3^\pm}^2 =  \lambda_4 (Y_4^2-Y_3^2)\[\frac{\kappa^2+\kappa'^2}
{\kappa^2-\kappa'^2}v_R^2+(\kappa^2-\kappa'^2) \]~.
\label{AppMH3pm}
\eeq
In the limit $\kappa'\to 0$ there is no LR mixing in the gauge
boson sector and so heavy and light $W_1^\pm$ and $W_2^\pm$
coincide with $W_R^\pm$ and $W_L^\pm$, respectively. In this case
the expressions for the Higgs boson mass eigenstates and
eigenvalues simplify significantly (see Sec.~\ref{sec:bubblepred}).

In the neutral $CP$--odd Higgs sector, $G_1^0=\chi_{Ri}^0$ is the
exact mass eigenstate with zero mass; it is the Goldstone boson
eaten by $Z_1^0\approx Z_R^0$. The rest of the mass matrix takes
the form [in the basis
($\phi_{1i}^0,\;\phi_{2i}^0,\;\chi_{Li}^0,\;\sigma_i$)]
\beq
M_{0i}^2=\(
\begin{array}{cccc}
A_0\kappa'^2 & A_0 \kappa\kappa' & 0 & 0\\
A_0\kappa\kappa' & A_0 \kappa^2 & 0 & 0\\
0 & 0& 2(\lambda_2-\lambda_1)v_R^2 & {}-m_D v_R/2\\
0 & 0 & {}-m_D v_R/2 & \tilde{M}_3^2+\lambda_6 v_R^2
\end{array}
\)~,
\label{imag}
\eeq
where
\beq
A_0\equiv \lambda_4 (Y_4^2-Y_3^2)\frac{v_R^2}{\kappa^2-\kappa'^2}
-2\lambda_7+\lambda_8/3 ~.
\label{A0}
\eeq
Since the gap equation for $\sigma$ is only trivially satisfied, it
decouples from the low--energy sector of the model, i.e.
$\tilde{M}_3^2$ is of the order of $\Lambda^2$ (see the discussion
in Sec.~\ref{sec:bubblepred}). It follows from eq.~(\ref{imag}) that
$\chi_{Li}^0$ is a mass eigenstate with the squared mass
$2(\lambda_2-\lambda_1)v_R^2$. The rest of the matrix has one zero
mass eigenstate
\beq
G_2^0=\frac{\kappa}{\sqrt{\kappa^2+\kappa'^2}}\phi_1^0 -
\frac{\kappa'}{\sqrt{\kappa^2+\kappa'^2}}\phi_2^0~,
\label{AppG20}
\eeq
which is the Goldstone boson eaten by $Z_2^0\approx Z_L^0$, and one
massive eigenstate
\beq
H_{3i}^0=\frac{\kappa'}{\sqrt{\kappa^2+\kappa'^2}}\phi_1^0 +
\frac{\kappa}{\sqrt{\kappa^2+\kappa'^2}}\phi_2^0 ~
\label{AppH3i0}
\eeq
with the mass
\beq
M_{H_{3i}}^2 =A_0(\kappa^2+\kappa'^2) \approx
\lambda_4 (Y_4^2-Y_3^2)v_R^2~.
\label{AppMH3i}
\eeq

The non--vanishing elements of the mass matrix of the $CP$--even
neutral Higgs bosons take the following form:
\bea
\(M_{0r}^2\)_{11}& = & \tilde{M}^2_{{3}}+{{\it v_{R}}}^{2}
\lambda_{{6}} \\
\(M_{0r}^2\)_{12}& = & \lambda_3 m_D v_R/2 \\
\(M_{0r}^2\)_{22}& = & 2(\lambda_2-\lambda_1)v_R^2 \\
\(M_{0r}^2\)_{33}& = & \lambda_4 (Y_4^2-Y_3^2)\frac{v_R^2}{\kappa^2-
\kappa'^2}\kappa^2+4\lambda_7 \kappa'^2+(\lambda_8-2\lambda_7)
\kappa^2+2\lambda_9 \kappa\kappa' \\
\(M_{0r}^2\)_{34}& = &{}-\lambda_4 (Y_4^2-Y_3^2)
\frac{v_R^2}{\kappa^2-
\kappa'^2}\kappa\kappa'+(\lambda_8+2\lambda_7) \kappa \kappa'
+2\lambda_9 (\kappa^2+\kappa'^2) \\
\(M_{0r}^2\)_{35} & = & 2(\lambda_4 Y_4 m_D+\lambda_5 \kappa')v_R \\
\(M_{0r}^2\)_{44}& = & \lambda_4 (Y_4^2-Y_3^2)\frac{v_R^2}{\kappa^2-
\kappa'^2}\kappa'^2+4\lambda_7 \kappa^2+(\lambda_8-2\lambda_7)
\kappa'^2+2\lambda_9 \kappa\kappa' \\
\(M_{0r}^2\)_{45} & = & 2(\lambda_4 Y_3 m_D+\lambda_5 \kappa)v_R \\
\(M_{0r}^2\)_{55}& = & 4\lambda_1 v_R^2
\label{real}
\eea
where, as before, the basis is
($\sigma_r,\;\chi^0_{Lr},\;\phi^0_{2r},\;\phi^0_{1r},\;\chi^0_{Rr}$).
Since the $\sigma$ field decouples, $\chi_{Lr}^0$ is an eigenstate
with squared mass $2(\lambda_2- \lambda_1)v_R^2$. The remaining
$3\times 3$ matrix can be diagonalized exactly, but the resulting
expressions for its eigenstates and eigenvalues are very cumbersome
and so we do not present them here. The simplified formulas for the
case $\kappa'=0$ are given in Sec.~\ref{sec:bubblepred}.

\section{Renormalization Group Equations}
\label{appbeta}

Here we present the one--loop \RG equations obtained in the
MS--scheme for Yukawa and gauge couplings in the the general
left--right symmetric model with $N_g$ generations of fermions
and the Higgs sector consisting of a bi--doublet $\phi$, two
doublets $\chi_L$ and $\chi_R$ and a singlet $\sigma$. We assume
that in addition to the usual fermions we have a gauge singlet
fermion $S_L$ in each generation. The presence of the singlet
scalar field $\sigma$ and singlet fermions $S_L$ will only affect
the \RG equations for the Yukawa couplings $Y_5$ and $Y_6$.

The relevant Yukawa interactions are given in eq.~(\ref{Laux}). All
the Yukawa couplings are matrices in  generation space (we suppress
the generation indices for brevity). Note that the discrete
left--right symmetry requires the Yukawa matrices $Y_1 \ldots Y_4$
and $Y_6$ to be hermitian. In addition, $Y_6$ is symmetric due to
the Majorana nature of the coupling. There are no restrictions on
the matrix $Y_5$ from symmetry arguments.

\noindent We find the following $\beta$--functions:
\bea
16 \pi^2 \beta(Y_1) & = &
2 Y_1^3 - \Big(Y_1 Y_2^2 + Y_2^2 Y_1\Big)
+ Y_1\Big[3\,\tr(Y_1^2+Y_2^2)+\tr(Y_3^2+Y_4^2) \Big]   \nonumber \\
 & &    + 2 Y_2\Big[3\,\tr (Y_1 Y_2)+\tr(Y_3 Y_4)\Big]
 - Y_1 \Big(8g_3^2+\frac{9}{2}g_2^2+\frac{1}{6}g_1^2\Big)
\label{beta1}\\
16 \pi^2 \beta(Y_2) & = &
2 Y_2^3 - \Big(Y_2 Y_1^2 + Y_1^2 Y_2\Big)
+ Y_2\Big[3\,\tr(Y_1^2+Y_2^2)+\tr(Y_3^2+Y_4^2) \Big]   \nonumber \\
 & &    + 2 Y_1\Big[3\,\tr(Y_1 Y_2) +\tr(Y_3 Y_4)\Big]
        - Y_2\Big(8g_3^2+\frac{9}{2}g_2^2+\frac{1}{6}g_1^2\Big)   \\
16 \pi^2 \beta(Y_3) & = &
2 Y_3^3 - \Big(Y_3 Y_4^2 + Y_4^2 Y_3\Big)
    +\frac{1}{2}\Big(Y_3 Y_5 Y_5^\dagger + Y_5 Y_5^\dagger Y_3\Big)
\nonumber \\
& & + Y_3\Big[3\,\tr(Y_1^2+Y_2^2)+\tr(Y_3^2+Y_4^2) \Big]
    + 2 Y_4\Big[3\,\tr (Y_1 Y_2) +\tr(Y_3 Y_4)\Big]  \nonumber \\
& & - Y_3\Big(\frac{9}{2}g_2^2+\frac{3}{2}g_1^2\Big)           \\
16 \pi^2 \beta(Y_4) & = &
2 Y_4^3 - \Big(Y_4 Y_3^2 + Y_3^2 Y_4\Big)
    +\frac{1}{2}\Big(Y_4 Y_5 Y_5^\dagger + Y_5 Y_5^\dagger Y_4\Big)
\nonumber \\
& & + Y_4 \Big[3\,\tr(Y_1^2+Y_2^2)+\tr(Y_3^2+Y_4^2) \Big]
    + 2 Y_3\Big[3\,\tr (Y_1 Y_2) +\tr(Y_3 Y_4)\Big]  \nonumber \\
& & - Y_4\Big(\frac{9}{2}g_2^2+\frac{3}{2}g_1^2\Big)         \\
16 \pi^2 \beta(Y_5) & = &
        \frac{3}{2}Y_5 Y_5^\dagger Y_5+Y_5(Y_5^\dagger Y_5)^T
+Y_5\tr(Y_5^\dagger Y_5)  \nonumber \\
& & +\Big( Y_3^2+Y_4^2 \Big) Y_5 + 2Y_5 Y_6^2
-Y_5\Big(\frac{9}{4}g_2^2+\frac{3}{4}g_1^2\Big) \\
16 \pi^2 \beta(Y_6) & = &
4 Y_6^3 + 2Y_6\tr(Y_6^2)+\Big(Y_6 \[Y_5^\dagger Y_5+
(Y_5^\dagger Y_5)^T\]+
\[Y_5^\dagger Y_5+(Y_5^\dagger Y_5)^T\] Y_6\Big)
\label{beta6}\eea
The gauge couplings in the model evolve according to
\bea
16 \pi^2 \beta(g_3) & = &\;\;g_3^3\Big( -11+\frac{4}{3}N_g \Big) \\
16 \pi^2 \beta(g_2) & = &
\;\;g_2^3\Big( -\frac{41}{6}+\frac{4}{3}N_g\Big) \\
16 \pi^2 \beta(g_1) & = &
\;\;g_1^3\Big(\frac{8}{9}N_g + \frac{1}{3}\Big)
\eea
where we set $N_g=3$ in the analysis of Sec.~\ref{sec:fixp}.


\end{document}